\documentclass[a4paper,11pt]{article}
\pdfoutput=1 

\usepackage{jheppub} 

\usepackage[T1]{fontenc} 
\usepackage[dvipsnames, table]{xcolor}
\usepackage{amsmath}
\usepackage{microtype}
\usepackage{enumitem}

\usepackage[T1]{fontenc} 
\usepackage{xspace}
\newcommand{\Py}{\textsc{Pythia}\xspace}

\newcommand{\parold}[1]{\textcolor{Violet}{\texttt{#1}}}
\newcommand{\parnew}[1]{\textcolor{purple}{\texttt{#1}}}

\usepackage{booktabs}
\usepackage{multirow}
\usepackage{array}
\usepackage{tabularx}
\usepackage[font=small]{caption}
\usepackage{xltabular}
\usepackage[nottoc]{tocbibind}

\usepackage{comment}
\usepackage[font=scriptsize]{subfig}
\usepackage[section,below]{placeins}
\renewcommand{\vec}{\boldsymbol}

\newcommand{\eqRef}[1]{eq.~\eqref{#1}\xspace}
\newcommand{\tabRef}[1]{tab.~\ref{#1}\xspace}
\newcommand{\appRef}[1]{appendix~\ref{#1}\xspace}

\newcommand{\secRef}[1]{section~\ref{#1}\xspace}

\newcommand{\SecsRef}[1]{Sections~\ref{#1}\xspace}

\newcommand{\ie}{{i.e.}\xspace}
\newcommand{\eg}{{e.g.}\xspace}

\usepackage{listings}
\usepackage{lmodern}
\title{\boldmath Closepacking effects on strangeness and baryon production at the LHC }

\author[a,1]{J. Altmann,\note{Corresponding author.}}
\author[b]{L. Bernardinis,}
\author[a]{P. Skands,}
\author[b]{V. Zaccolo,}

\affiliation[a]{School of Physics and Astronomy, Monash University, VIC-3800, Australia}
\affiliation[b]{Dipartimento di Fisica dell’Università and Sezione INFN, Via Alfonso Valerio 2, Trieste, 34127, Italy}

\emailAdd{javira.altmann@monash.edu}
\emailAdd{lorenzo.bernardinis@phd.units.it}
\emailAdd{peter.skands@monash.edu}
\emailAdd{valentina.zaccolo@units.it}

\abstract{
Data from the LHC show a rise in strange-hadron production with charged-particle multiplicity in $pp$ collisions. 
The Monte-Carlo event generator \Py, using its default Monash 
tune, instead predicts constant strangeness. We investigate a mechanism invoked during hadronization called ``string closepacking'', where overlapping strings generate a background field, here assumed to be predominantly aligned with the beam axis. This increases the effective string tension, reducing strangeness suppression and thus enhancing strangeness production.
We tune this model to LHC data and contrast it with several alternatives. We comment specifically on the challenge of simultaneously describing the non-strange $p/\pi$ ratio, and introduce a mechanism which may act to suppress this. Many of the salient particle-production ratios can be qualitatively described by this model, although the $\Xi_c/D$ ratio and the shape of $p_\perp$ spectra remain challenging to account for. 
}

\begin{document} 
\maketitle
\flushbottom
\clearpage
\section{Introduction}

The Lund string model for hadronization~\cite{Andersson:1983ia}, as implemented in the Monte Carlo event generator \Py~\cite{Sjostrand:2006za,Bierlich:2022pfr}, has been shown to well describe data from $e^+e^-$ colliders such as LEP and SLC~\cite{OPAL:1994zan,Hamacher:1995df,DELPHI:2000ahn,SLD:2003ogn,ALEPH:2003obs,L3:2004cdh,Buckley:2009bj,Skands:2014pea,Amoroso:2018qga,Jueid:2022qjg,Korneeva:2024oho} as well as having had some success in describing $pp$ collisions~\cite{Skands:2010ak,Skands:2014pea,TheATLAScollaboration:2014rfk,CMS:2022awf,Korneeva:2024oho}. Recent data from the LHC however has highlighted discrepancies between \Py predictions and $pp$ collision events, prompting further investigations into string modelling, particularly in small systems. 

The baseline Lund string model (LSM) describes the fragmentation of a string of arbitrary invariant mass $\gg \Lambda_\mathrm{QCD}$. The string is taken to be in vacuum and isolated from any (other) sources of colour fields. This applies especially well to the context of hadronic $Z$ decays, which is also the main reference case for constraining the nonperturbative modelling  parameters of the LSM. The main generalization when going to $pp$ collisions is that multiple parton-parton interactions (MPI)~\cite{Sjostrand:1987su,Sjostrand:2004ef} as well as the colour-octet nature of gluon exchanges intrinsically lead to a picture of  multiple such strings fragmenting simultaneously. Both the number and the lengths (invariant masses) of these strings can vary significantly from event to event, and when this is furthermore allowed to depend upon the proton-proton impact parameter~\cite{Sjostrand:1986ep}, one can successfully account~\cite{Sjostrand:1987su} for the very wide ($\sim$ negative-binomial) multiplicity distributions that are observed in minimum-bias collisions~\cite{UA5:1985fid}, with large long-range correlations~\cite{Uhlig:1977dc,UA5:1988osh}, as well as for the comparatively large underlying events with a hard trigger (pedestal effect)~\cite{UA1:1983hhd,AxialFieldSpectrometer:1984fhw,Field:2000dy,CDF:2004jod,ATLAS:2010kmf,CMS:2010rux,ALICE:2011ac}.

It is worth pausing to note that, in the context of the baseline LSM, these successes can be achieved while remaining within the comparatively simple paradigm that the fragmentation of a system of multiple strings can be described as the summed fragmentations of each of its constituent strings, with universal parameters --- taken to be the same as those measured in hadronic $Z$ decays. 

This, however, leads to predictions, \eg, of approximately universal ratios of the production of each hadron spin and flavour type relative to the total amount of hadron production. Such universal behaviours have been convincingly ruled out by more recent measurements, in particular by the ALICE experiment~\cite{Aamodt:2008zz,ALICE:2017jyt,ALICE:2018pal,ALICE:2023wbx} whose excellent particle identification (PID) capabilities down to very low track momenta have played a crucial role in reconstructing an almost complete set of ground-state ($L=0$) hadrons\footnote{In the sense of having measurements of at least one isospin state for almost every distinct spin and strangeness combination, sometimes more. The main exception is the $\Delta$ family of baryons whose strong decays make them particularly challenging to reconstruct.}. We also note that analogous and sometimes even more striking conclusions are reached in the heavy-flavour sector~\cite{Altmann:2024kwx}.

This has raised questions about whether hadronization in $pp$ collisions, especially at high particle densities, differs in some more fundamental way from hadronization in 
$Z$ decays. If one wishes to remain within a string context, one relatively obvious possibility (with some benefit of hindsight) is that it is the above-mentioned assumption of independently fragmenting strings which is violated. 

This question of so-called collective effects is one that has already been posed and explored in models such as Rope hadronization~\cite{Bierlich:2014xba, Bierlich:2017sxk} and string shoving~\cite{Bierlich:2016vgw, Bierlich:2020naj, Bierlich:2024odg}, which model effects on the string tension and string kinematics respectively. An interesting complementary proposal is that, in a high-density ``core'' of high-multiplicity $pp$ collisions, the string picture may break down and small droplets of quark-gluon plasma (QGP) be formed instead, while strings continue to fragment more or less as in vacuum in a low-density ``corona''~\cite{Werner:2007bf,Pierog:2013ria,Pierog:2019opp}.

Interest in collective nonperturbative effects has recently been reignited by the above-mentioned measurements of clear breakdowns of universality, such as the ratios of strange hadrons to pions from the ALICE collaboration~\cite{ALICE:2017jyt, ALICE:2018pal}, which show an increase in strange hadron production relative to charged multiplicity in $pp$ collisions. 
This rise in strangeness gives a good indication that nonperturbative strange-quark production (\eg, via string breaks) is altered in correlation with overall particle production, an effect which the baseline LSM fails to capture. 


The existing Rope hadronization model has shown promising success in describing such strangeness-enhancement effects, doing so by implementing string-tension modifications as an extension to the baseline LSM.
More specifically, the Rope model increases the string tension in higher-density string systems, resulting in a reduced strangeness suppression (\ie,~strangeness enhancement) in string breaks.
The Rope hadronization model construction however requires spacetime modelling of string breaks, which is computationally expensive and in contrast to the standard LSM fragmentation procedure which is implemented entirely in momentum space. 
As an alternative, we update and extend a simpler momentum-space model for string-tension modifications, called ``closepacking’’ which was first proposed in the context of a thermal string-breaking model~\cite{Fischer:2016zzs}; in this work, we generalize the implementation to Schwinger-type string breaking and introduce sensitivity to relative flux orientations on the basis of Casimir scaling. The model employs the same basic premise of an increased string tension as the more elaborate Rope model and hence should share a number of its features. But where the Rope model attempts to trace the explicit evolution of the fragmenting parton systems in spacetime, with multi-string ropes breaking strand by strand, the closepacking model currently only attempts to describe a time-averaged ``effective background'' for each string, as a function of rapidity along the beam axis. The main consequence is a rapidity-dependent effective tension that depends on each string's environment (relative to the beam axis).

As such, our closepacking model is currently only applicable to $pp$ collisions and not to heavy-ion ones where spacetime distances between string systems can be large compared to a typical hadron size. Such systems would be better addressed by the Rope model or by QGP-inspired models such as EPOS~\cite{Pierog:2013ria,Pierog:2019opp}. In its current formulation, the closepacking model can also not be applied to systems like $e^+e^-\to W^+W^-\to \mathrm{hadrons}$ where the beam axis plays no special role, nor will it produce any strangeness-enhancement effects among high-$p_\perp$ hadrons inside jets. We plan to lift some of these restrictions in future work but for the time being focus on minimum-bias $pp$ collisions where effects along the beam axis should dominate.

Interestingly, though strangeness-enhancement modelling has previously been studied by the Rope hadronization model, no models of string collectivity have thus far attempted to address the overprediction of the proton-to-pion ratio observed by ALICE in $pp$ collisions~\cite{ALICE:2017jyt}.
The default Monash tune~\cite{Skands:2014pea}, despite having its diquark production probabilities tuned to LEP data, overpredict the proton-to-pion ratio seen by ALICE by around 20\%, prompting the question of how the $p/\pi$ ratio appears lower at the LHC than at LEP and thus hinting at a flaw in the modelling of baryon production in denser string environments. 
This overprediction only becomes further exacerbated with the inclusion of diquark enhancement via colour ropes or so-called junction baryons in the QCD Colour Reconnection (CR) model~\cite{Christiansen:2015yqa}.
Production of these junction baryons via so-called string junctions~\cite{Sjostrand:2002ip,Altmann:2024odn} has been shown to be important in describing increased baryon-to-meson ratios at LHC compared to LEP, such as the ratio $(\Lambda + \bar{\Lambda}) / 2K_S^0$  ($0.191\pm 0.004$ at 91~GeV LEP~\cite{ParticleDataGroup:2024cfk}, and $0.267\pm 0.002$ at 7~TeV LHC~\cite{CMS:2011jlm}),
and the heavy-flavoured $\Lambda^{\pm}_c/D^0$ ($0.130\pm 0.013$ at 91~GeV LEP~\cite{ParticleDataGroup:2024cfk}, and $0.440\pm 0.026$ at 13~TeV LHC~\cite{ALICE:2023sgl}).
Given the apparent importance of junction baryons, this begs the question of whether modelling of diquark production is to blame for the overpredicted $p/\pi$ ratio.
Building upon the existing popcorn mechanism~\cite{Eden:1996xi, Andersson:1984af} (which provides a more sophisticated model for diquark production than what is used by the baseline string model), in this paper we introduce a novel mechanism to decrease diquark formation in denser string environments and thus reduce proton production in $pp$ collisions.



The remainder of the paper will be structured as follows; first we will provide a brief overview of the Lund string model in \secRef{sec:LSM}, with particular focus on modelling of flavour and baryon production.
\SecsRef{sec:closePacking},  \ref{sec:strangeJunc}, and \ref{sec:popDestr} explore the theory behind the newly implemented models, with \secRef{sec:closePacking} focusing on the closepacking model, \secRef{sec:strangeJunc} the so-called ``strange junctions'' model, and \secRef{sec:popDestr} introduces the novel popcorn destructive interference model. 
Following this, in \secRef{sec:results} we will provide a more detailed outline of the aforementioned existing models which we shall use for comparisons, and highlight the key relevant \Py parameters. Then we provide a first tuning effort implementing the new models introduced in \secRef{sec:theory}, with comparisons focusing on results from the ALICE collaboration~\cite{ALICE:2017jyt, ALICE:2018pal}. Proceeding the tuning efforts, we briefly explore the implications on the heavy-flavour sector. 

For ease of reference, throughout the text technical parameters in the code are emphasized with \texttt{typewriter font}, using \parold{Violet} for existing \Py parameters while new ones introduced in this work are highlighted in \parnew{Burgundy}. 

\section{Theoretical modelling}
\label{sec:theory}

\subsection{Basics of the Lund String Model
\label{sec:LSM}}
The Lund String model (LSM) is a semi-classical hadronization model based on QCD confinement and relativistic string dynamics.
Lattice QCD results~\cite{Bali:1992ab} show a linear potential between static colour charges at non-perturbative scales, which can be parameterized phenomenologically, \eg, by the Cornell potential~\cite{Eichten:1978tg}. 
Such a linear potential is equivalently interpreted as a string with a constant tension, thus allowing the colour confinement field to be represented as a 1+1 dimensional string obeying relativistic string dynamics~\cite{Andersson:2001yu} with characteristic string tension $\kappa \sim 1$~GeV/fm. 
These confining potentials -- or so-called Lund strings  -- are expected to form between colour-connected charges, \ie, colour charges that form overall colour-singlet states.

\begin{figure}[t]
\centering
        \includegraphics[width = 0.9\textwidth]{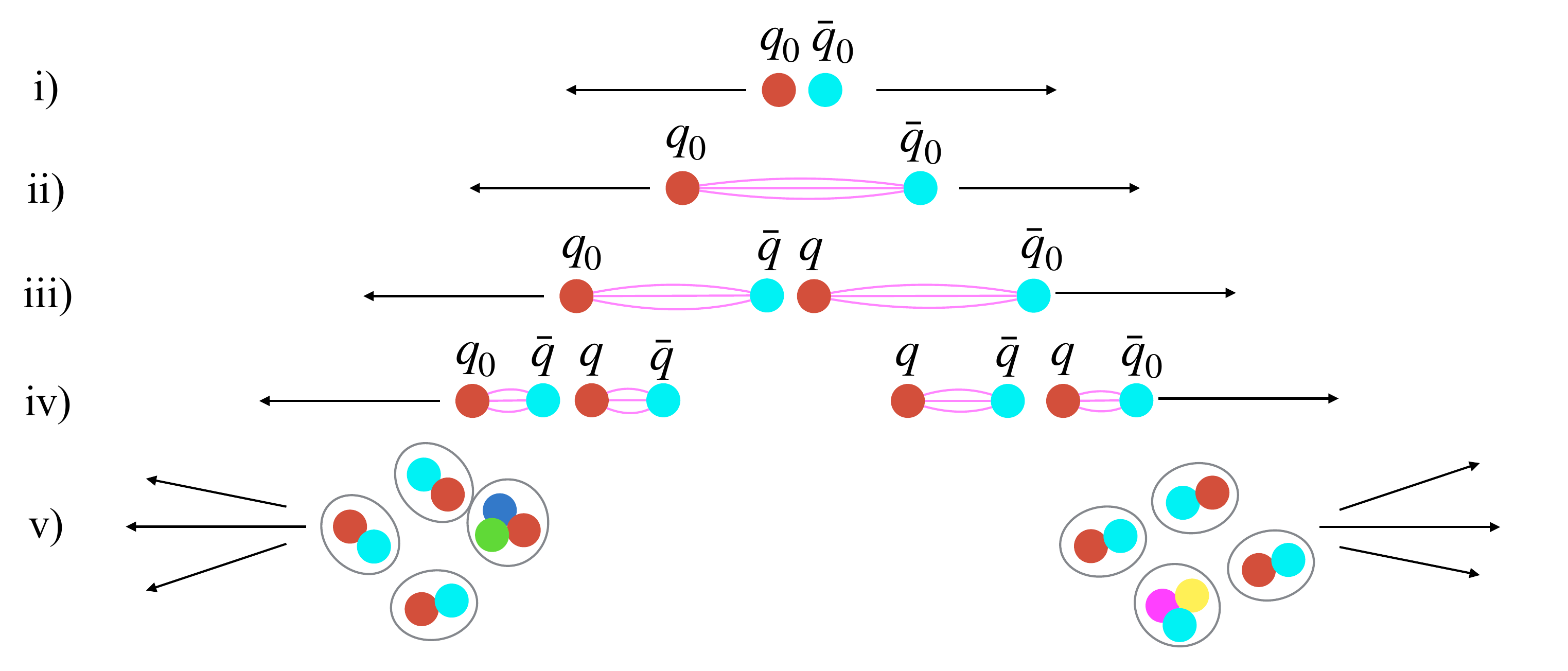}
    \caption{Schematic illustration of hadronization given a string representation of the colour confinement field. (i) The initial red-antired quark-antiquark pair, $q_0\bar{q}_0$, moving apart at high energies. (ii) The string forms between the $q_0\bar{q}_0$ pair. (iii) The initial string breaks and a red-antired $q\bar{q}$ pair is created at the site of the string break. (iv) Further string breaking occurs. (v) Final jet of hadrons formed.}
    \label{frag}
\end{figure}

In high-energy collision processes such as those at the LHC, these colour-connected partons move apart from one another at high energies, in turn stretching the colour fields. 
Given a sufficiently large confining potential, it can become energetically favourable to break the confinement field and spontaneously pair create a quark-antiquark (or diquark-antidiquark) from the vacuum. This pair creation process from string breaking is what we call Lund string fragmentation, and is used to model the hadronization process. 
A given string will break — or “fragment” — till there is no longer sufficient energy to keep breaking the string, resulting in observable final-state hadrons. 
Fig.~\ref{frag} shows a simple schematic depiction of hadronization given the string breaking picture described above. 

To model the (di)quark pair creation process from string breaks, quantum tunnelling analogous to the Schwinger mechanism for QED~\cite{Schwinger:1951nm} is used. The Schwinger mechanism in QED is used to model $e^+e^-$ spontaneous pair creation in the presence of a strong electric field in a vacuum, with the tunnelling probability following a Gaussian distribution in transverse mass. The analogous QCD tunnelling probability used by the baseline LSM is given by,

\begin{equation}
    \exp\left( \frac{-\pi m_q^2}{\kappa}\right) \exp\left( \frac{-\pi p_{\perp q}^2}{\kappa}\right) = \exp\left( \frac{-\pi m_{\perp q}^2}{\kappa}\right), 
\label{eq:schwinger}
\end{equation}
where $m_{\perp q}$ is the transverse quark mass, $p_{\perp q}$ is the transverse momentum, and $m_q$ is the effective quark mass.
Given QCD is a non-Abelian SU(3) gauge theory in contrast to the Abelian QED, the Schwinger mechanism here serves only as an analogy. Exact solutions have been explored, \eg, in~\cite{Nayak:2005pf}.

The transverse mass dependence of the Schwinger mechanism leads to a Gaussian transverse momentum distribution along with a suppression of larger mass quark flavours.
Effectively this leads to reduced production rates of strange quarks and diquarks relative to up/down quark production.
Note that given a string tension of $\kappa \sim 1$~GeV/fm, string breaks can only pair-create light-flavoured (di)quarks, \ie, up-, down-, or strange-flavoured. Charm and bottom quark production via string breaks would be heavily suppressed in the Schwinger picture ($\lesssim10^{-11}$), and thus in \Py their production probabilities (in string breaks) are set to zero. This agrees with the expectation that heavy quark flavours (and high-$p_\perp$ excitations) should only be produced via perturbative processes, and not from non-perturbative mechanisms such as string breaks. 

Due to the ambiguity around both the quark masses and the precision of the string tension, flavour selection in \Py does not directly implement mass-suppression factors according to the Schwinger mechanism Gaussian. 
Instead, the mass suppression is parameterized in terms of a tuneable ratio of probabilities, \ie,  \parold{StringFlav:probStoUD} ($P(s:u/d)$) and \parold{StringFlav:probQQtoQ} ($P(qq:q)$), where the former is the probability of strange quark production relative to up/down production, and the latter is the diquark-to-quark probability. 

Important to note is that the form of \eqRef{eq:schwinger} means that an increase in the string tension value will have two key consequences on the tunnelling probabilities -- a broadening of the $p_\perp$ spectrum, and a reduction of the mass suppression. This relationship between string tension and flavour will be foundational for the closepacking mechanism in Section \ref{sec:closePacking}.

\subsubsection*{Mechanisms for baryon production: diquarks and junctions}
\label{sec:baryonProd}

Within the LSM, there are two primary mechanisms for baryon production; diquark-antidiquark pair production via string breaks, and junction string topologies. 

In the baseline LSM, diquark-antidiquark pair production is modelled on equal footing as quark-antiquark production -- \ie, spontaneously pair created out of the vacuum as a triplet-antitriplet pair. 
This means that the mass suppression of diquark production is directly governed by the diquark masses and the Schwinger mechanism.

\begin{figure}[t]
\centering
    \includegraphics[width = 0.6\textwidth]{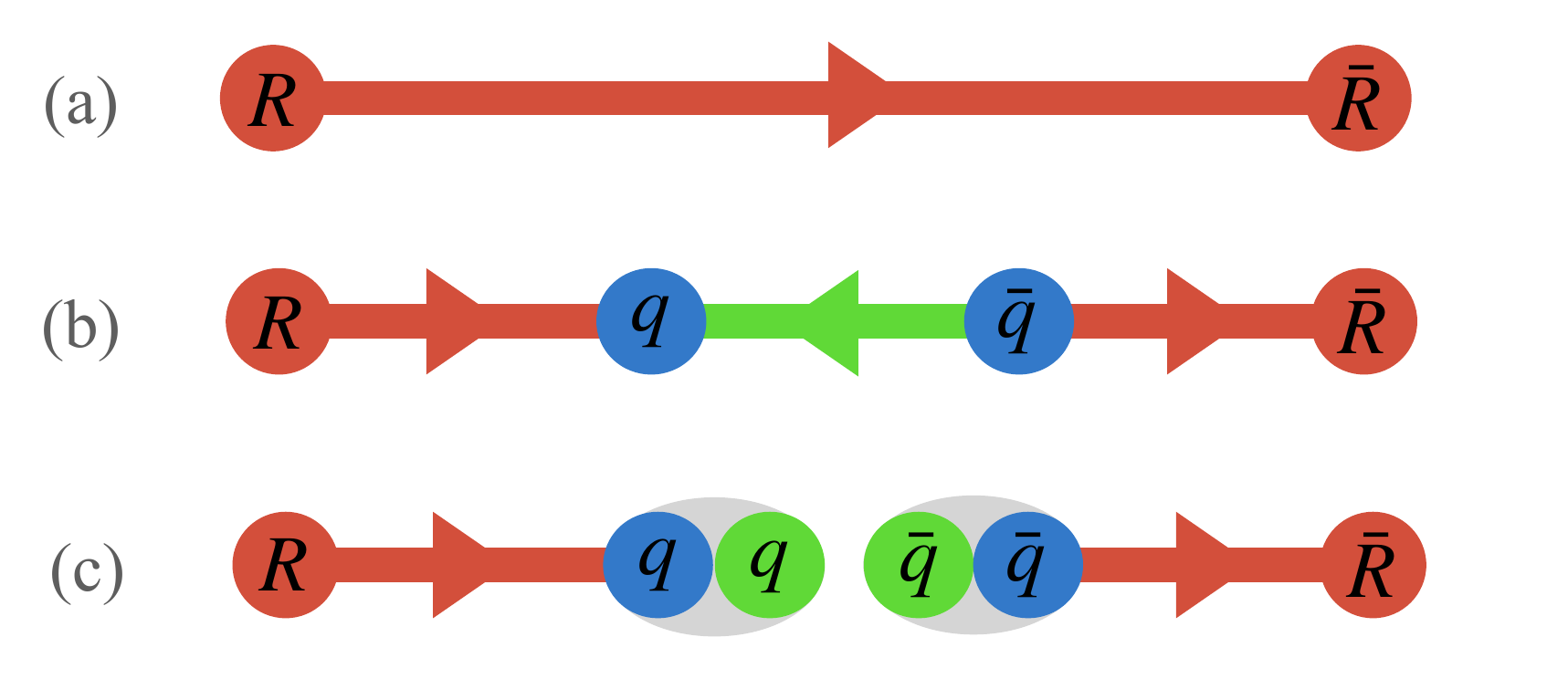}
    \caption{An illustration of the stepwise process of the popcorn mechanism given a red-antired string. (a) the unbroken red-antired string. (b) a blue-antiblue virtual colour fluctuation on the red-antired string, which results in a short segment of ``green-coloured" string. (c) a green-antigreen quark pair creation that combined with the blue-antiblue quark fluctuations forms a red-antired diquark-antidiquark pair, breaking the string.}
    \label{fig:popcorn}
\end{figure}

An alternative interpretation of diquark production is the so-called ``popcorn mechanism’’~\cite{Andersson:1984af, Eden:1996xi}. 
Instead of considering diquarks as directly tunnelling from the vacuum as constituent particles, the popcorn mechanism supposes that diquarks are formed via the combination of two successive colour fluctuations on a string. 
An example demonstrating this colour-fluctuation based argument for diquark formation can be seen in fig.~\ref{fig:popcorn}. 
Per this example, consider a red-antired $q\bar{q}$ pair with a red confining field between them (fig.~\ref{fig:popcorn}(a)). 
This red confining field is characterized by two defining features -- its red colour, and by its ``flux'' direction as indicated by an arrow on the string. 
Here the term ``flux’’ refers to the direction of the colour flow, conventionally orientated as flowing from the triplet to the antitriplet colour charge. 
Suppose then we allow arbitrarily coloured fluctuations to form on this red string, \eg a blue-antiblue coloured $q\bar{q}$ fluctuation. 
Unlike conventional string breaks which require the tunnelling of a red-antired $q\bar{q}$ pair to break the confining field, a virtual blue-antiblue pair would not initially ``break'' the string as there would remain a non-zero colour field between the blue-antiblue pair. 
More specifically, the combination of the red and blue colour charges results in an overall antigreen charge, thus producing a green colour field oriented in the opposite flux direction to the initial red string as can be seen in fig.~\ref{fig:popcorn}(b).
From this point, tunnelling of a green-antigreen $q\bar{q}$ pair is able to break the green string segment, resulting in the formation of a blue-green diquark and an antiblue-antigreen antidiquark as shown in fig.~\ref{fig:popcorn}(c).

\begin{figure}[t]
\centering
    \includegraphics[width = 0.8\textwidth]{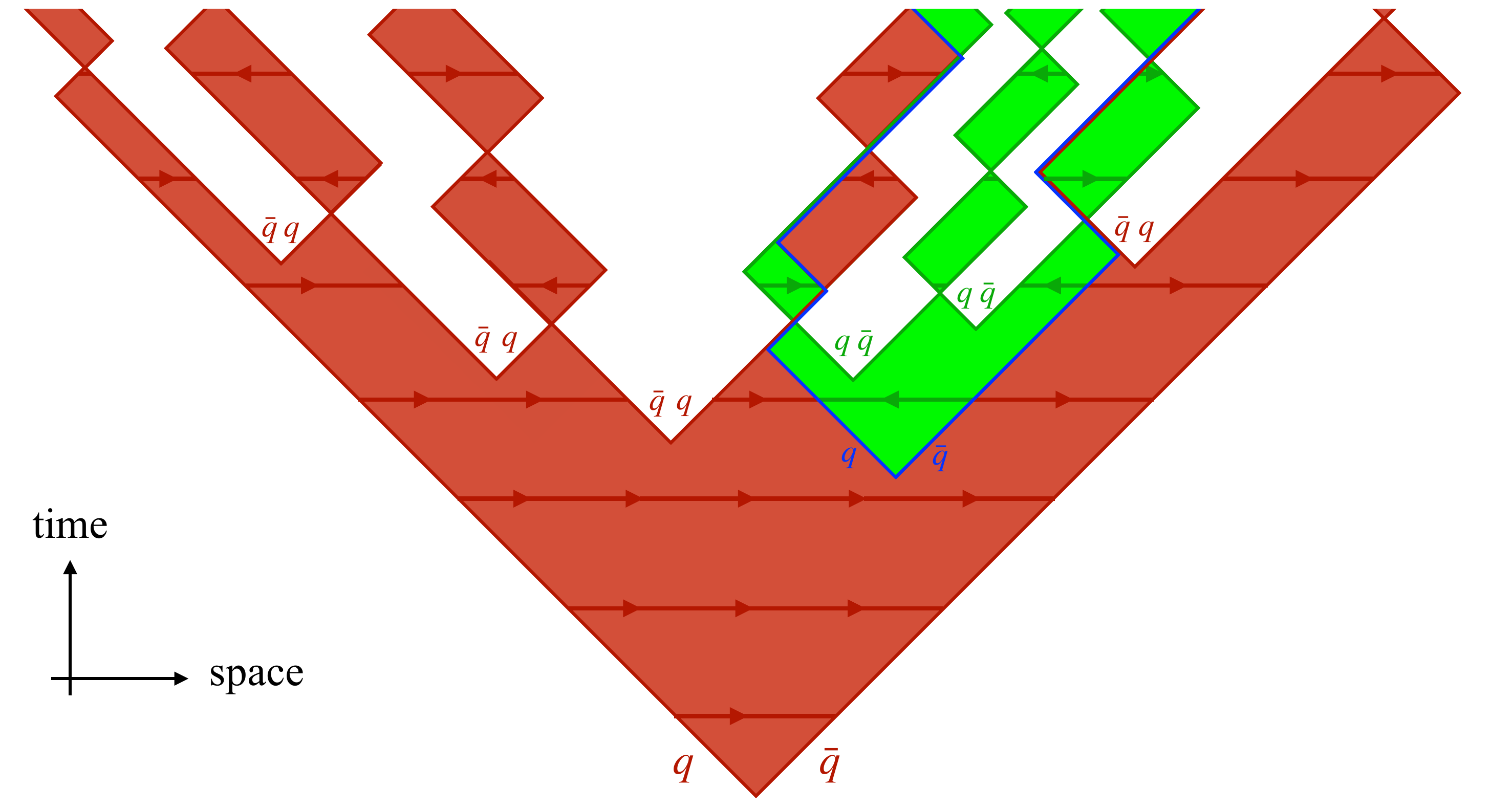}
    \caption{Minkowski diagram of the popcorn mechanism showing diquark-antidiquark formation and a single meson produced between the baryon and antibaryon \ie, $BM\bar{B}$ configuration. The flux direction of the strings are indicated by arrows, with colour flow from the triplet to antitriplet colour charge. In this picture, we assume the green-antigreen fluctuation forms with zero energy and momentum.}
    \label{fig:popcornMink}
\end{figure}

A key consequence of modelling diquark formation using the ``popcorn mechanism'' is the possibility of meson production occurring between diquark-antidiquark pairs. 
After an initial colour fluctuation on a given string, rather than a single $q\bar{q}$ pair breaking the intermediate string per fig.~\ref{fig:popcorn}, suppose multiple colour fluctuations were allowed. Given the same colour structure as the example in fig.~\ref{fig:popcorn}, should we allow for multiple green-antigreen fluctuations to occur on the intermediate green string, it would result in the formation of one or more green-antigreen mesons between the resulting diquark-antidiquark pair. Fig.~\ref{fig:popcornMink} depicts an example of this process in the form of a Minkowski space-time diagram.
Consequently this reduces the local correlation of the baryon-antibaryon pair formed, as it allows $BM\bar{B}$ configurations alongside $B\bar{B}$ (where $B$ and $M$ denotes baryons and mesons respectively).
Contrastingly, in the standard string breaking picture diquark-antidiquark pairs are always created next to one another at string breaks, and hence only $B\bar{B}$ configurations are permitted.

In \Py, these successive colour fluctuations used by the popcorn mechanism are not explicitly dynamically modelled. Rather, an additional parameter \parold{StringFlav:popcornRate} is introduced which represents the probability of the $BM\bar{B}$ configuration occurring relative to the $B\bar{B}$ configuration (with configurations with multiple intermediate mesons -- $BMM\bar{B}$, $BM^3\bar{B}$, etc. -- neglected as they are expected to be suppressed).
Note that the diquark production probability is assumed to be constant along a string regardless of whether one uses the popcorn or Schwinger mechanism for diquark production, and hence diquark production remains set by the tuneable parameter \parold{StringFlav:probQQtoQ}. 

A reexamination of diquark production via the popcorn mechanism given high density string systems will be the focus of Section~\ref{sec:popDestr} of this paper, with a particular focus on the effect on the $p/\pi$ ratio.

\begin{figure}[t]
\centering
        \includegraphics[width = 0.9\textwidth]{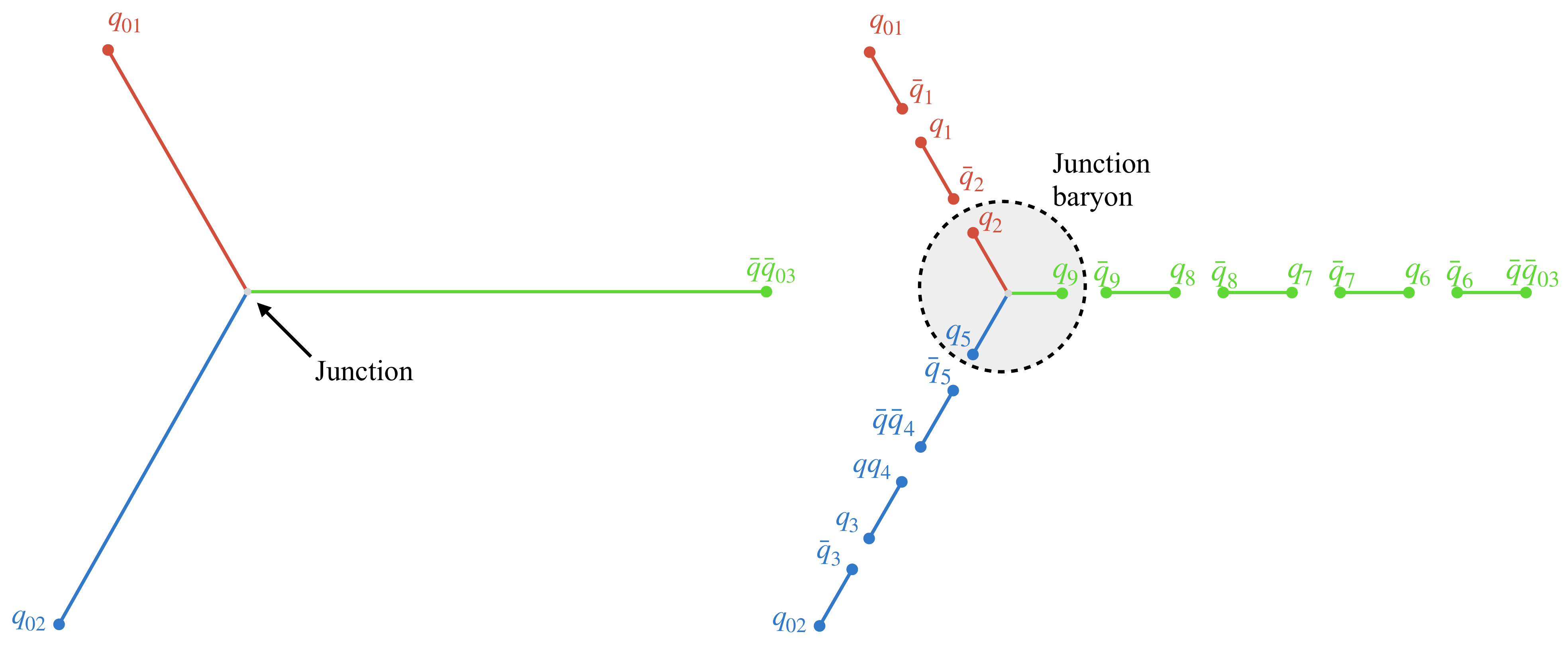}
    \caption{Junction configuration before (left) and after (right) fragmentation, with quarks $q_2q_9q_5$ together forming the so-called junction baryon.}
    \label{fig:juncFrag}
\end{figure}

Additional to diquark-antidiquark pair production, baryon formation can also occur via so-called string junctions. 
Thus far, the representations of strings depicted above have all been ``dipole'' string topologies, defined by a string spanning between a triplet and antitriplet endpoint forming the overall colour-anticolour singlet state. 
However given an SU(3) representation of colour, one should also consider the confining potential formed due to the red-green-blue colour singlet -- a junction string~\cite{Sjostrand:2002ip, Altmann:2024odn}. 
Such junction configurations are typically represented by a Y-shaped string as shown in the left image of fig. \ref{fig:juncFrag}, consisting of three junction ``legs'' with the ``junction'' itself being the meeting point of the three legs. 
Importantly, the baryon number of junction configurations are conserved, meaning that after fragmentation, the quarks surrounding the (anti)junction itself form a (anti)baryon. An example of a junction topology before and after fragmentation is shown in the left and right images of fig.~\ref{fig:juncFrag} respectively. For more detail on the modelling of such junctions and the fragmentation procedure, see~\cite{Altmann:2024odn}.

Though the field towards the endpoints of each junction leg would be expected to behave similar to a dipole string, there remains ambiguity about field properties directly around the junction itself. For example, one may ask whether the energy density around a junction should be treated the same as a dipole string in vacuum.
Such a question and implementation of this ambiguity will be further explored in Section~\ref{sec:strangeJunc}.

\subsection{Closepacking}
\label{sec:closePacking}

The closepacking model was first introduced in an earlier study in the context of thermodynamical string breaks~\cite{Fischer:2016zzs}. In this paper, we generalize the model to conventional (Schwinger-type) string fragmentation, introduce sensitivity to relative flux orientations, and explore a few further optional implementation modifications. 

Closepacking describes the collective effect of multiple strings in the near vicinity to one another. 
Typically in the LSM, the confining potential is approximated as a 1+1 dimensional field line, which fragments independently. However, realistically one should treat them more as flux tubes with non-zero thickness. 
Analogous to parallel conducting wires exerting an influence on one another -- or perhaps more accurately vortex lines in a superconductor or superfluid --
one would expect nearby strings to also affect each other. 
We suppose this background of surrounding strings manifests itself as an altered energy density, \ie, an effective enhanced string tension, $\kappa_{\text{eff}}$.
Given the Gaussian form of the Schwinger mechanism, \eqRef{eq:schwinger}, an increased effective string tension would correspond to reduced mass suppression of the quark-antiquark pairs produced in string breaks, and thus would lead to enhanced strange-quark production.

To motivate how the effective string tension should scale with the density of surrounding strings, we start by following in the footsteps of the Rope model, noting that the ``total colour charge'' of an irreducible representation of SU(3) with $p$ fundamental (colour) indices and $q$ antifundamental (anticolour) ones is given by the quadratic Casimir operator, 
\begin{equation}
    C_{2}(p,q) = \frac{p^2 + pq + q^2 + 3p + 3q}{3}\,.
\end{equation}
For example, a quark (which has one free colour index) has $C_2(1,0) = C_F = 4/3$, while a gluon (which has one free colour index and one free anticolour one) has $C_2(1,1) = C_A = 9/3 = 3$. For reference, we give explicit colour factors for irreducible SU(3) representations with up to 8 free indices in \appRef{app:casimirs}. 
Assuming that a preceding stage of colour reconnections has acted to reduce (screen) colour charges so that only ones that cannot be further screened remain, the assumption that the colour charges defining the hadronizing string configurations constitute irreducible representations seems reasonable, at least in a first approximation. The assumption that the linear part of the QCD potential --- and hence the total effective string tension --- scales with this effective colour charge is referred to as Casimir scaling. We note that lattice studies support this assumption, though still with relatively large error bars~\cite{Bali:2000un}.

\begin{figure}[t]
\centering
        \includegraphics[width = 0.7\textwidth]{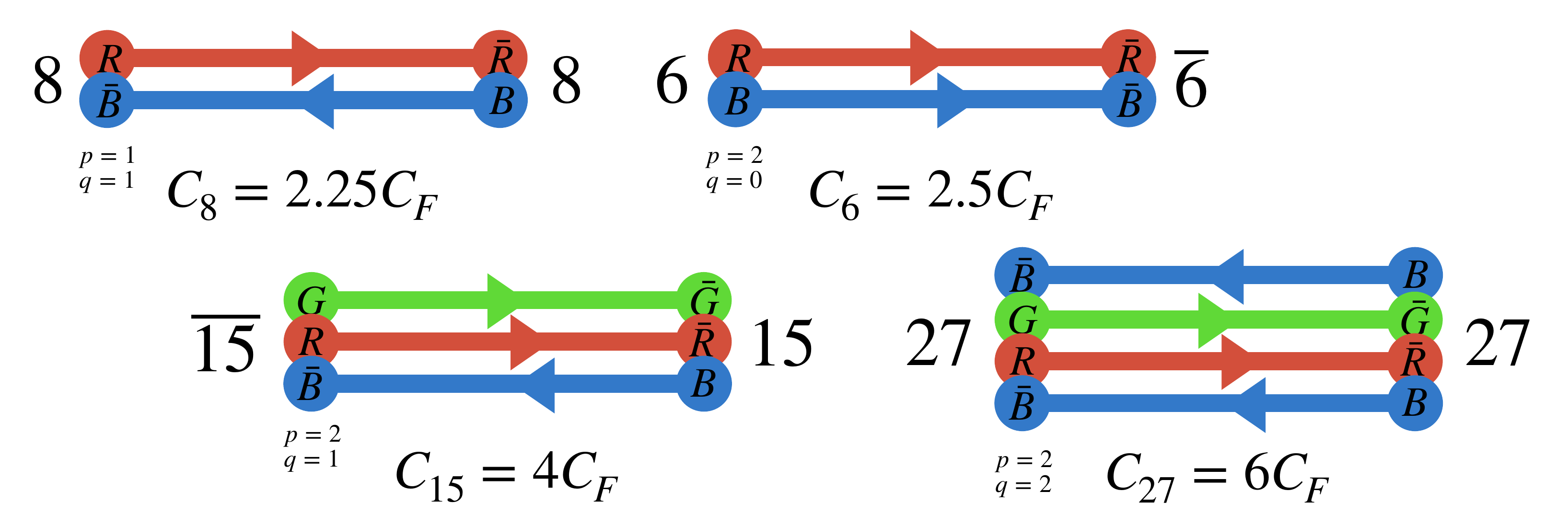}
    \caption{Examples of QCD multiplets in overall colour-singlet states and their colour factors according to Casimir scaling. The total number of parallel and antiparallel strings, labelled by $p$ and $q$ respectively, correspond to the number of free indices on the left-hand ``side'' of the multi-string systems that are contracted with those on the right-hand side, with $p>q$. See also \appRef{app:casimirs}.}
    \label{fig:closepacking}
\end{figure}

Since the fundamental (quark-antiquark) string constitutes the baseline case for the LSM, it is useful to parameterize the colour factors of higher QCD representations with respect to that of the triplet, $C_F = 4/3$.
Fig.~\ref{fig:closepacking} provides a simple pictorial representation of a some examples of such multiplets with their corresponding Casimir-scaled colour factors relative to $C_F$. 

Let us suppose the net tension of the colour system scales proportional to these colour factors. 
Assuming that each string in a given multiplet shares an equal contribution to this modified colour factor -- \ie, we should scale each strings tension by a factor of $(C_R/C_F)/(p+q)$ -- one can then simply parameterize the effective string tension in terms of the $p$ and $q$ values, 
\begin{equation}
    \kappa_{\text{eff}} ~=~ \frac{C_R}{(p+q)C_F}\kappa_0 ~=~ \left( 1+ 0.25 p + 0.125 q \right) \kappa_0.
    \label{eq:casimir}
\end{equation}
For the purposes of calculating the multiplet representation, $R$, for a given string, the ``flux orientation'' will refer to the direction relative to the string in question being fragmented. 
To make this clear, take the example of an octet configuration with colour factor $C_8=C_A=2.25C_F$. As the octet is comprised of two strings, each will contribute a factor of $1.125C_F$, and hence we assume the effective tension for a string in an octet configuration is given by $1.125\kappa_0$, making the total tension (summed over the two strings) $2.25\kappa_0 = (C_A/C_F)\kappa_0$. See \tabRef{tab:casimirs} in \appRef{app:casimirs} for further examples.

\begin{figure}[t]
\centering
        \includegraphics[width = 0.9\textwidth]{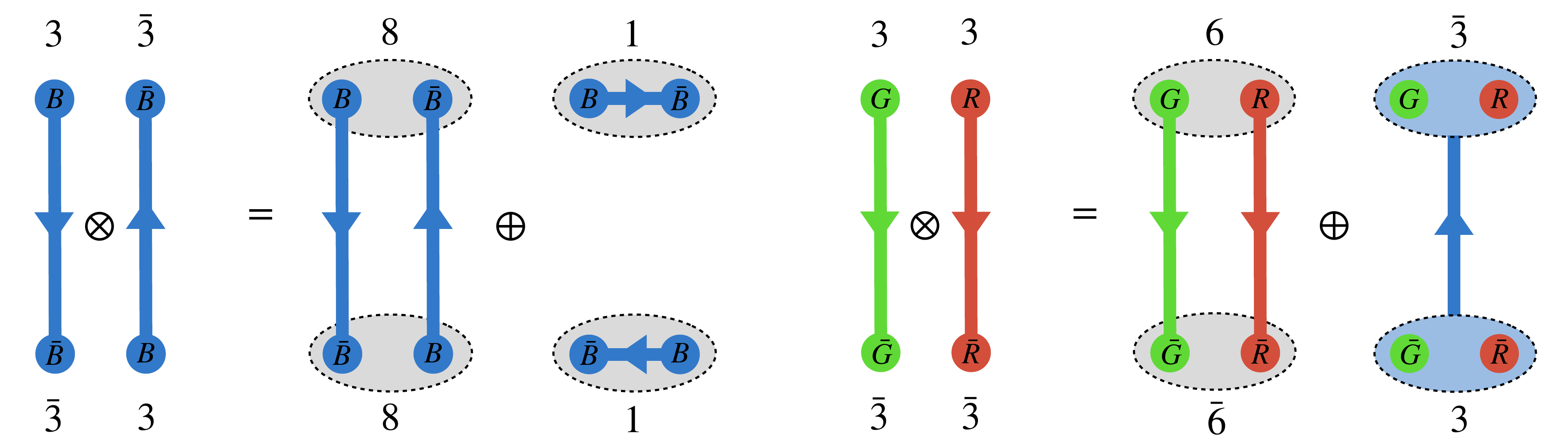}
    \caption{Pictorial representations possible multiplet representations for the case of two nearby strings, as dictated by SU(3) algebra. Left image:  $3\otimes\bar{3} = 8\oplus 1$. Right image: $3\otimes 3 = 6\oplus \bar{3}$.}
    \label{fig:CR}
\end{figure}

Important to note is that the colour-factor calculation with $p$ and $q$ here assumes the collective of strings is in the highest multiplet representation. More specifically, according to $SU(3)$ algebra the multiplet representation of a parallel and an antiparallel string (\ie, $p=1$, $q=1$) does not necessarily always result in an octet string configuration. This is as $SU(3)$ algebra dictates that there is a 1/9 probability of a colour and anticolour charge in combination ``screening'' each other and resulting in a singlet state, \ie, $3\otimes\bar{3} = 8\oplus 1$. What this translates to in terms of string configurations can be seen in fig.~\ref{fig:CR}. As mentioned above,  we work under the assumption that QCD CR~\cite{Christiansen:2015yqa}, which proceeds the fragmentation procedure, effectively models any screening effects such that $p$ and $q$ post CR have already been reduced to their smallest irreducible-representation equivalents. 

Though Casimir scaling provides a well-motivated foundation for string-tension modifications for pointlike static colour charges, we must also consider how this relates to the dynamical environment of collision systems, in which the total effective colour charges are made up of triplet quarks and octet gluons whose spatial separations are changing at or near light speed. In small systems like hadronic $e^+e^-$ collisions and $pp$ collisions, the individual colour charges originate from a space-time region that can be taken to be small compared with a typical hadron size. Taking a point-like origin as our starting point (which excludes heavy-ion collisions from consideration in the present formulation of the model), the main effect we shall seek to account for is the fact that the individual fragmenting strings do not sit right on top of each other (as in the lattice case) but instead will be expanding along axes that are determined by their individual momenta. 

In this work, we shall not seek to account for the relative orientations (and boosts) of strings in detail. Instead, we allow for dampening factors that reduce the effect of the baseline Casimir scaling, to reflect that:
\begin{enumerate}
\item on average, strings will be some distance apart by the time they fragment; 
\item high-$p_\perp$ hadrons are likely to be produced in spacetime regions well away from large string overlaps.
\end{enumerate}
The first change we implement by generalizing \eqRef{eq:casimir} with the introduction two adjustable constants, 
\begin{equation}
\frac{\kappa_{\text{eff}}}{\kappa_0} ~=~ 1 + c_p \left( p + \omega q \right)~,
    \label{eq:casimirMod}
\end{equation}
where $c_p$ now governs the overall (average) closepacking strength and $\omega\in[0,1]$ governs the relative strength of contributions from strings with flux orientations antiparallel to the one being fragmented. 
Full-strength Casimir scaling corresponds to $c_p=0.25$ and $\omega = 0.5$. 

In reality, we expect $c_p < 0.25$ since, as discussed above, strings presumably fragment some distance apart, so that 0.25 should be considered a limiting maximum value (at least within the paradigm of Casimir scaling). A lower effective strength parameter may also account for imperfections in the colour-reconnection modelling leaving some representations in the modelling which could have been further reduced. 
Since we expect closepacking to impact both the flavour (strange and diquark production rates) and $p_\perp$ distributions of string breaks (detailed discussions on this follow below), in the \Py implementation we allow for the strength of closepacking effects on each of these distributions to be set independently. This is done by allowing three separate $c_p$ values, one for each corresponding physics effect,
\begin{equation}
c_p ~=~ \left\{~\begin{array}{l} 
\parnew{ClosePacking:enhanceStrange} \\   \parnew{ClosePacking:enhanceDiquark} \\  \parnew{ClosePacking:enhancePT}
\end{array}\right.\vspace*{2mm}
\end{equation}
For $\omega$, the \Py implementation uses a single universal value,
\begin{equation}
\omega ~=~ \parnew{ClosePacking:fluxRatio}~.
\end{equation}
As mentioned above, the ratio for Casimir scaling is $\omega = 0.5$. The old closepacking model~\cite{Fischer:2016zzs} which was insensitive to the flux direction would correspond to using $\omega = 1$, while for $\omega = 0$ only strings with flux directions parallel to each other would be allowed to increase each others' effective string tension, while strings with fluxes antiparallel to each other would be insensitive to each other. We believe this range of options gives a reasonable amount of flexibility to explore the modelling space.

\begin{figure}[t]
\centering
        \includegraphics[width = 0.98\textwidth]{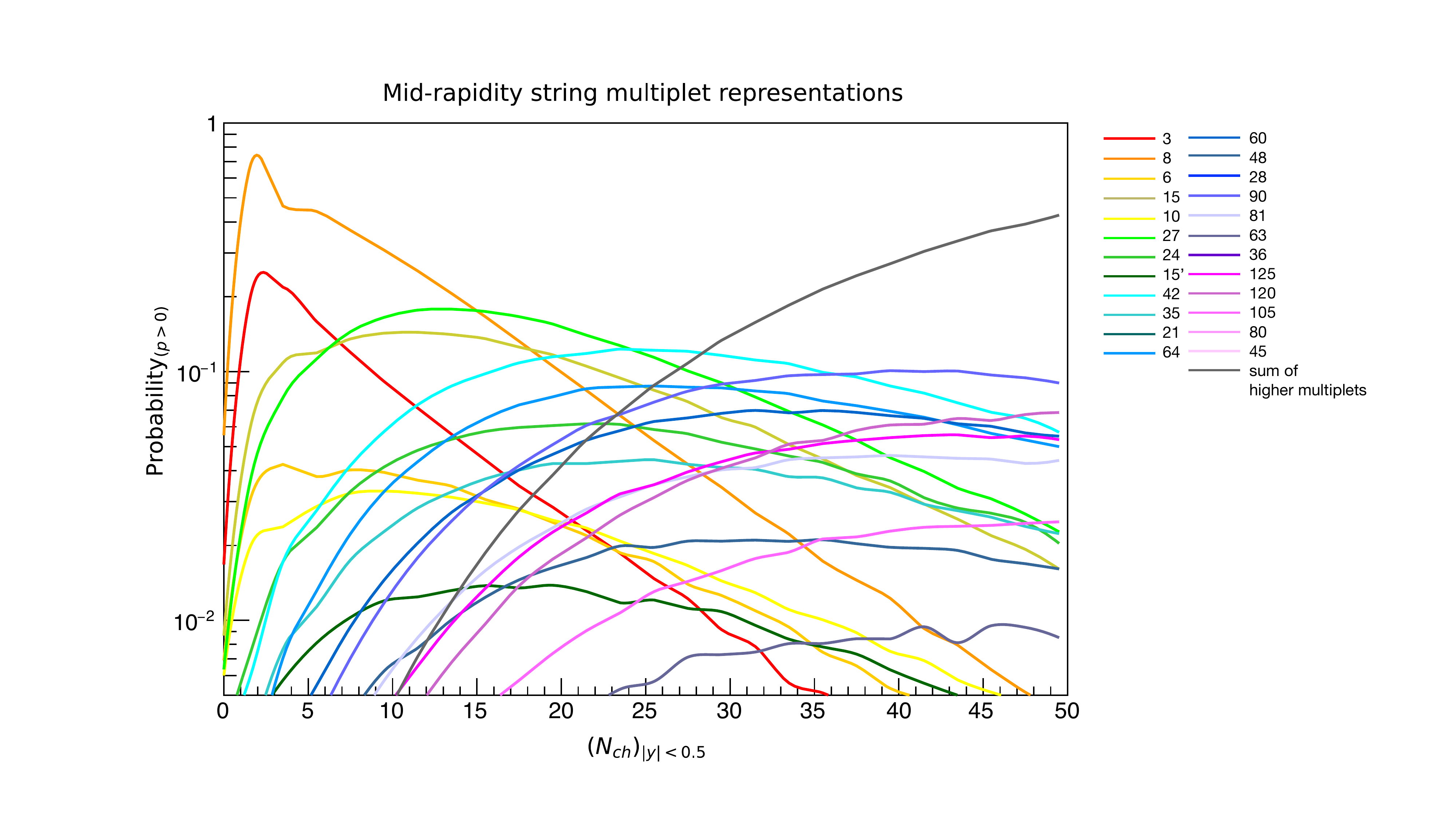}
    \caption{The probability of colour multiplet representations at y=0 with respect to the midrapidity charged multiplicity for 7~TeV inelastic events using QCD CR ($m_\tau = 2$)~\cite{Christiansen:2015yqa} parameters. Only events with $p>0$ at $y=0$ have been included. 
    Flux configurations and Casimir-scaling factors for each multiplet are given in  \tabRef{tab:casimirs}. The highest multiplet shown here, the {\bf 125}, can be reached with just 4 gluon exchanges, if none of their colour charges screen each other.}
    \label{fig:multiplet}
\end{figure}

The next question one must address is how to measure the multiplet representation a string is in given the distribution of colour charges in collision events is not comprised of a neat collection of (anti)parallel strings with well-defined colour-multiplet endpoints as in fig.~\ref{fig:closepacking}.
This means that we need some measure of the number of nearby surrounding strings, alongside a way to take into account partial string overlaps.
To do so, in the following we use rapidity with respect to the beam axis to determine whether two strings overlap. This variable choice is a logical approximation for $pp$ collisions as the majority of strings are expected to be oriented 
along the beam axis due to beam remnants and low-$p_\perp$ multi-parton interactions. 
Given this rapidity measure for calculating $p$ and $q$, and under the assumption that QCD CR takes into account any screening effects, fig.~\ref{fig:multiplet} shows the likelihood of different multiplet representations at midrapidity in 7~TeV inelastic $pp$ collisions, in fully simulated \Py events using the QCD CR model (with time-dilation mode $m_\tau = 2$). For high-multiplicity events, we see that very high multiplets dominate, meaning that high-multiplicity events correspond to dense string systems (as one would reasonably expect), and one would hence expect closepacking effects to be particularly significant at these multiplicities. 
Note however that using rapidity with respect to the beam axis to measure string overlaps would no longer hold as a good approximation in hadronic $e^+e^-$ events, for which the beam axis plays no special role; alternative approaches allowing for overlaps along generic axes will be explored in future studies. 

To take into account partial string overlaps, the effective multiplet representation is determined on a per-string-break basis -- \ie, using the rapidity associated with a given string break and comparing to the rapidity extents of the other strings in the collision system. 
This begs the question of how to determine the rapidity of a given string break. For this, we use a simple approximation of the rapidity of the produced hadron rather than the rapidity of the string break itself\footnote{This, again, is an approximation that could be lifted in a future study, but it would require solving the string world-sheet equations for a varying tension, which would be a nontrivial research project in its own right.}.
The flavour and transverse momentum of the produced hadron however are themselves dependent on the string tension (details of the dependence outlined in following sections), and thus sensitive to the number of surrounding strings. 
To navigate around this issue, we follow the original closepacking implementation~\cite{Fischer:2016zzs} and create a trial hadron with an average hadronic mass and transverse-momentum selection given the standard vacuum string-tension probabilities. The rapidity of this trial hadron is then used to determine the number of overlapping strings at that rapidity, which is thence used to calculate $\kappa_{\text{eff}}$ to be used for the actual string break.

To further capture the dynamical nature of collisions, a concept of the separation of strings needs to be introduced. 
Under the assumption that we are working with hadron collisions, the bulk of strings will be aligned predominantly along the beam axis.
However, hard interactions may also give rise to high-$p_\perp$ jets. 
The leading partons in such jets will move far from the beam axis before fragmenting and thus we don't expect them to experience as much closepacking tension enhancement as partons formed near the beam axis. Conversely, partons formed near the beam axis will be nearest the bulk collective of strings and thus experience maximal closepacking effects. To implement this effect, we use the $p_\perp$ of the corresponding hadrons as a measure of whether that hadron was produced ``far away'' from the beam axis ($p_\perp \gg {\mathcal{O}}(\Lambda_{\mathrm{QCD}}) $) or ``near'' the beam axis ($p_\perp \sim {\mathcal{O}}(\Lambda_{\mathrm{QCD}})$). 
To do so, we require the functional form for $\kappa_{\text{eff}}$ to scale such that when $p_{\perp} \rightarrow \infty$, the tension returns to the vacuum value (\ie, $\kappa_{\text{eff}} \rightarrow \kappa_0$), and when $p_{\perp} \rightarrow 0$ there are maximal closepacking effects. 
We choose a simple functional dependence inversely proportional to $1+p_\perp^2/p_{\perp 0}^2$, with regularization parameter $p_{\perp 0}$ (\parnew{ClosePacking:PT0}), and similarly to determining rapidity overlaps, the $p_\perp$ used here is that of the trial hadron, $p_{\perp \text{Had}}$. 
This allows very soft hadrons produced in such jets to still ``feel'' collective effects, while the leading hadrons will not. 
The use of $p_\perp$ here is admittedly not perfect and can be challenged. Nevertheless, it allows us to interpolate between ``soft'' fragmentation aligned with the beam axis and ``hard'' fragmentation (along jet axes), the latter of which we presume still takes place in a good approximation to vacuum. 

This gives us our final, fully generalized effective string-tension formula,
\begin{equation}
    \kappa_{\text{eff}} = \left( 1+ c_p \left( \frac{p+\omega q}{1+p^2_{\perp \text{Had}}/p^2_{\perp 0}}\right) \right)^{2r} \kappa_0, 
    \label{eq:kappaEff}
\end{equation}
with we remind that exact (unsuppressed) Casimir scaling corresponding to overall strength parameter $c_p=0.25$ and antiparallel flux-orientation suppression factor $\omega=0.5$. The optional non-linear scaling exponent $r$ (\parnew{ClosePacking:expNSP}, with default value 0.5) is a leftover from the original thermal-model implementation~\cite{Fischer:2016zzs}. It is not extensively examined here but has been left in the implementation for generality and for consistency with~\cite{Fischer:2016zzs}.

\subsubsection*{Closepacking effects on flavour}
Given an effective string tension, the next question then becomes how does this effect flavour probabilities used in string fragmentation by \Py. The main set of parameters that will be impacted by a modified string tension are \parold{StringFlav:ProbStoUD} ($P(s:u/d)$, the probability of a strange quark relative to up/down), \parold{StringFlav:ProbQQtoQ} ($P(qq:q)$, the probability of a diquark relative to quark), \parold{StringFlav:ProbSQtoQQ} ($P(sq:qq)$, additional suppression of strange flavour production for diquarks), and\\ \parold{StringFlav:ProbQQ1toQQ0} ($P(qq_1:qq_0)$, the probability of a spin-1 diquark relative to a spin-0 diquark, additional to the factor 3 enhancement of spin-1 production from state counting). \footnote{In \Py, up and down quarks are treated on the same footing \ie, with equal probabilities.}

\begin{figure}[t]
\centering
        \includegraphics[width = 0.7\textwidth]{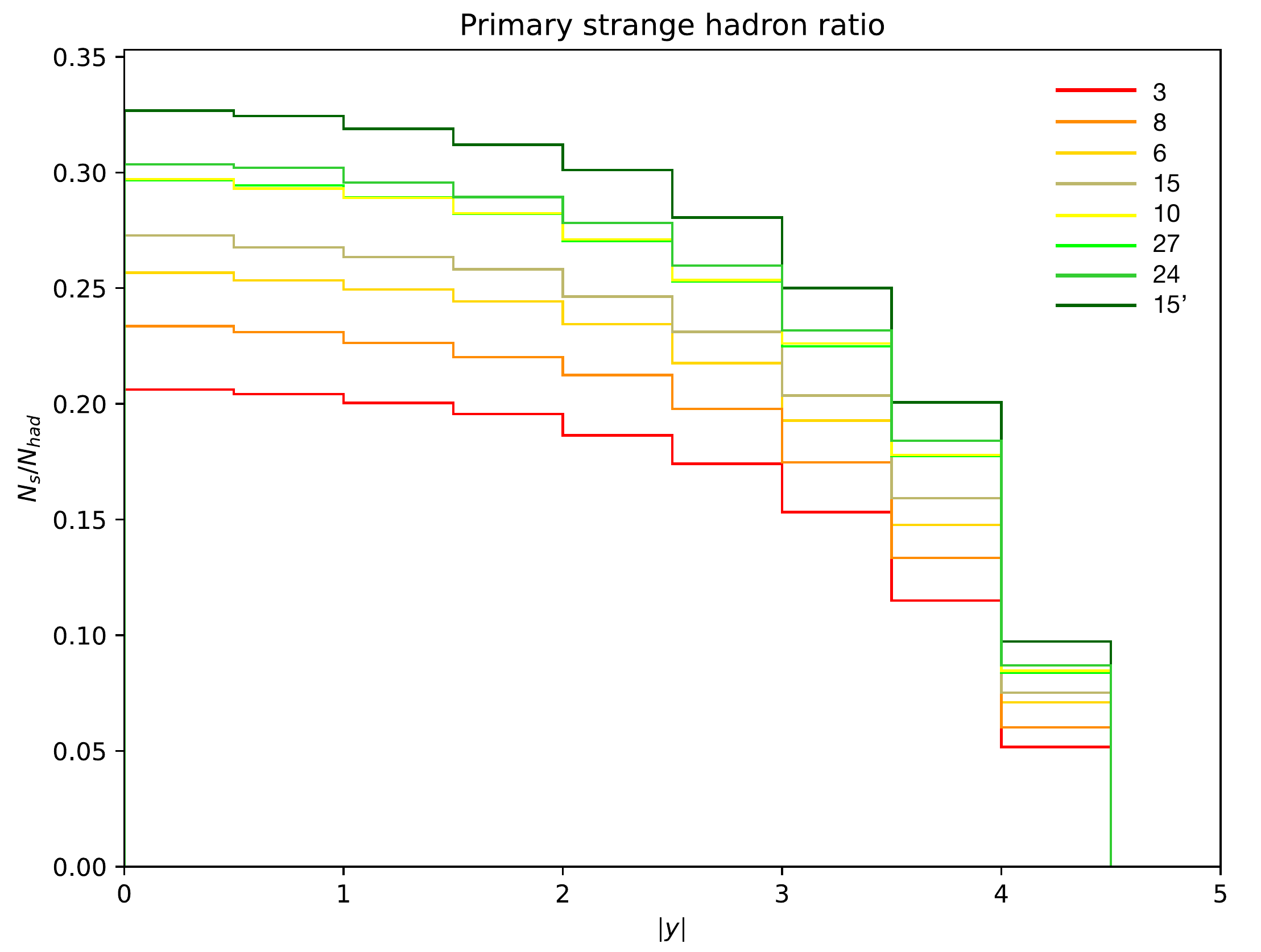}
    \caption{Strange hadron probabilities with respect to absolute rapidity for different string multiplet configurations. These multiplet configurations are constructed using parallel  $u\bar{u}$ strings along the beam axis, with endpoint energies of 20~GeV. }
    \label{fig:CPproof}
\end{figure}

Let us first consider how the strangeness probability, $P(s:u/d)$, scales according to the string tension. 
Given the Gaussian form of the Schwinger mechanism as per \eqRef{eq:schwinger}, the strange-to-up/down quark production probability takes the form, 

\begin{equation}
	P(s:u/d) = \frac{P \left( m_s^2 \right)}{P\left(m_{u/d}^2\right)} = \frac{\exp\left( -\frac{\pi m_s^2}{\kappa_0}\right)}{\exp\left( -\frac{\pi m_{u/d}^2}{\kappa_0}\right)} = \exp\left( -\frac{\pi (m_s^2-m_{u/d}^2)}{\kappa_0}\right).
	\label{probStoUD}
\end{equation}
Similarly, given an effective string tension, the modified probability $P'(s:u/d)$ is then given by,

\begin{equation}
	P'(s:u/d) = \exp\left( -\frac{\pi (m_s^2-m_{u/d}^2)}{\kappa_{\text{eff}}}\right).
	\label{probStoUDeff}
\end{equation}
This allows us to rewrite the probability $P'(s:u/d)$ in terms of $P(s:u/d)$ and the ratio of string tensions $\kappa_0/\kappa_{\text{eff}}$, 

\begin{equation}
	P'(s:u/d) = \exp\left( -\frac{\pi (m_s^2-m_{u/d}^2)}{\kappa_{0}} \frac{\kappa_0}{\kappa_{\text{eff}}} \right) = P(s:u/d)^{\kappa_0/\kappa_{\text{eff}}}.
	\label{eq:probStoUDmod}
\end{equation}
Importantly, the effective string tension $\kappa_{\text{eff}}$ as given in \eqRef{eq:kappaEff} is proportional to the vacuum string tension $\kappa_0$, meaning that the ratio $\kappa_0/\kappa_{\text{eff}}$ is independent of the vacuum string tension and hence this scaling of the strangeness probability is parameterized insensitive to both the uncertainty in the string tension value and the quark masses. The implications of eq~\eqref{eq:probStoUDmod} and the modified string tensions is evident in fig.~\ref{fig:CPproof}, which shows the fraction of strange hadrons given various multiplet representations. Interestingly, the 10-plet ($p=3,~q=0$) and 27-plet ($p=2,~q=2$) representations result in the same strange hadron fractions. Though being different colour multiplets with differing colour factors ($C_{10}=4.5C_F$ and $C_{27}=6C_F$), the per string contribution to the effective tension for both configurations is 1.5, and thus the same level of strangeness enhancement is expected as evident in fig.~\ref{fig:CPproof}.

The probabilities \parold{StringFlav:ProbSQtoQQ} and \parold{StringFlav:ProbQQ1toQQ0} also are expected to be modified in the same manner as per \eqRef{eq:probStoUDmod}. The strength of the closepacking mechanism, $c_p$ in \eqRef{eq:kappaEff}, for the aforementioned flavour probabilities is controlled by parameter \parnew{ClosePacking:enhanceStrange}.

The modification of the diquark probability \parold{StringFlav:ProbQQtoQ} is less straightforward however. 
There remains ambiguity whether the probability should be scaled assuming the diquark is treated as a constituent formed via the Schwinger mechanism, or whether it should be modelled via the popcorn mechanism. In Section~\ref{sec:popDestr} we will explore further the physics of the popcorn mechanism given dense string environments, however here let us first consider how the probability would be scaled assuming Schwinger-type diquark production. 

The diquark probability $P(qq:q)$ can be parameterized in terms of probabilities $\mathcal{P}_{ud0}$, $\mathcal{P}_{u}$, and a multiplicative factor $\alpha$ which is a function of the probabilities $P(s:u/d)$, $P(sq:qq)$, and $P(qq_1:qq_0)$. The full functional form is seen in Appendix~\ref{app:probQQtoQ}. 

\begin{equation}
	P(qq:q) = \alpha \frac{\mathcal{P}_{ud0}}{\mathcal{P}_u}.
	\label{eq:probQQtoQ}
\end{equation}

Assuming the term $\mathcal{P}_{ud0}/\mathcal{P}_{u}$ scales with the a power of $\kappa_0/\kappa_{\text{eff}}$, the modified diquark probability $P'(qq:q)$ can be written as

\begin{equation}
	P'(qq:q) = \tilde{\alpha} \left( \frac{P(qq:q)}{\alpha} \right) ^{\frac{\kappa_0}{\kappa_{\text{eff}}}},
	\label{eq:probQQtoQmod}
\end{equation}
with $\tilde\alpha$ having the same functional form as $\alpha$ whilst taking into account the effective string tension. 
In principle, then one can scale $P(qq:q)$ using the same $\kappa_0/\kappa_{\text{eff}}$ used to modify the aforementioned probabilities such as $P(s:u/d)$. 
However given the additional ambiguity afforded to diquark creation -- for example the uncertainty between the Schwinger-type production and popcorn mechanism -- we allow the closepacking strength on diquark production -- \ie, $c_P$ in eq.\eqref{eq:kappaEff} -- to not match that used by the other flavour probabilities.
Thus we introduce additional parameter \parnew{ClosePacking:enhanceDiquark}, which sets $c_P$ used in calculating $\kappa_{\text{eff}}$ for modifying \parold{StringFlav:ProbQQtoQ}.
Note that even with \parnew{ClosePacking:enhanceDiquark} set to zero, due to $\tilde{\alpha}$ having dependence on the other flavour probabilities, the diquark probability will still be altered by closepacking. 
Thus to further capture the ambiguity of diquark production modification, we introduce parameter \parnew{ClosePacking:doEnhanceDiquark}, which allows for any modification of the diquark probability due to closepacking to be switched off entirely.

\subsubsection*{Closepacking effects on $\mathbf{p_\perp}$}

Due to the Gaussian form of the Schwinger mechanism with respect to transverse mass, the transverse momentum spectra for string breaks should also be effected by string tension changes. In \Py, the width of the $p_\perp$ spectrum of string breaks is given by $\sigma^2$, with $\sigma$ set by parameter \parold{StringPT:sigma}. In terms of the string tension, the average transverse momentum can thus be written as 

\begin{equation}
	\sigma^2 = \langle p_\perp^2 \rangle = \frac{\pi}{\kappa_0} \int_0^\infty p_\perp^2 \exp\left( \frac{-\pi p_\perp^2}{\kappa_0} \right) dp_\perp^2 = \frac{\kappa_0}{\pi}.
	\label{eq:pTavg}
\end{equation}

Given this simple dependence on the string tension, the modified width can be written as per \eqRef{eq:sigmaEff} in terms of the ratio of string tensions, effectively scaling \parold{StringPT:sigma} by a multiplicative factor of $\sqrt{\kappa_{\text{eff}}/\kappa_0}$.

\begin{equation}
	\sigma'^{2} = \frac{\kappa_{\text{eff}}}{\pi} = \frac{\kappa_{0}}{\pi} \frac{\kappa_{\text{eff}}}{\kappa_0} = \sigma^2 \frac{\kappa_{\text{eff}}}{\kappa_0}.
    \label{eq:sigmaEff}
\end{equation}

To calculate the effective tension used for string-break $p_\perp$-distribution modifications, \parnew{ClosePacking:enhancePT} is used to set the strength parameter $c_P$ used in \eqRef{eq:kappaEff}.

\subsection{Strange junctions}
\label{sec:strangeJunc}

The apparent surplus of proton production in \Py given the overprediction of $p/\pi$ ratio in conjunction with the enhancement of multi-strange baryons seen by ALICE~\cite{ALICE:2017jyt} is suggestive of strangeness enhancement effects possibly being focused in the baryon sector. Evidently though, the strangeness enhancement effects are not exclusively in the baryon sector as can be seen by the strange meson ratio increases (\eg $K_s/\pi$ and $\phi/\pi$ ratio increases with multiplicity), however the level of increase in the multi-strange baryon sector certainly brings into question whether there are different considerations for strange baryon production. 
In this section, we explore the possibility of strange baryon enhancement being sourced from junction-baryon formation specifically.

\begin{figure}[t]
\centering
        \includegraphics[width = 0.7\textwidth]{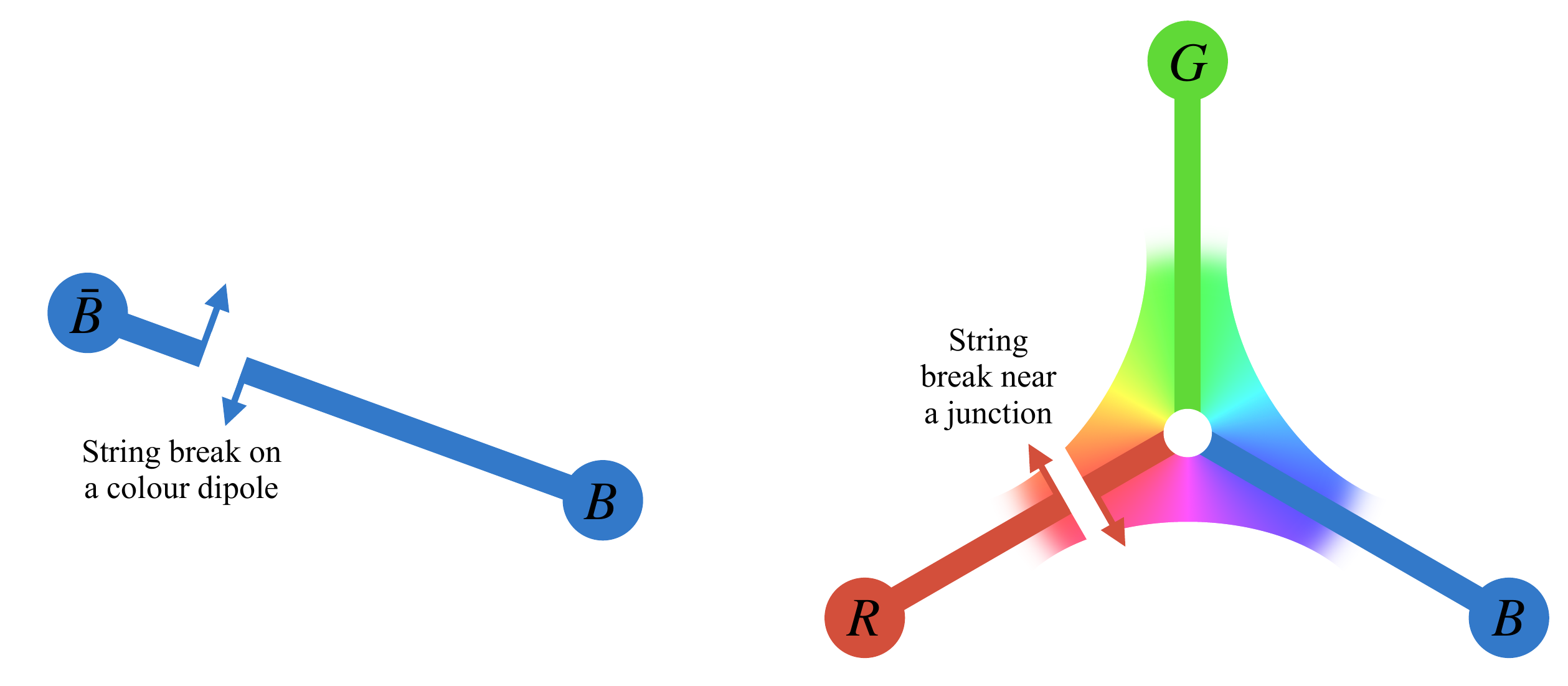}
    \caption{Here we show side-by-side a dipole and junction string to demonstrate the motivation for the strange-junction model. On the left is a simple colour dipole string in vacuum with a single string break. On the right is a junction string topology, with a string break near the junction. The ambiguity of the colour field description surrounding the junction itself is pictorially represented by the continuous colour field between junction legs nearby the junction. This of course is only a pictorial example to convey such ambiguity with possible unknown properties close to the junction, and by no means should be taken as an exact representation of what the junction field should look like.}
    \label{fig:strangeJunc}
\end{figure}

One approach is to exploit the ambiguity around what the properties of the colour field near an SU(3) junction are, a question which has recently been explored, \eg, in \cite{Komargodski:2024swh}. The simple sketch in fig.~\ref{fig:strangeJunc} illustrates that one may question whether the properties of the confinement field around a junction are identical to those in a simple dipole string in vacuum. Within the closepacking paradigm, one may suppose that the large observed enhancements of strange baryons in particular could in some way be related to an enhanced energy density per unit length around a junction, which would primarily affect string breaks ``near'' the junction. This in turn would focus strangeness enhancement around junctions, and hence show up primarily in the baryon sector. 

Of course in order to create a $s\bar{s}$ pair from a string break next to a junction, if the junction baryon contains the $s$ quark then the next adjacent hadron must contain the $\bar{s}$ antiquark, meaning this mechanism would enhance not only junction-baryon strangeness but also meson strangeness ``near'' junctions. Nonetheless it does ensure a focus on strangeness in the baryon sector, and the relative scaling of the production of different strange hadrons species (\eg, with respect to variables such as charged multiplicity) will be varied comparative to other strangeness enhancement models such as closepacking.

We will refer to this ambiguity of the junction field strength as ``strange junctions'' throughout the remainder of this text. 
In the implementation of strange junctions, we allow the strangeness enhancement to only be effective for the first string break next to the junction itself. 
In \Py, this can be scaled by the parameter $J_{s}\in [0,1]$\\ (\parnew{StringFragmentation:enhanceStrangeJunction}).  $J_{s}=1$ corresponds to maximal strangeness around junctions, while $J_{s}=0$ corresponds to no (additional) strangeness enhancement around junctions.
As such, the strangeness probability for breaks next to the junction are modified according to,
\begin{equation}
    P'(s:u/d) = P(s:u/d)^{1-J_{s}}.
    \label{eq:strangeJunc}
\end{equation}
This additional strangeness enhancement has been implemented subsequent to closepacking effects on the strangeness probability. 


\subsection{Popcorn destructive interference}
\label{sec:popDestr}

In this section we explore the expected modification of diquark production in dense string environments given the popcorn mechanism~\cite{Eden:1996xi, Andersson:1984af}. 
Contrary to the closepacking assumption of diquark probabilities scaling upwards in higher-density string systems, here we consider a model which drives the diquark production rate in the opposite direction. 
Building upon the existing popcorn mechanism, we introduce a novel mechanism which instead reduces diquark formation via a combination of physics and colour-algebra arguments. 

In the popcorn mechanism as outlined in Section~\ref{sec:baryonProd}, diquark production is described by successive colour fluctuations on a string that are allowed to form real diquarks given the necessary colour configurations. 
Given multiple strings in close proximity, suppose these virtual colour fluctuations are allowed to colour-connect to other nearby strings prior to diquark formation occurring. This would mean that a colour fluctuation on one string could allow for the formation of a real quark-antiquark break on another. 
Effectively this would prevent a portion of diquarks from forming, reducing the overall diquark -- and hence baryon -- production rate. 

\begin{figure}[t]
\centering
    \includegraphics[width = 0.55\textwidth]{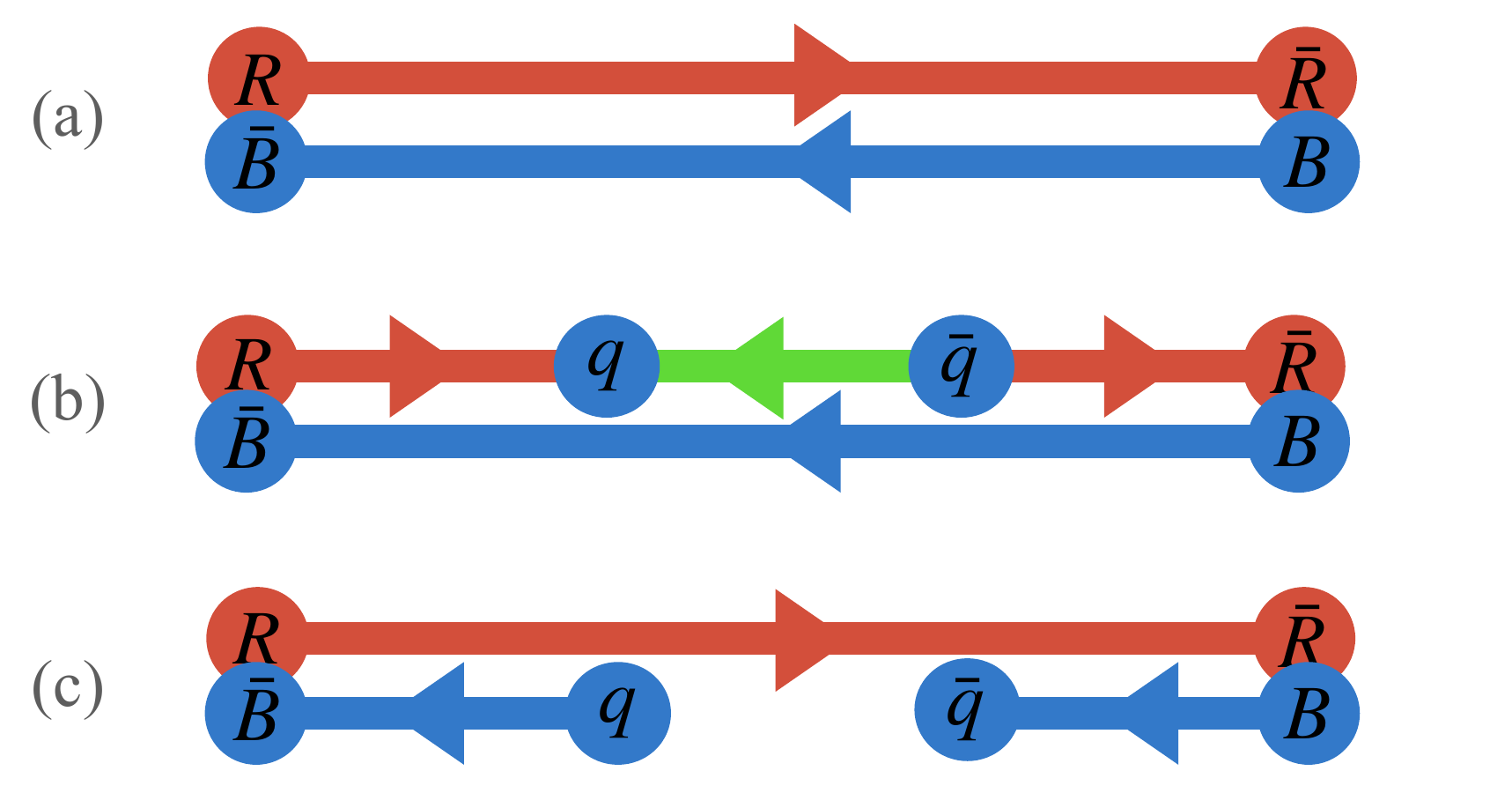}
    \caption{An illustration of the stepwise process of the popcorn mechanism given a red-antired string. (a) the unbroken red-antired string next to a blue-antiblue string with oppositely oriented flux. (b) a blue-antiblue virtual colour fluctuation on the red-antired string, which results in a short segment of ``green-coloured" string. (c) the blue-antiblue fluctuation breaks the nearby blue string.}
    \label{fig:altmannMech}
\end{figure}

To illustrate this notion, let us consider the same example as per fig.~\ref{fig:popcorn}; a red-coloured string with an initial blue-antiblue colour fluctuation. 
Now consider the same scenario but also with a blue-colour string piece nearby as shown in fig.~\ref{fig:altmannMech}(a)~-~(b). 
Should the nearby blue string be able to form a colour singlet with the blue-antiblue colour fluctuation formed on the red string, suppose the fluctuated quark pair was allowed to form a real break the nearby blue-colour string with some probability.  This processes is illustrated in fig.~\ref{fig:altmannMech}(c), and will thence prevent a fraction of diquark formation on the red string.

To implement this so-called popcorn destructive interference effect in \Py, rather than dynamically modelling the colour fluctuations themselves, the net effect is modelled via a reduced diquark production probability that scales with the density of the string environment. 
To do so, we require some probability that a given virtual colour fluctuation will colour-connect with an overlapping string provided the correct colour configuration. 
Parameters \parnew{ClosePacking:baryonSup} ($\mathcal{P}_{q}$) and \parnew{ClosePacking:parallelBaryonSup} ($\mathcal{P}_{p}$), give this interference probability for antiparallel and parallel string orientations respectively. 
The setting of such probabilities for each flux orientation is included as flux sensitivity of this mechanism is reasonably expected. 
As evident in the scenario outlined in fig.~\ref{fig:altmannMech}, in order for the blue-antiblue colour fluctuation to become a real break on the blue string, the favourable orientation is for the blue string to be antiparallel to the red.

Important to note is that in the above thus far we have simply represented colour as red, green, or blue. However more realistically one needs to consider accurately describing SU(3) colour-space probabilities. 
According to SU(3) colour algebra we have $3\otimes \bar{3} = 8 \oplus 1$, which effectively gives a 1/9 probability of a given colour-anticolour pair forming a singlet state. This means that for $n$ nearby strings, we expect on average $n/9$ strings to have the correct colour configuration to form a colour singlet with a given virtual colour fluctuation. 
Per nearby string with the necessary colour configuration, the survival probability of a given fluctuation is $(1-\mathcal{P})$. 
As the fluctuation needs to survive each nearby string, with the inclusion of flux sensitivity the modified diquark formation probability is therefore defined by,

\begin{equation}
	P'(qq:q) = P(qq:q)(1-\mathcal{P}_q)^{n_q/9} (1-\mathcal{P}_p)^{n_p/9},
	\label{eq:modPopcorn}
\end{equation}
with $n_p$ and $n_q$ being the total number of overlapping parallel and antiparallel strings respectively.

\begin{figure}[t]
\centering
        \includegraphics[width = 0.7\textwidth]{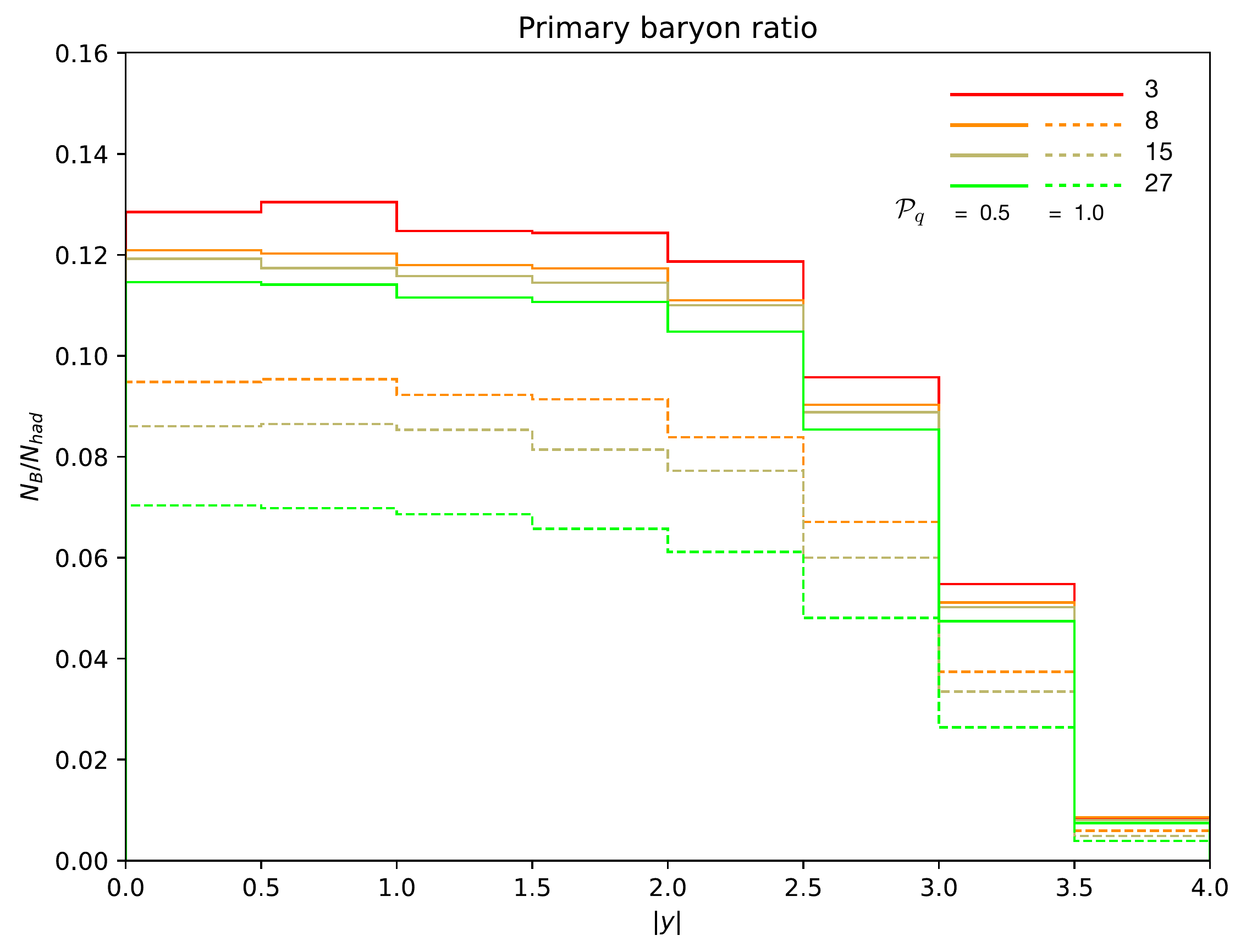}
    \caption{Baryon probabilities with respect to absolute rapidity for various string multiplet configurations, with string configurations constructed using parallel $u\bar{u}$ strings along the beam axis, with endpoint energies of 20~GeV.}
    \label{fig:QQproof}
\end{figure}

Additionally, similarly to closepacking effects being lessened when strings are further apart from one another, one would expect the same notion here with the probability of a fluctuation's colour-connection to a nearby string reducing with separation. Using the same trial-hadron method as for closepacking, with  the same $p_\perp$ suppression form on the connection probabilities $\mathcal{P}_{q/p}$, we get the full form of the modified diquark probability according to popcorn destructive interference,

\begin{equation}
	P'(qq:q) = P(qq:q)\left(1-\frac{\mathcal{P}_q}{1+\frac{p_{\perp had}}{p_{\perp 0}}}\right)^{n_q/9} \left(1-\frac{\mathcal{P}_p}{1+\frac{p_{\perp had}}{p_{\perp 0}}}\right)^{n_p/9}.
	\label{eq:modPopcornPT}
\end{equation}

Fig.~\ref{fig:QQproof} provides proof of concept for the implementation of the popcorn destructive interference effect, clearly demonstrating that the baryon fractions reduce for higher multiplet representations, and more so for larger values of $\mathcal{P}_q$.

\section{Constraints and Results}
\label{sec:results}
In this section, we will refine the parameters used for the models introduced in the preceding text, with the primary aim to fit the ALICE strange hadron-to-pion ratios simultaneously with the typically overpredicted proton-to-pion ratio~\cite{ALICE:2017jyt}. 
We will make comparisons to the existing Monash 2013 tune~\cite{Skands:2014pea}, QCD CR (time dilation mode $m_\tau = 2$)~\cite{Christiansen:2015yqa}, the Rope hadronization model~\cite{Bierlich:2014xba, Bierlich:2017sxk}, and the closepacking tune from~\cite{LorenzoThesis} which we label the ``Trieste 1'' tune, or T1 for short. Below we shall give brief overviews of each of these models and key aspects that shall be pertinent to the following comparison and tuning discussions. We then 
 elaborate on some of the constraints imposed on the aforementioned new models, and outline the methodology used to drive our tuning procedure. 
 
 We perform three separate tunes, all using QCD CR ($m_\tau = 2$) as the foundation; one with only the strange junctions model turned on (which we shall label the ``Strange junctions'' tune), one with only closepacking turned on (``Closepacking (simple)''), and a third tune with closepacking, popcorn destructive interference, and strange junctions all used in combination, which we label the Closepacking Trieste 2 tune (``Closepacking (T2)'').

\subsection{Models for comparison}
\label{sec:modelsForComparison}

\begin{figure}[t]
    \centering
    \subfloat[]{\includegraphics[width=0.49\linewidth]{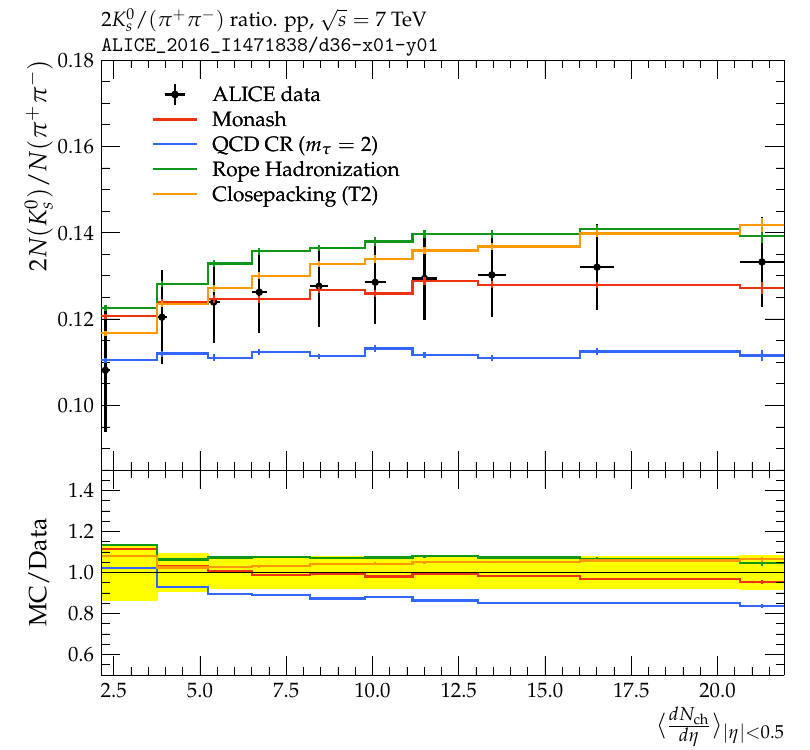}\label{fig:KsPi_old}} 
    \hfill
    \subfloat[]{\includegraphics[width=0.49\linewidth]{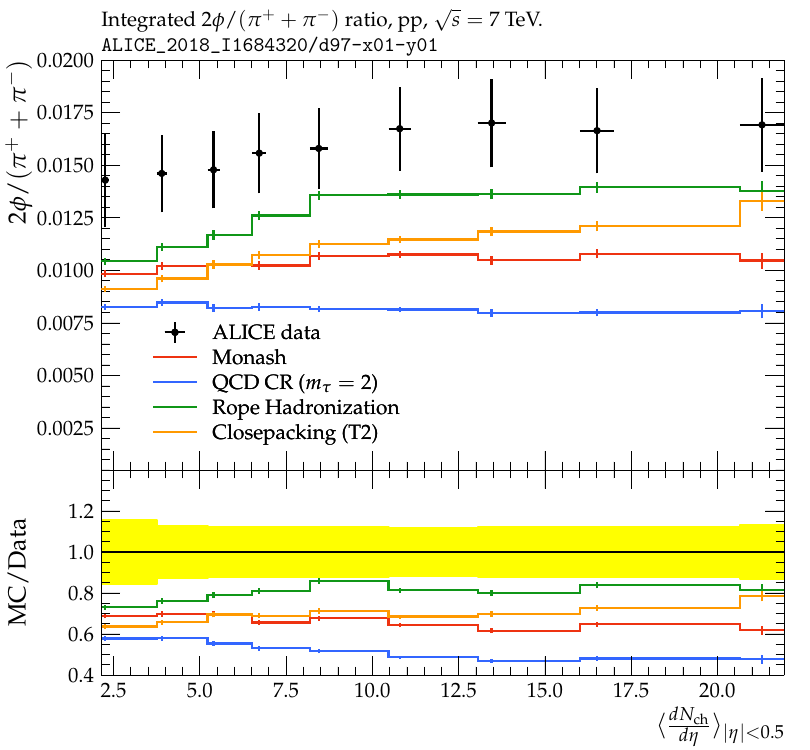}\label{fig:phiPi_old}} \\
    \subfloat[]{\includegraphics[width=0.49\linewidth]{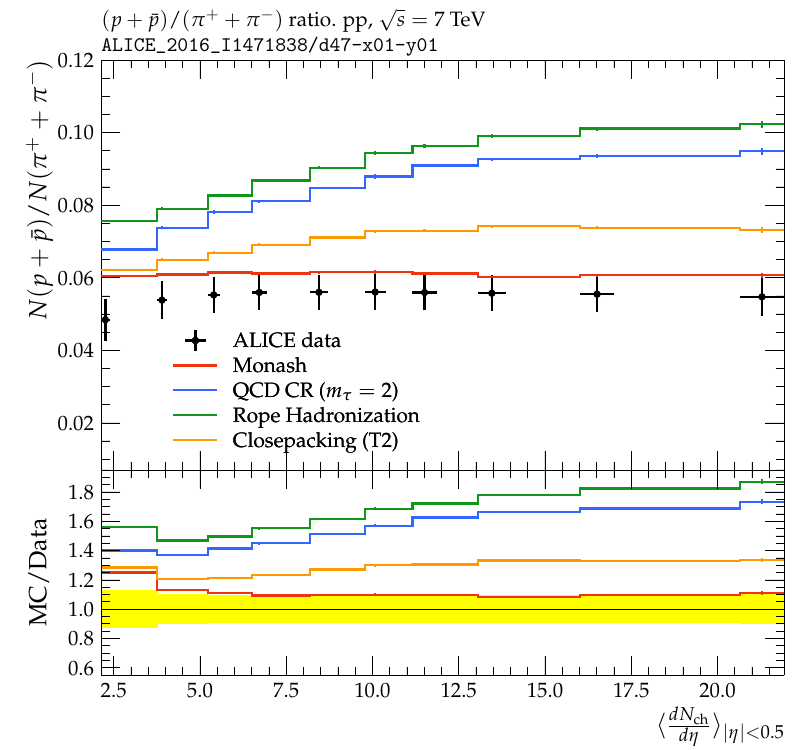}\label{fig:pPi_old}} \hfill
    \subfloat[]{\includegraphics[width=0.49\linewidth]{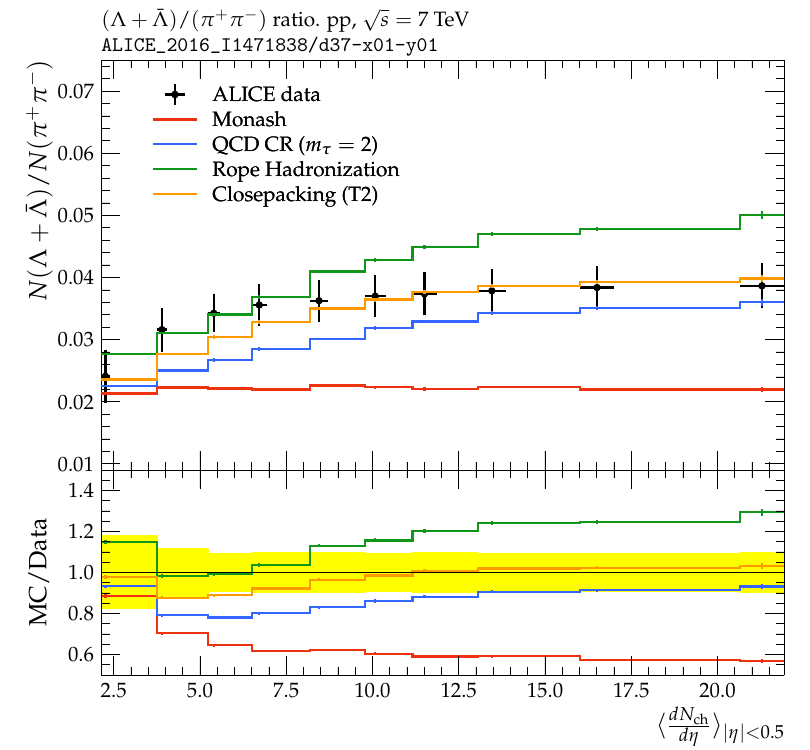}\label{fig:lambdaPi_old}} \\

    \caption{Hadron-to-pion ratios with respect to ALICE midrapidity charged multiplicity classes. \Py 8.316 simulations for parameters in \tabRef{tab:command_cards} of 7~TeV INEL>0 events are run with a lifetime cut of $\tau_{max} = 10$mm/c, and no $p_\perp$ cuts on final-state particles. Top left: $2K_S^0/\pi$~\cite{ALICE:2017jyt}. Top right: $2\phi/\pi$~\cite{ALICE:2018pal}. Bottom left: $p/\pi$~\cite{ALICE:2017jyt}. Bottom right: $\Lambda/\pi$~\cite{ALICE:2017jyt}.}
    \label{fig:ratios_old}
\end{figure}

The Monash tune is the default parameter set for \Py 8.3~\cite{Bierlich:2022pfr} and uses the standard LSM~\cite{Andersson:1983ia}. It utilizes a simple (so-called ``MPI-based'') colour-reconnection model, for which the modelling of colour space remains leading colour (\ie, number of colours $N_C\rightarrow\infty$). Accordingly, this CR model does not lead to the creation of any (new) colour junctions. With the exception of baryon beam remnants~\cite{Sjostrand:2004pf,Fieg:2023kld} (of which implications will not be elaborated on within the scope of this paper), the only source of baryon formation is diquark-antidiquark pair production in string breaks. 
The fragmentation parameters in the Monash tune are tuned to  measurements of identified-hadron yields and spectra in hadronic $Z$ decays at LEP, of which it describes quite well~\cite{Skands:2014pea}. Nonetheless, the model falls short at describing $pp$ collision data. 
Of important notice is the constant diquark and strangeness probabilities along a string, which are fixed by parameters \parold{StringFlav:ProbQQtoQ} and \parold{StringFlav:ProbStoUD} respectively.
This consequently results in relatively constant distributions of strange hadron-to-pion and baryon-to-meson ratios with respect to both $p_\perp$ and charged multiplicity as would be expected from the baseline LSM. 
This no longer remains adequate in $pp$ collision environments however, evidenced by many LHC results. A clear example of such discrepancy in the strange sector is given by the strangeness enhancement with respect to charged multiplicity observed by the ALICE collaboration~\cite{ALICE:2017jyt, ALICE:2018pal}, as demonstrated by fig.~\ref{fig:ratios_old}.
Interestingly, despite the diquark production rate \parold{StringFlav:ProbQQtoQ} being tuned LEP events, the Monash tune overpredicts the proton-to-pion ratio in $pp$ collision environments as seen in fig.~\ref{fig:pPi_old}.

The next model we compare to in this paper is called the QCD CR model~\cite{Christiansen:2015yqa}. The key defining feature of this model is the stochastic reintroduction of colour-space ambiguities according to SU(3) algebra, including the formation of so-called junction string topologies~\cite{Sjostrand:2002ip,Christiansen:2015yqa,Altmann:2024odn}. 
Details of the implementation of this beyond-leading-colour model will not be outlined in this paper, however for a simple review on QCD CR see~\cite{Altmann:2024kwx}. 
In particular, we will compare to a variation of the tune called ``CR (time-dilation mode $m_\tau = 2$)’’, with parameters listed in Appendix~\ref{app:modelParm} \footnote{The parameter values for \parold{StringFlav:probQQ1toQQ0join} used throughout this paper are varied from the initial tuning on QCD CR ($m_\tau = 2$) as per~\cite{Christiansen:2015yqa}. Instead we return to the default \Py values for \parold{StringFlav:probQQ1toQQ0join} in order to better fit the $\Sigma^{0,+,++}_c/D^0$ and $\Lambda^+_c(\leftarrow\Sigma^{0,+,++}_c)/\Lambda^+_c$ ratios (see \cite{Altmann:2024kwx}).}.

In $e^+e^-$ collision events, the corrections of including colour-space ambiguities are expected to be approximately $1/N_C^2\sim 10\%$.
In $pp$ collision events however, denser partonic systems allow for more colour space overlaps, thus resulting in enhanced QCD CR effects.
Of particular interest is the significance of the inclusion of junction strings as an additional baryon production mechanism.
Junction baryons have proven effective in the description of the $\Lambda/K_S$ ratio~\cite{CMS:2011jlm} in $pp$ collisions, alongside well describing heavy flavour baryon distributions such as the $\Lambda_c/D^0$ ratio~\cite{ALICE:2021rzj} and the $\Lambda_b$ asymmetry~\cite{LHCb:2021xyh} with respect to transverse momentum. 
Despite the success of describing the aforementioned baryon distributions, there remain several shortcomings to the model.
Given the Monash tune already overpredicts the $p/\pi$ ratio in $pp$ collisions, the inclusion of junction baryons only further pronounces such an overprediction as demonstrated in fig.~\ref{fig:pPi_old}. 
Additionally, QCD CR does not provide any strangeness-enhancement mechanism. 
CR effects -- including junction-baryon formation -- are correlated with the density of the partonic environment, of which the charged multiplicity ($N_{ch}$) is also correlated. 
This means that the model results in an enhancement of baryon production with respect to charged multiplicity, however such baryon enhancement cannot describe the level of enhancement required for multi-strange baryon-to-meson ratios, nor can the observed enhancement of strange meson-to-pion ratios (\eg $K_S/\pi$ and $\phi/\pi$ shown in fig.~\ref{fig:KsPi_old} and fig.~\ref{fig:phiPi_old} respectively) be explained. 

Another model we shall compare to here is Rope hadronization~\cite{Bierlich:2014xba, Bierlich:2017sxk}. 
The Rope hadronization model is designed to produce strangeness enhancement and thus is important for comparison here. In this model, a collection of nearby strings is treated as a coherent system -- namely a ``rope’’ -- which then fragments with an enhanced string tension correlated to the number of strings. 
Consequently the model results in enhancement of both strangeness and diquark production with respect to charged multiplicity. Given the default values of the Rope hadronization model, sufficient levels of strangeness enhancement are present in order to fit both the strange meson $K_S/\pi$ ratio (see fig.~\ref{fig:KsPi_old}) alongside the multi-strange baryon $\Omega/\pi$ ratio.
Nonetheless, there remains a problem with the overprediction of the $p/\pi$ ratio (see fig.~\ref{fig:pPi_old}).

One should also note that the Rope hadronization model takes a similar approach to strangeness enhancement which we have taken in our closepacking model outlined in Section~\ref{sec:closePacking}, however closepacking aims to be a somewhat simplified model of the same principle. 
The key difference in the two approaches lies in the treatment of the surrounding strings.
The Rope hadronization model assumes the formation of coherent colour-rope structures, whereas the closepacking model simply treats the surrounding strings as an effective background on a given fragmenting string. 
This variance in approaches means that whilst the closepacking model remains a fully momentum-space implementation, Rope hadronization requires space-time mappings of string breaks. Such a fully momentum-space fragmentation procedure for closepacking allows for both a more natural compatibility with junction topologies, alongside a reduced execution speed. 

The final tune we shall compare to is the earliest closepacking tune obtained in~\cite{LorenzoThesis}, using a version of the closepacking implementation from \Py 8.311. We will label this tune ``Closepacking (T1)'' throughout the following. 

For completeness, parameter specifications for each of the reference models mentioned above are given in \appRef{app:modelParm}.

\begin{figure}[t]
    \centering
    \subfloat{\includegraphics[width=0.53\linewidth]{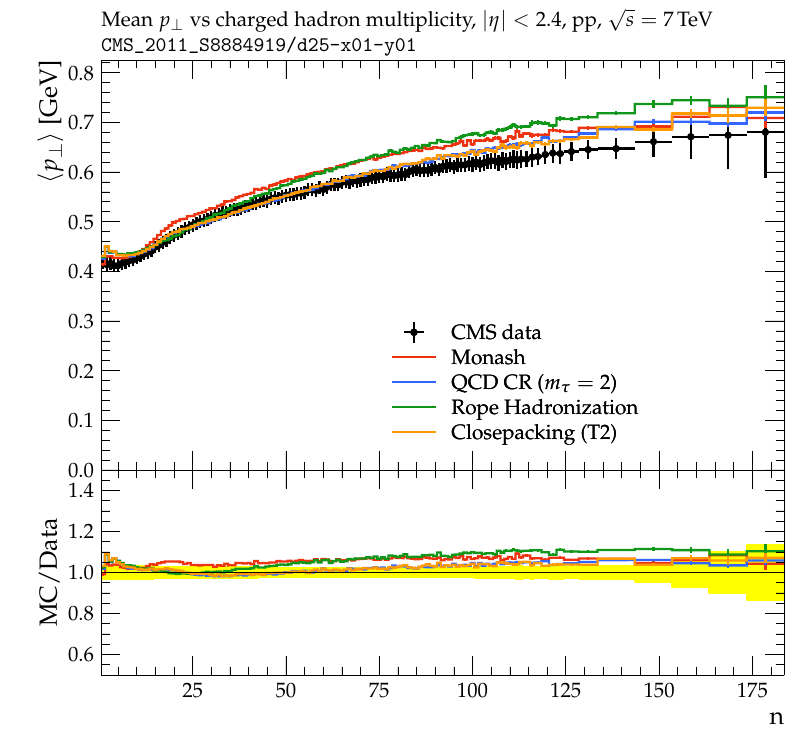}} 
    \caption{Average $p_\perp$ with respect to charged multiplicity for 7~TeV NSD events at CMS~\cite{CMS:2010qvf}. \Py 8.316 simulations for tunes listed in \tabRef{tab:command_cards} are run with a lifetime cut of $\tau_{max} = 10$mm/c, and no $p_\perp$ cuts on final-state particles.}
    \label{fig:pTavg_old}
\end{figure}

\subsection{Tuning methodology}
\label{sec:tuning}

\begin{figure}[tp]
    \centering
    \subfloat{\includegraphics[width=0.6\linewidth]{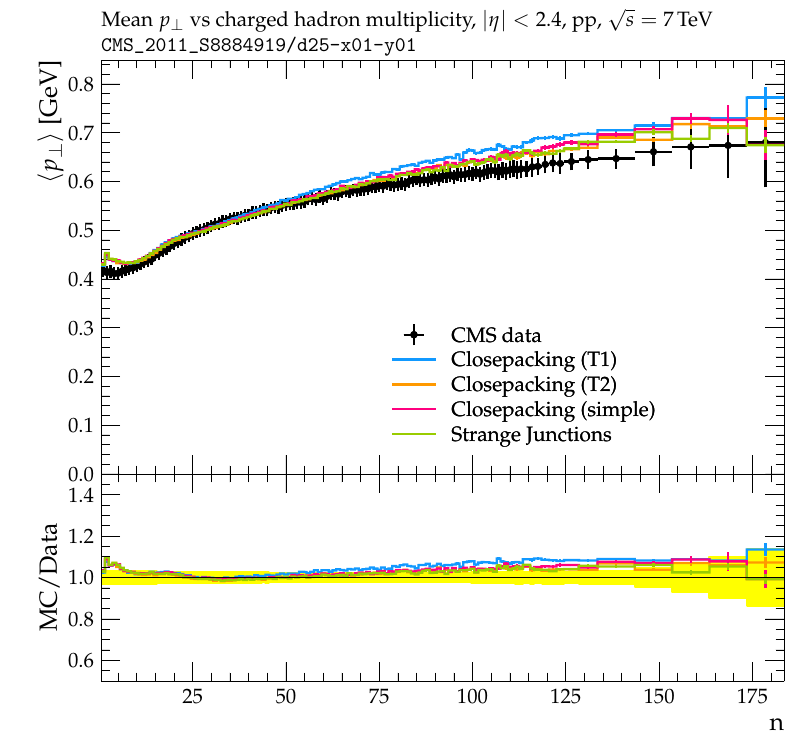}} 
    \caption{Average $p_\perp$ with respect to charged multiplicity for 7~TeV NSD events at CMS~\cite{CMS:2010qvf}. \Py 8.316 simulations for tunes listed in \tabRef{tab:tunes} are run with a lifetime cut of $\tau_{max} = 10$mm/c, and no $p_\perp$ cuts on final-state particles.}
    \label{fig:pTavg_new}
\end{figure}

Given the key interest of the above collectivity models is to reproduce effects expected in $pp$ collisions, the QCD CR (time-dilation mode $m_\tau = 2$)~\cite{Christiansen:2015yqa} will be the basis for the proceeding tune. 
We will focus on describing the ALICE strange hadron-to-pion ratios simultaneously with the proton-to-pion ratio, whilst maintaining consistency with other global $pp$ observables such as $\langle p_\perp\rangle$ (see fig.~\ref{fig:pTavg_old} and~\ref{fig:pTavg_new}) and $dn_{ch}/d\eta$. This means that we require the tuning of the additional new parameters introduced for closepacking modelling, additional to the retuning of MPI and CR parameters. 

In the following, we will use Professor~\cite{Buckley:2009bj} to find best fit values for parameters.
The workflow with Professor can be divided into four steps; sampling, event generation, interpolation, and $\chi^2$ minimization. The first one consists in \textit{sampling} the parameter space of the $P$ parameters that one wants to tune, defining the corresponding variation ranges $\vec{p_\textup{min}}$ and $\vec{p_\textup{max}}$ (using the same notation of the original paper~\cite{Buckley:2009bj}). By default, sampling is performed uniformly within the $P$-dimensional hypercube $[\vec{p_\textup{min}},\vec{p_\textup{max}}]$. At each sampled point, the MC generator is run for the desired Rivet \cite{Rivet3} analyses, producing histograms saved as YODA files~\cite{Buckley:2023xqh}. Upon conclusion of the simulations, the Professor tool can read all the Rivet histograms corresponding to the different parameter values. It then performs a fit of the MC response $f_b(\vec{p})$, specifically for the same bin $b$ of each observable $\mathcal{O}$, using polynomial functions of the desired order. This process, known as \textit{interpolation}, yields a continuous parameterization of the MC response, as a function of the tuning parameters. The tuning results presented in this work were obtained using $4^\textup{th}$- and $5^\textup{th}$-order polynomials for the interpolation step. The higher the degree of the polynomial and the number of parameters to tune, the higher the number of sampled points required  for the interpolation step. For example, in a $4$-dimensional parameter space, the minimum number of sampled points required is 35 with $3^\textup{rd}$-order polynomials. This number increases to 70 and 126 using $4^\textup{th}$ and $5^\textup{th}$-order polynomials, respectively. Moreover, \cite{Buckley:2009bj} suggests an oversampling by a factor of 2.

The parameterization of the MC is subsequently used to construct a weighted $\chi^2$ along with the experimental data:
\begin{equation}
    \label{eqn:prof_chi2}
    \chi^2 = \sum_\mathcal{O} w_\mathcal{O}^2 \sum_{b \,\in\, \mathcal{O}} \frac{\bigl(d_b-f_b(\vec{p})\bigr)^2}{\sigma_{d_b}^2+\sigma_{f_b}^2},
\end{equation}
where $d_b$ are the data for the bin $b$ of the observable $\mathcal{O}$, which are automatically retrieved by Professor from Rivet \cite{Rivet3}. Variables $\sigma_{d_b}$ and $\sigma_{f_b}$ are the errors associated to the experimental measurement and to the MC parameterization, respectively, and $w_\mathcal{O}$ is the weight associated to the observable $\mathcal{O}$ as chosen by the user.

To prevent overfitting a flat $5\%$ ``theory uncertainty''~\cite{Skands:2014pea,Amoroso:2018qga,Jueid:2022qjg,Jueid:2023vrb} is applied to the $\chi^2$ as follows:
\begin{equation}
    \label{eqn:gof}
    \chi^2_{5\%} = \sum_\mathcal{O} w_\mathcal{O}^2 \sum_{b \,\in\, \mathcal{O}}  \frac{\bigl(d_b-f_b(\vec{p})\bigr)^2}{\sigma_{d_b}^2+\sigma_{f_b}^2+(0.05\,f_b)^2},
\end{equation}
For the purpose of this tuning, $\chi^2_{5\%}$ is employed. The minimization of this function performed with Professor returns the vector of the best parameter values $\vec{p}_\textup{best}$.
Throughout this work YODA 1.9.10, Rivet 3.1.10 \cite{Rivet3}, Professor 2.4.2 \cite{Buckley:2009bj} are used. Plots comparing different tunes predictions are obtained using the \texttt{rivet-mkhtml-mpl} script with some additions. Moreover, for each plot a goodness of fit is computed and shown for each observable using a similar definition implemented in CERN MCPLOTS \cite{Karneyeu:2013aha, Korneeva:2024oho},
\begin{equation}
    \chi^2_\textup{GoF} = \frac{1}{N_\textup{bins}} \sum_{b=1}^{N_\textup{bins}} \frac{(d_b - \mathrm{MC}_{b})^2} {\sigma_{d_b}^2 + \sigma_{\mathrm{MC}_b}^2 + (0.05\,\mathrm{MC}_{b})^2},
\end{equation}
where $N_\textup{bins}$ is the number of bins of the histogram, $\mathrm{MC}_{b}$ the value of the MC prediction in bin $b$ and $\sigma_{\mathrm{MC}_b}$ its corresponding uncertainty. A $5\%$ ``theory uncertainty'' is also added in this case. 
In ~\tabRef{tab:GOF}, the goodness of fit level is shown for various observables used in the tuning procedure for each of the tunes, using a gradient colour scheme from green to purple/black, with green corresponding to a low $\chi^2_\textup{GoF}$ value, and purple a high $\chi^2_\textup{GoF}$ value. 

We note that, since most of the experimental measurements we compare to are dominated by systematics, we expect $\chi^2_\textup{GoF} \ll 1$ for a model that goes ``straight through'' the data, rather than $\chi^2_\textup{GoF} \sim 1$ as would be the case if the data were statistically scattered within their uncertainty bands. However, our main focus here will be on which cases exhibit $\chi^2_\textup{GoF} \gg 1$. We do not mean to imply that a model that achieves $\chi^2_\textup{GoF} \sim 0$ is much better than one that achieves $\chi^2_\textup{GoF} \sim 1$ although the $\chi^2_\textup{GoF}$ minimiser will of course select for the lowest possible values.


To constrain the dimensionality of the problem, the tuning procedure has been split into several stages with each targeting a focused subset of observables. After each stage in this tuning procedure, parameters are then either fixed according to the best fit or allowed to remain floating with a more tightly constrained range in the proceeding tuning stage. 
Each stage of this tuning methodology has been chosen in order to aim to minimize the correlation between parameters across stages, with the target areas as follows (listed in order of approach); MPI and CR levels, baryons and strangeness, and $p_\perp$ dependence. 

Preceding the explanation of the tuning scheme, below we outline which parameters are assumed as fixed in our tune, and the justification for such settings. 
Firstly, in order to remain consistent with strange/diquark production rates expected in $e^+e^-$ collisions, flavour probabilities such as \parold{StringFlav:probStoUD} and \parold{StringFlav:probQQtoQ} will remain fixed according to the QCD CR ($m_\tau = 2$) tune. In principle, closepacking may effect a subset of $e^+e^-$ events that involve nearby jets, however we expect these effects to be minimal. Regardless, the existing model for both closepacking and popcorn destructive interference is designed only for application to hadronic collisions given the beam axis dependence which is based on the assumption that the bulk of strings are oriented along the beam axis. 
This therefore means the models are not yet useful for describing $e^+e^-$ collisions. 
Generalization of the model applications to arbitrary collision environments remains work for a future study.

We also acknowledge that the newly introduced parameters that dictate flux sensitivity of the models -- \parnew{ClosePacking:parallelBaryonSup} and \parnew{ClosePacking:fluxRatio} -- cannot be finely tuned given the lack of flux-sensitive experimental observables available. In order to tune such flux sensitivity, one would need to be able to differentiate between multiplet configurations in cleaner collision environments, such as $e^+e^-$ collisions, and then compare hadronic distributions sourced from the fragmentation of \eg an octet vs a sextet configuration. 
Given we do not have such experimental observables at hand, we instead make reasonable default assumptions for the values of the flux-sensitive parameters. 
For the popcorn destructive interference modelling, we assume \parnew{ClosePacking:parallelBaryonSup} ($\mathcal{P}_p$ per \eqRef{eq:modPopcornPT}) is always set to zero, so that the contribution of the interference effect is isolated to antiparallel string configurations. This is physically sensible as antiparallel strings are the favourable orientation for the mechanism to occur and would thus be expected to be the dominating contribution. For tension modifications via closepacking, \parnew{ClosePacking:fluxRatio} ($\omega$ per \eqRef{eq:kappaEff}) is fixed at 0.5, which is aligned to the strength of flux sensitivity expected by Casimir scaling. 

Additionally, we set \parnew{ClosePacking:doEnhanceDiquark~=~off}. This prevents the string-tension modifications due to closepacking from scaling the diquark probability upwards in a manner assuming Schwinger-type production of diquarks per \eqRef{eq:probQQtoQmod}. 
Instead, as we wish to simultaneously implement closepacking alongside the popcorn destructive interference mechanism introduced in \secRef{sec:popDestr}, we must then assume popcorn-type diquark formation which one would not necessarily expect to scale in the same way as \eqRef{eq:probQQtoQmod}. 
How exactly popcorn diquark formation probabilities should scale with an increased string tension remains somewhat ambiguous.
Regardless, if one were to allow the diquark probability to increase due to enhanced tensions, this effect would then be directly counteracted by popcorn destructive interference, and thus it would not make sense to allow both scalings to occur in conjunction. 
As the popcorn destructive interference has not been dynamically implemented, but rather the effectively implemented via a suppression factor on the diquark probability controlled by \parnew{ClosePacking:baryonSup}, we allow this parameter to holistically capture the effect of surrounding strings on the diquark production rate — incorporating the combination of enhanced string tensions from closepacking and popcorn destructive interference effects. This of course assumes a net decrease in the diquark production rates. Should one wish to assume a net enhancement, one should set 
\parnew{ClosePacking:baryonSup~=~0}, and use \parnew{ClosePacking:doEnhanceDiquark~=~on} whilst allowing \parnew{ClosePacking:enhanceDiquark} to control the effective diquark enhancement strength. In the following however, we will assume the diquark probability decreases with the density of the string system. 

We also fix the closepacking regularization parameter $p_{\perp 0}$ (\parnew{ClosePacking:PT0}) to a value of 1.75~GeV. This is partially as we do not possess experimental results for observables that would be particularly sensitive to this parameter. Ideally for tuning of this parameter, one would want two (anti)parallel strings with varying separations. For the $p_\perp$ observables we do have, sensitivity to the exact value of this parameter is not large. Given the hadronic $p_\perp$ distributions are not particularly well describe in general, we do not fine-tune this parameter. Expecting it to sit somewhere between 1-3~GeV, we fix it 1.75~GeV in this work, implying that the additional tension modification would halve for a $p_{\perp\text{Had}}$ of 1.75~GeV.

The final parameter which we fix is \parnew{ColourReconnection:mPseudo} ($m_\text{pseudo}$). This is a newly introduced parameter as of \Py 8.311. Formerly, both of the parameters \parold{ColourReconnection:m0} and \parnew{ColourReconnection:mPseudo} were set by the same parameter labelled \parold{ColourReconnection:m0}. These have been since factorized given their physical meanings should be treated as different. \parold{ColourReconnection:m0} is now only used as a regularization parameter in the string length measure, and in this tuning scheme is fixed to 0.5. The parameter \parnew{ColourReconnection:mPseudo} imposes a cut on the invariant mass of pseudo-particles that are not colour reconnected, and is primarily constrained by the $\langle p_\perp\rangle$ vs $n_{ch}$ distribution. In the following tuning methodology, $m_\text{pseudo}$ was fixed to the set of values 0.2, 0.4, and 0.6 in tuning steps 1 and 2. The best of these values, $m_\text{pseudo}$~=~0.4, was then selected and fixed throughout the remainder of tunes. This parameter was not allowed to float throughout the entirety of the tuning as we do not have observables capable of splitting the degeneracy between $m_\text{pseudo}$ and parameter \parold{ColourReconnection:timeDilationPar}.

With the above assumptions in place, we can now proceed with tuning the remaining parameters from the newly introduced models. 
Below we will outline each stage of the tuning scheme, including which parameters are allowed to float, which are fixed, and which observables each parameter is primarily constrained by.

\subsubsection*{Step 1: MPI and CR levels}

In this step, we aim to more tightly constrain parameters \parold{MultipartonInteractions:pT0Ref} and \parold{ColourReconnection:timeDilationPar}. A description of each of these parameters and observables to be constrained to is given below:

\begin{itemize}[itemsep=3pt]
    \item \parold{ColourReconnection:timeDilationPar}: controls the strength of CR by suppressing high-boost reconnections, constrained largely by $\langle p_\perp\rangle$ vs $n_{ch}$.
    \item \parold{MultipartonInteractions:pT0Ref}: regularization scale controlling the amount of low $p_\perp$ MPIs, strongly correlated to the total multiplicity and thus allowing tuning to $dn_{ch}/d\eta$.
\end{itemize}

Though the primary aim is to tune closepacking effects to strange hadron and baryon production rates, one must also revisit more global observables given these modified strange/diquark production rates will impact MPI and CR-sensitive distributions such as total hadronic multiplicities and average transverse hadronic momenta.
Thus a retuning of the relevant MPI and CR parameters with closepacking effects switched on is required. 
Of course as we have yet to tune the closepacking parameters themselves, in this step of the tuning procedure we instead fix these parameters to reasonable guesses. 
Doing so should yield relatively stable results in proceeding tuning steps despite $\langle p_\perp\rangle$ and $dn_{ch}/d\eta$ distributions having some dependence on closepacking effects. 
The starting parameters were chosen as follows: 
\begin{itemize}[noitemsep]
\item \parold{ColourReconnection:junctionCorrection}~=~1.5, 
\item \parnew{ClosePacking:enhanceStrange}~=~0.010, 
\item \parnew{ClosePacking:enhancePT}~=~0.040, 
\item \parnew{StringFragmentation:enhanceStrangeJunction}~=~0.5,  \item 
\parnew{ClosePacking:baryonSup}~=~0.8. 
\end{itemize} 
These starting values were chosen as reasonable guesses given previously conducted rough tuning attempts.

\subsubsection*{Step 2: Baryons and Strangeness}

In the second step of tuning, we aim to tune the strange hadron and baryon ratios such as $K_S/\pi$, $\Lambda/\pi$, $p/\pi$, etc. (see ~\appRef{app:observables} for full list). In particular, we ensure to have a balance of both strange baryon-to-pion ratios alongside strange meson-to-pion ratios. This ensures we can tune the difference in effects of closepacking via \parnew{ClosePacking:enhanceStrange} alongside strange junctions via \parnew{StringFragmentation:enhanceStrangeJunction}. Of particular note is the ratio $\phi/\pi$. As the $\phi$~meson is a double strange meson, the strange junctions mechanism will contribute lesser to the $\phi$~meson enhancement comparatively to closepacking. This is as strange junctions only contribute to strange quark enhancement for a single quark in the meson formed next to the junction baryon. Contrastingly, closepacking will contribute strangeness enhancement to both quarks equally for $\phi$~meson production. The parameters
\parold{ColourReconnection:timeDilationPar} and \parold{MultipartonInteractions:pT0Ref} are fixed from the previous stage of tuning. The parameters tuned in this step alongside the observables they are sensitive to are as follows:

\begin{itemize}[itemsep=3pt]
    \item \parold{ColourReconnection:junctionCorrection}: modifies how easily junction topologies form via a correction to the string length measure used in colour reconnections, and hence alters the baryon production rates. 
    \item \parnew{ClosePacking:baryonSup}: popcorn destructive interference strength parameter, $\mathcal{P}_q$ as per \eqRef{eq:modPopcornPT}, sensitive to baryon production rates. Particularly will be tuned to the $p/\pi$ ratio, which is a non-strange baryon-to-meson ratio that the LEP-tuned diquark production rates overpredicts. 
    \item \parnew{ClosePacking:enhanceStrange}: closepacking strength parameter $c_P$ defined in \eqRef{eq:kappaEff}, used to calculate the effective tension for modifications to the strangeness production rate, $P(s:u/d)$, per \eqRef{eq:probStoUDmod}. This is sensitive to all strange hadron ratios. 
    \item \parnew{StringFragmentation:enhanceStrangeJunction}: strength parameter for strangeness enhancement around a junction, $J_s$, as per \eqRef{eq:strangeJunc}. This parameter will be particularly sensitive to multi-strange baryon ratios \eg $\Omega/\pi$.
\end{itemize}

\subsubsection*{Step 3: $p_\perp$ dependence}
Here we aim to constrain the $p_\perp$ dependence of closepacking \parnew{ClosePacking:enhancePT}, alongside retuning the CR parameters \parold{MultipartonInteractions:pT0Ref} and \\ \parold{ColourReconnection:timeDilationPar}. The floating parameters from step 2 of the tuning procedure are fixed. 

Description of the final parameter to be tuned is below:
\begin{itemize}
    \item \parnew{ClosePacking:enhancePT}: closepacking strength parameter $c_P$ defined in \eqRef{eq:kappaEff}, used to calculate the effective tension for modifications to the width of the $p_\perp$ distribution of string breaks, $\sigma$, per \eqRef{eq:sigmaEff}. 
\end{itemize}


After step 3 of the tuning procedure, steps 2 and 3 can be iterated until the parameter values converge, or when no further significant improvement in the $\chi^2$ (and the overall description of the data) is observed. 

\subsection{Tuning outcomes}

\begin{figure}[tp]
    \centering
    \subfloat[]{\includegraphics[width=0.49\linewidth]{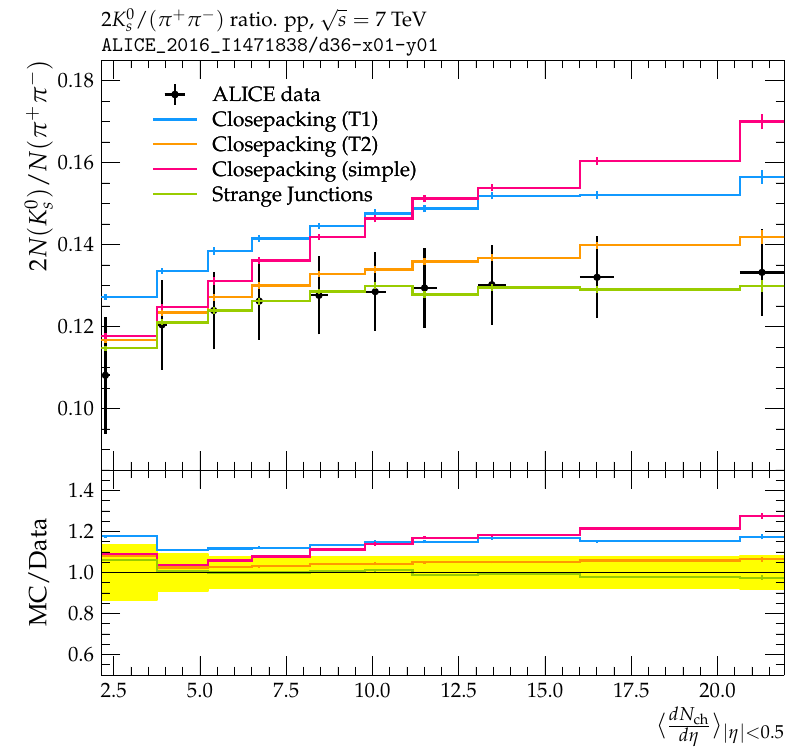}\label{fig:KsPi_new}} \hfill
    \subfloat[]{\includegraphics[width=0.49\linewidth]{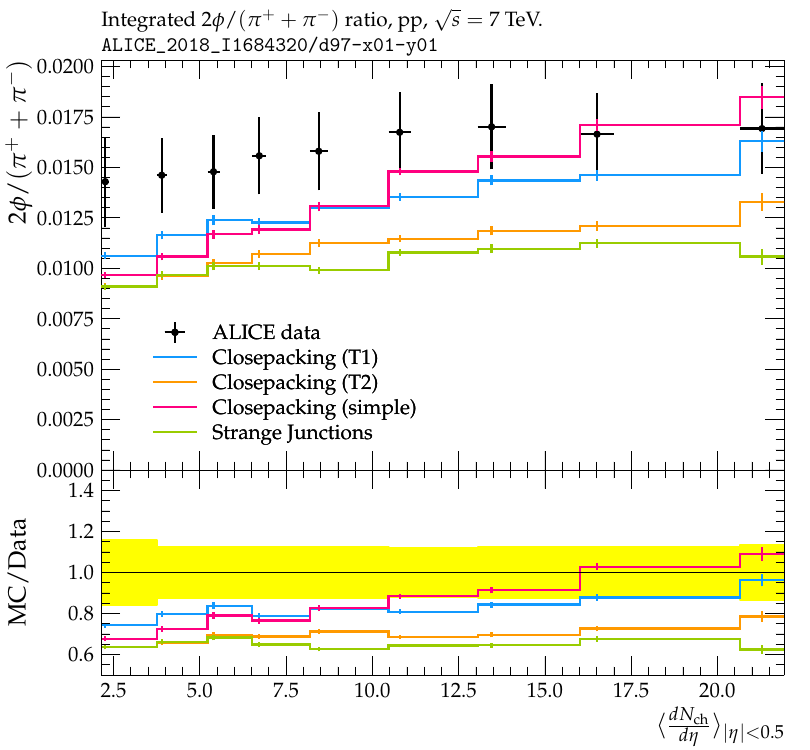}\label{fig:phiPi_new}} \\
    \subfloat[]{\includegraphics[width=0.49\linewidth]{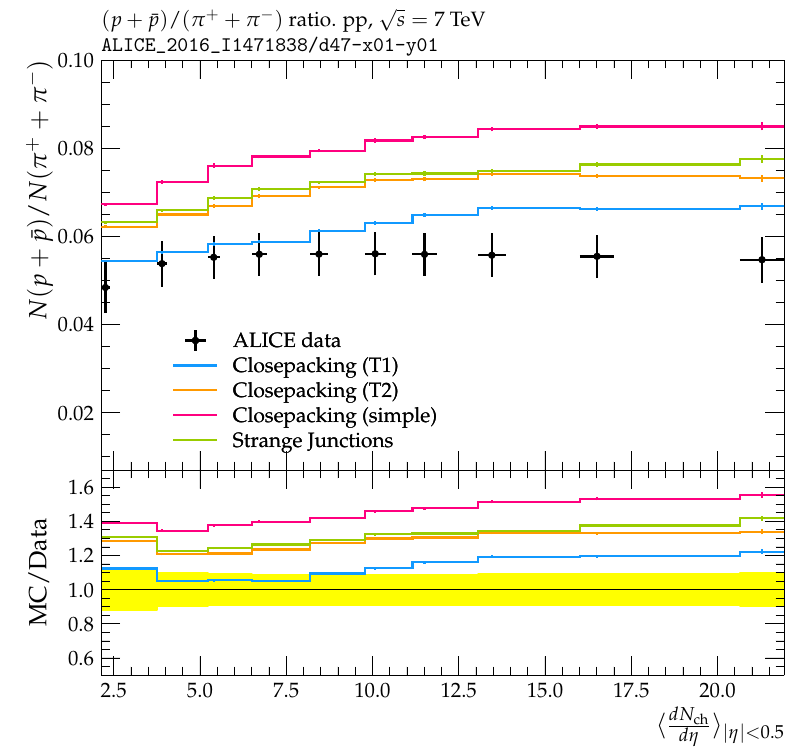}\label{fig:pPi_new}} \hfill
    \subfloat[]{\includegraphics[width=0.49\linewidth]{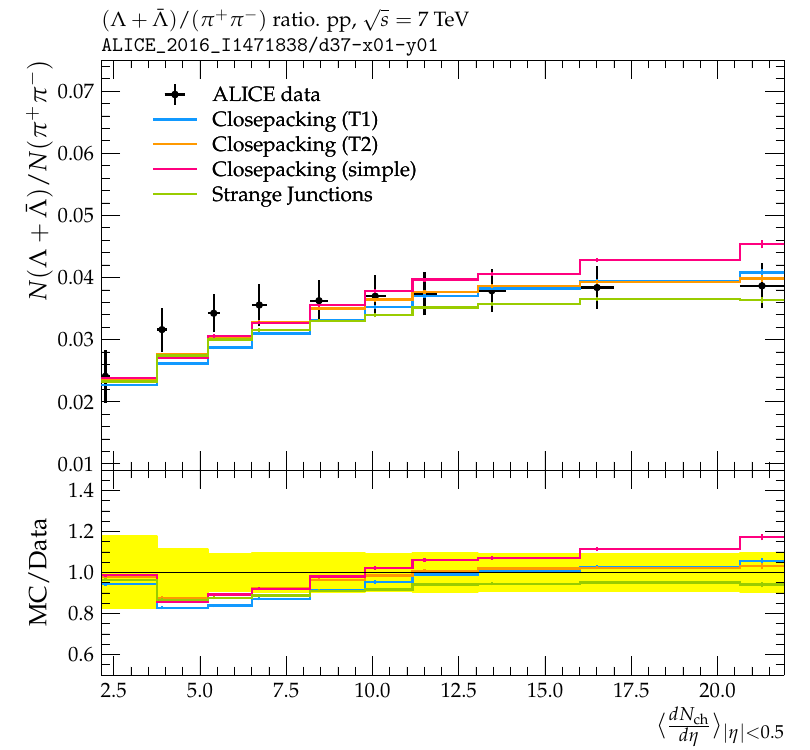}\label{fig:lambdaPi_new}} \\

    \caption{Hadron-to-pion ratios with respect to ALICE midrapidity charged multiplicity classes. \Py 8.316 simulations for parameters in \tabRef{tab:tunes} of 7~TeV INEL>0 events are run with a lifetime cut of $\tau_{max} = 10$mm/c, and no $p_\perp$ cuts on final-state particles. Top left: $2K_S^0/\pi$~\cite{ALICE:2017jyt}. Top right: $2\phi/\pi$~\cite{ALICE:2018pal}. Bottom left: $p/\pi$~\cite{ALICE:2017jyt}. Bottom right: $\Lambda/\pi$~\cite{ALICE:2017jyt}.}
    \label{fig:ratios_new}
\end{figure}

As stated above, we ran the above tuning methodology
\footnote{Throughout the tuning procedure, there were various bug fixes and improvements were made in the closepacking implementation in \Py. 
The versions of \Py, the weight configurations for the observables, and the polynomial order used for interpolation for each step in the tuning procedure can be found in app.~\ref{app:observables}.}
on three different configurations -- Strange junctions only (``Strange junctions''), closepacking only (``Closepacking (simple)''), and a combination of closepacking, popcorn destructive interference, and strange junctions (``Closepacking Trieste 2 (T2)''). The parameter values for each of these tunes can be seen in \tabRef{tab:tunes}. 

\begin{table}[tp]
    \small
    \centering
    \begin{tabular}{p{7.8cm}|lll}
       & \multicolumn{2}{c}{\centering \textbf{Closepacking}}  & \multicolumn{1}{c}{\centering \textbf{Strange}} \\
       & \textbf{Trieste 2} & \textbf{Simple}~~~ 
       & \textbf{Junctions} \\
       \toprule
       \parold{MultiPartonInteractions:pT0Ref} & = 2.22 & = 2.26 & = 2.23\\
       \parold{ColourReconnection:mPseudo}  & = 0.4 & = 0.4 & = 0.4\\
       \parold{ColourReconnection:m0}  & = 0.5 & = 0.5 & = 0.5\\
       \parold{ColourReconnection:junctionCorrection}  & = 1.67 & = 1.68 & = 0.96 \\
       \parold{ColourReconnection:timeDilationPar}  & = 0.17 & = 0.16 & = 0.19 \\
       \hline
       \parnew{ClosePacking:doClosePacking}  & = on & = on  & = off  \\
       \parnew{ClosePacking:PT0}             & = 1.75 & = 1.75 & -- \\
       \parnew{ClosePacking:enhanceStrange}  & = 0.010 & = 0.038 & -- \\
       \parnew{ClosePacking:enhancePT}  & = 0.005 & = 0.013 & -- \\
       \parnew{ClosePacking:doEnhanceDiquark}  & = off & = off  & = off  \\
       \hline
       \parnew{StringFragmentation:doStrangeJunctions}  & = on & = off  & = on  \\
       \parnew{StringFragmentation:enhanceStrangeJunction}  & = 0.57 & --  & = 0.70  \\
       \hline
       \parnew{ClosePacking:baryonSup} & = 0.54 & = 0.00 & -- \\
       \parnew{ClosePacking:parallelBaryonSup} & = 0.00 & = 0.00 & -- \\
       \bottomrule
    \end{tabular}
    \caption{Parameter values for tunes using the methodology outlined in \secRef{sec:tuning}. All other \Py parameters for each of these tunes are set according to QCD CR ($m_\tau = 2$)~\cite{Christiansen:2015yqa} with the  \parold{StringFlav:probQQ1toQQ0join} values reverted back to the default Monash tune values. Parameters for the other models investigated in this work are given in the appendix, \tabRef{tab:command_cards}.}
    \label{tab:tunes}
\end{table}

To allow for comparison of each models performance, \tabRef{tab:GOF} and \tabRef{tab:GOF2} show the goodness of fit values ($\chi^2_\textup{GoF}$ per the definition in \eqRef{eqn:gof}) for each of the model tunes for various observables used in the tuning process. 
As one can see, the $\chi^2_\textup{GoF}$ values show that the Monash and QCD CR tunes fail to describe the majority of strange hadron ratios, and though the Rope hadronization model well describes the strange hadron-to-meson ratios, the $p/\pi$ ratio is overpredicted. The remainder of the $\chi^2_\textup{GoF}$ values also reflect observations pointed out in \secRef{sec:modelsForComparison}.

Additional to the goodness of fit tables, fig.~\ref{fig:ratios_new} shows various hadron-to-pion ratios given the tune parameters in \tabRef{tab:tunes} alongside Closepacking (T1), with further plots shown in app~\ref{app:morePlots}.
The Closepacking (simple) tune performs well for most strange hadron-to-pion ratios, however underpredicts the triple-strange baryon-to-pion ratio $\Omega/\pi$ whilst overpredicting the $K_S/\pi$ ratio (seen in fig.~\ref{fig:KsPi_new}). This is somewhat expected given the strangeness enhancement is being driven equally across the mesonic and baryonic sectors, and thus to fit the multi-strange baryons, the single-strange mesons are overpredicted. Contrastingly, the Strange Junctions tune performs particularly poorly at describing the double-strange mesonic $\phi/\pi$ ratio (see fig.~\ref{fig:phiPi_new}). This is as the strange junctions mechanism contributes strangeness enhancement around junctions only, and thus applies strangeness enhancement to only a single quark in the meson adjacent to the junction. Therefore the enhancement to the production of double-strange $\phi$ meson is minimal. 
Interestingly, the two Trieste tunes (T1 and T2)  describe the observables in a contrasting manner despite implementing the same strangeness enhancement models. The T1 tune describes the $p/\pi$ and $\phi/\pi$ ratios better than the T2 one, but does less well for other ratios such as $K_S^0/\pi$ and $\Lambda/K^0_S$.

\begin{table}[tp]
    \centering
    
    \begin{tabular}{l| >{\centering\arraybackslash}p{1.1cm} | >{\centering\arraybackslash}p{1.1cm} | >{\centering\arraybackslash}p{1.1cm} | >{\centering\arraybackslash}p{1.1cm} | >{\centering\arraybackslash}p{1.1cm} | >{\centering\arraybackslash}p{1.7cm}}
        & $K^0_S/\pi^\pm$ & $\Lambda/\pi^\pm$
        & $\Xi^\pm/\pi^\pm$ & $\Omega^\pm/\pi^\pm$
        & $\Lambda/K^0_S$ & $(p+\bar{p})/\pi^\pm$ \\[3pt]
        \hline
       Monash                      & \cellcolor[HTML]{56BB46} 0.12
                                   & \cellcolor[HTML]{FF0000} 14.18
                                   & \cellcolor[HTML]{B92D5D} \textcolor{white}{20.11}
                                   & \cellcolor[HTML]{000000} \textcolor{white}{31.83}
                                   & \cellcolor[HTML]{EF4A24} 12.32
                                   & \cellcolor[HTML]{81C342} 1.17 \\

        QCD CR ($m_\tau = 2$)      & \cellcolor[HTML]{BDCD30} 1.91
                                   & \cellcolor[HTML]{BDCD30} 1.98
                                   & \cellcolor[HTML]{EF4A24} 12.22
                                   & \cellcolor[HTML]{B92D5D} \textcolor{white}{18.92}
                                   & \cellcolor[HTML]{6CBD45} 0.58
                                   & \cellcolor[HTML]{B92D5D} \textcolor{white}{20.97} \\

        Rope Hadronization         & \cellcolor[HTML]{6CBD45} 0.58
                                   & \cellcolor[HTML]{BDCD30} 2.40
                                   & \cellcolor[HTML]{6CBD45} 0.70
                                   & \cellcolor[HTML]{6CBD45} 0.92
                                   & \cellcolor[HTML]{81C342} 1.11
                                   & \cellcolor[HTML]{000000} \textcolor{white}{28.42} \\

        Closepacking (T1)          & \cellcolor[HTML]{BDCD30} 2.14
                                   & \cellcolor[HTML]{6CBD45} 0.75
                                   & \cellcolor[HTML]{56BB46} 0.32
                                   & \cellcolor[HTML]{BDCD30} 1.98
                                   & \cellcolor[HTML]{BDCD30} 2.58
                                   & \cellcolor[HTML]{81C342} 1.65 \\

        Closepacking (T2)          & \cellcolor[HTML]{56BB46} 0.24
                                   & \cellcolor[HTML]{56BB46} 0.31
                                   & \cellcolor[HTML]{40B748} 0.11
                                   & \cellcolor[HTML]{6CBD45} 0.94
                                   & \cellcolor[HTML]{6CBD45} 0.50
                                   & \cellcolor[HTML]{F79717} 6.37 \\

        Closepacking (simple)      & \cellcolor[HTML]{BDCD30} 2.41
                                   & \cellcolor[HTML]{6CBD45} 0.76
                                   & \cellcolor[HTML]{6CBD45} 0.80
                                   & \cellcolor[HTML]{BDCD30} 2.08
                                   & \cellcolor[HTML]{81C342} 1.07
                                   & \cellcolor[HTML]{FF0000} 14.67 \\

        Strange Junctions          & \cellcolor[HTML]{40B748} 0.04
                                   & \cellcolor[HTML]{6CBD45} 0.64
                                   & \cellcolor[HTML]{56BB46} 0.15
                                   & \cellcolor[HTML]{6CBD45} 0.54
                                   & \cellcolor[HTML]{6CBD45} 0.57
                                   & \cellcolor[HTML]{F78220} 7.72 \\
\hline
    \end{tabular}
    \caption{Goodness of fit values, $\chi^2_\textup{GoF}$, given tunes for various observables used in the tuning procedure. The colour scheme gradients (see \tabRef{tab:chi2_colours}) range from green to purple/black, with green corresponding to low $\chi^2_\textup{GoF}$ and purple/black to high $\chi^2_\textup{GoF}$ values.    \label{tab:GOF}
}
\end{table}

\begin{table}[tp]
    \centering
    
    \begin{tabular}{l| >{\centering\arraybackslash}p{1.3cm} | >{\centering\arraybackslash}p{1.3cm} | >{\centering\arraybackslash}p{1.3cm} | >{\centering\arraybackslash}p{1.7cm} | >{\centering\arraybackslash}p{2cm}}
        & $\frac{K^+ + K^-}{\pi^++\pi^-}$
        & $\frac{K^{*0}+\bar{K}^{*0}}{\pi^++\pi^-}$
        & $\frac{2\phi}{\pi^++\pi^-}$
        & $\frac{dN_{ch}}{d\eta}$ vs $\eta$
        & $\langle p_\perp \rangle$ vs $n_{ch}$ \\[5pt]
        \hline
        Monash                     & \cellcolor[HTML]{56BB46} 0.12
                                   & \cellcolor[HTML]{6CBD45} 0.74
                                   & \cellcolor[HTML]{F79717} 6.76
                                   & \cellcolor[HTML]{40B748} 0.02
                                   & \cellcolor[HTML]{6CBD45} 0.95 \\

        QCD CR ($m_\tau = 2$)      & \cellcolor[HTML]{81C342} 1.67
                                   & \cellcolor[HTML]{56BB46} 0.19
                                   & \cellcolor[HTML]{FF0000} 13.98
                                   & \cellcolor[HTML]{56BB46} 0.24
                                   & \cellcolor[HTML]{56BB46} 0.24 \\

        Rope Hadronization          & \cellcolor[HTML]{6CBD45} 0.53
                                   & \cellcolor[HTML]{6CBD45} 0.89
                                   & \cellcolor[HTML]{BDCD30} 2.19
                                   & \cellcolor[HTML]{40B748} 0.05
                                   & \cellcolor[HTML]{81C342} 1.19 \\

        Closepacking (T1)          & \cellcolor[HTML]{BDCD30} 1.80
                                   & \cellcolor[HTML]{81C342} 1.63
                                   & \cellcolor[HTML]{81C342} 1.73
                                   & \cellcolor[HTML]{40B748} 0.05
                                   & \cellcolor[HTML]{6CBD45} 0.66 \\

        Closepacking (T2)          & \cellcolor[HTML]{56BB46} 0.19
                                   & \cellcolor[HTML]{6CBD45} 0.65
                                   & \cellcolor[HTML]{FDB314} 5.25
                                   & \cellcolor[HTML]{40B748} 0.02
                                   & \cellcolor[HTML]{56BB46} 0.24 \\

        Closepacking (simple)      & \cellcolor[HTML]{BDCD30} 1.89
                                   & \cellcolor[HTML]{BDCD30} 2.58
                                   & \cellcolor[HTML]{BDCD30} 1.97
                                   & \cellcolor[HTML]{40B748} 0.11
                                   & \cellcolor[HTML]{56BB46} 0.31 \\

        Strange Junctions          & \cellcolor[HTML]{40B748} 0.05
                                   & \cellcolor[HTML]{56BB46} 0.33
                                   & \cellcolor[HTML]{F79717} 7.09
                                   & \cellcolor[HTML]{40B748} 0.02
                                   & \cellcolor[HTML]{56BB46} 0.23 \\
\hline
    \end{tabular}
    \caption{Goodness of fit values, $\chi^2_\textup{GoF}$, given tunes for various observables used in the tuning procedure. The colour scheme gradients (see \tabRef{tab:chi2_colours}) range from green to purple/black, with green corresponding to low $\chi^2_\textup{GoF}$ and purple/black to high $\chi^2_\textup{GoF}$ values.    \label{tab:GOF2}
}
\end{table}
\FloatBarrier

The stability and robustness of the new tunes has been tested using a subsampling procedure. 
This subsampling procedure randomly selects about two thirds of the simulations and performs a tuning procedure on this subset which can then be compared to the tune including all simulations. 
Results were shown to be stable for most parameters, demonstrating that the tuning is well-constrained and independent of the specific
choice of sampled points.
Of note is the baryon parameters (\parold{ColourReconnection:junctionCorrection}, \parnew{ClosePacking:baryonSup}, and \parnew{StringFragmentation:enhanceStrangeJunction}), which had slightly larger dispersions in tuned values between various random subsets. This however is somewhat expected given the level of uncertainties on the baryon data, alongside the lack of available experimental observables which are able to distinguish the junction and diquark baryon sources.
Overall however, the results indicate that while some parameters related to baryon production have a broader stability envelope, the overall parameter set is robust and represents a well-defined global minimum.

\subsection{Heavy flavour}
\label{sec:heavyFlav}

\begin{figure}[tp]
    \centering
    \subfloat{\includegraphics[width=0.49\linewidth]{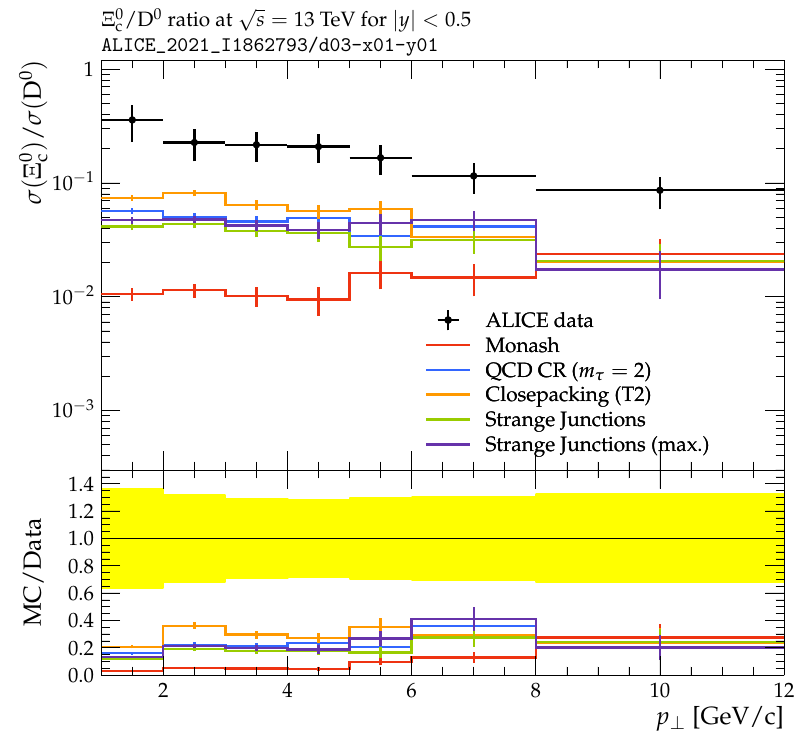}\label{fig:Xic0D0}} \hfill
    \subfloat{\includegraphics[width=0.49\linewidth]{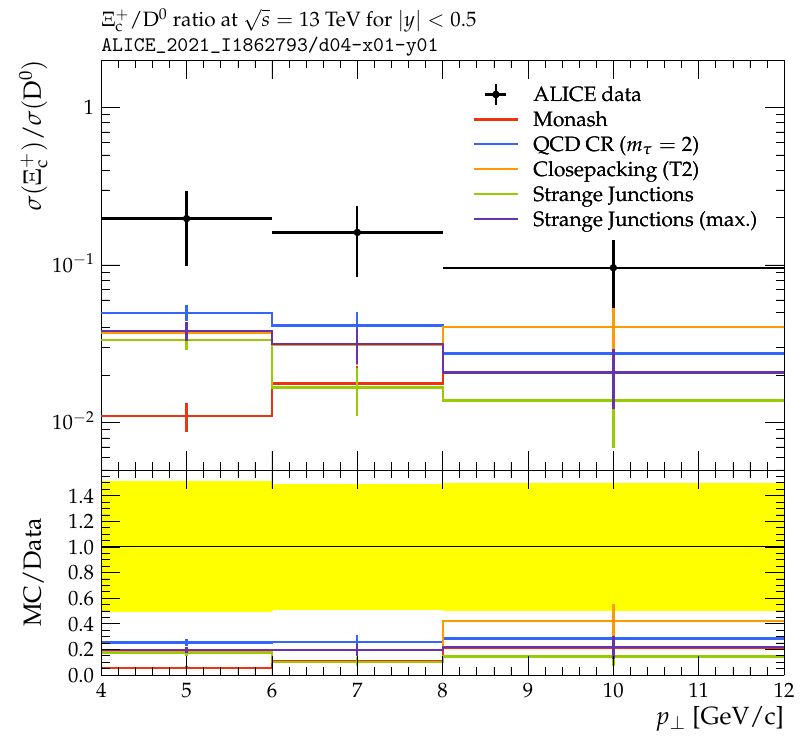}\label{fig:XicpD0}} \\
    \subfloat{\includegraphics[width=0.49\linewidth]{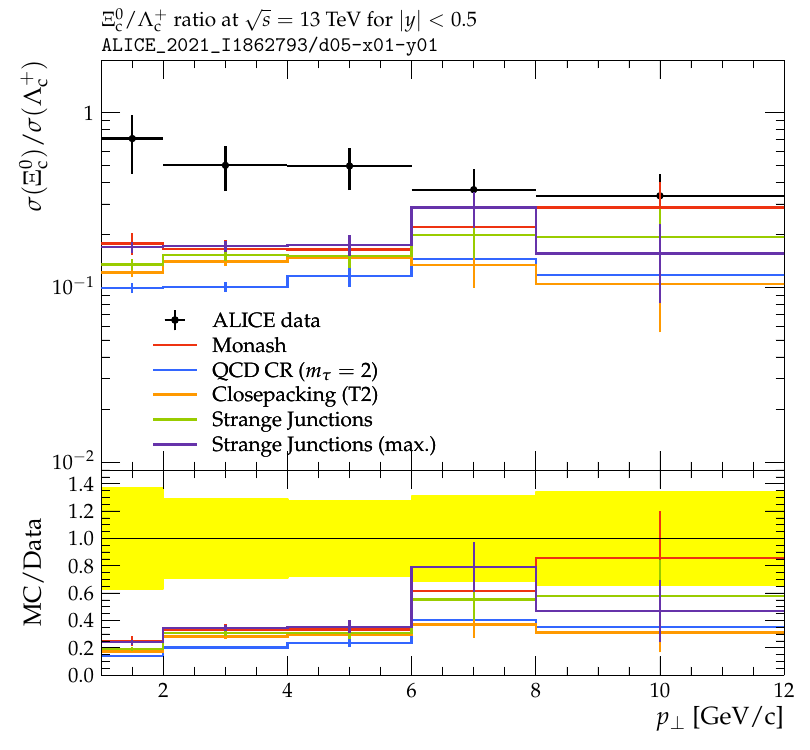}\label{fig:Xic0Lambdac}} \hfill
    \subfloat{\includegraphics[width=0.49\linewidth]{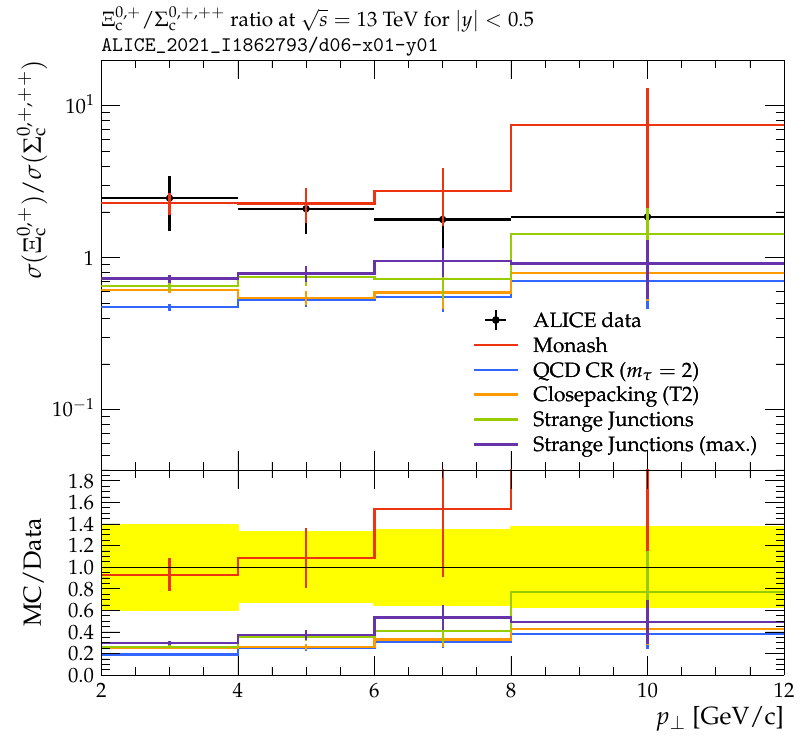}\label{fig:XicSigmac}} \\

    \caption{Ratios of $\Xi_c$ to various hadronic species with respect to $p_\perp$, measured by ALICE~\cite{ALICE:2021bli} for $\sqrt{s}=13$~TeV inelastic events. Measurements are conducted at midrapidity, $|y|<0.5$, for prompt charm hadrons. Top left: $\Xi_c^0/D^0$. Top right: $\Xi_c^+/D^0$. Bottom left: $\Xi_c^0/\Lambda_c^+$. Bottom right: $\Xi_c^{0,+}/\Sigma_c^{0,+,++}$.
    \label{fig:xiRatios}}
\end{figure}

A full-fledged tuning including heavy-flavour hadron production is not carried out in this work, however in the following we will briefly touch on the effects of the aforementioned strangeness enhancement mechanisms on heavy flavour hadrons and demonstrate how well the tunes perform on key heavy-flavour ratios. In particular, we will focus on examining the large underprediction of the $\Xi_c$ ratios in $pp$ collision environments (see fig.~\ref{fig:xiRatios}), particularly at low-$p_\perp$ values. 

\begin{figure}[tp]
    \centering
    \subfloat{\includegraphics[width=0.49\linewidth]{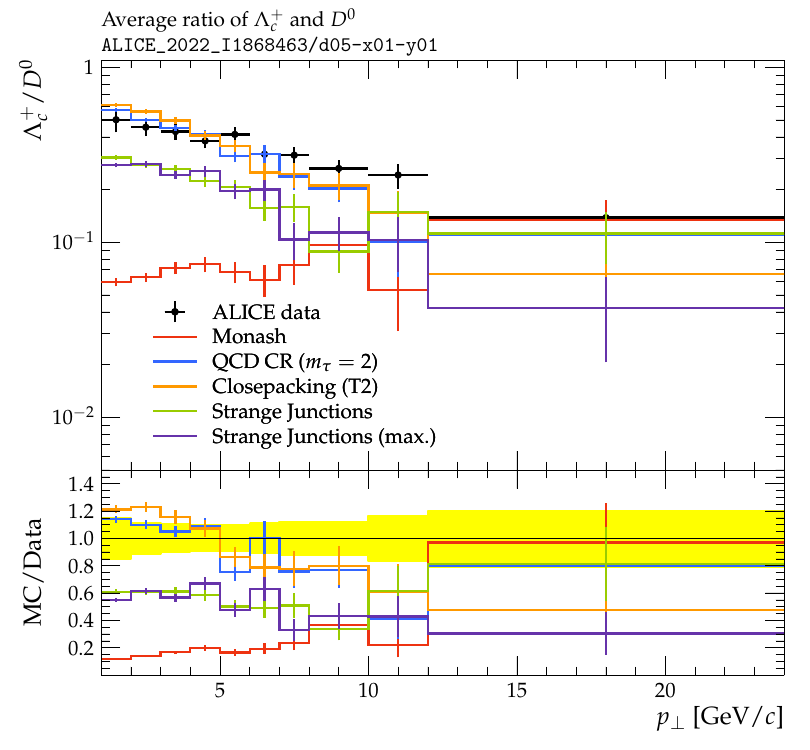}\label{fig:LcD0}} \hfill
    \subfloat{\includegraphics[width=0.49\linewidth]{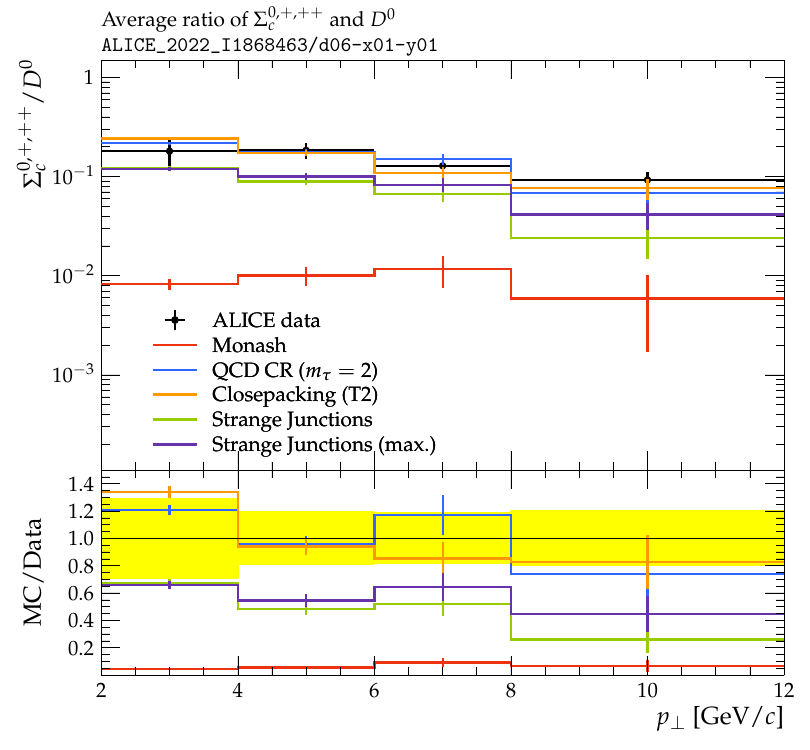}\label{fig:SigcD0}} \\
    \subfloat{\includegraphics[width=0.49\linewidth]{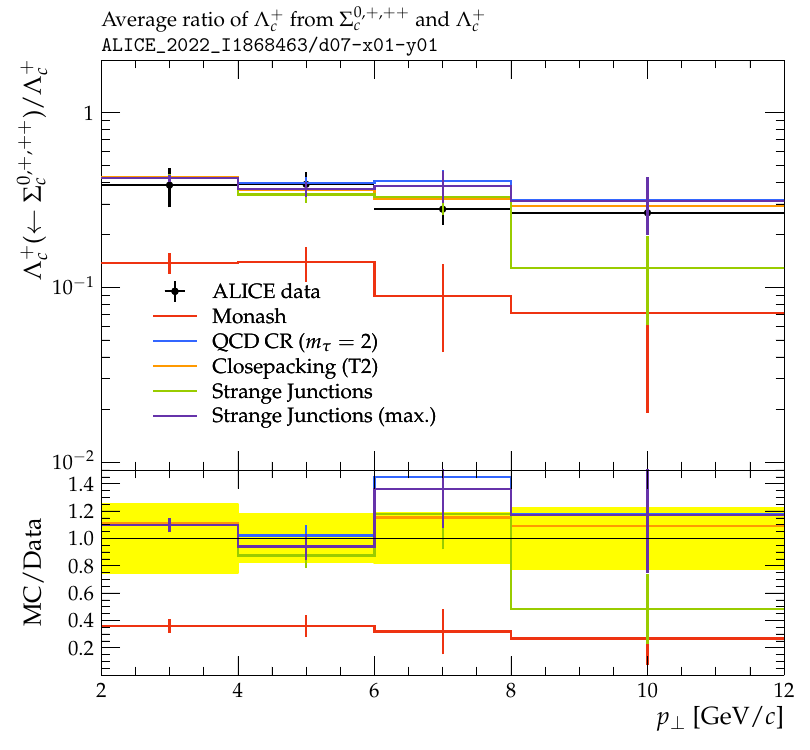}\label{fig:LcFromSigc}}  \\

    \caption{Non-strange charm hadronic ratios, measured by ALICE~\cite{ALICE:2021rzj} for $\sqrt{s}=13~\mathrm{TeV}$ $pp$ collision inelastic events. Measurements are conducted at midrapidity, $|y|<0.5$, for prompt charm hadrons. Top left: $\Lambda_c^+/D^0$. Top right: $\Sigma_c^{0,+,++}/D^0$. Bottom: $\Lambda_c^+(\leftarrow\Sigma_c^{0,+,++})/\Lambda_c^+$.
    \label{fig:lambdaCratios}}
\end{figure}

Though the inclusion of junction baryons has proven to be particularly useful in describing low-$p_\perp$ heavy-flavour baryon production, such as the $\Lambda_c^+/D^0$ ratio shown in fig.~\ref{fig:lambdaCratios}, $\Xi_c$ ratios remain described very poorly. A key example is the 80\% underprediction of the $\Xi^0_c/D^0$ ratio~\cite{ALICE:2021bli, Altmann:2024odn}.
Of note is the strangeness content of the $\Xi_c$ baryon, which contains both a charm and strange quark in contrast to the $\Lambda_c$ and $D^0$ hadrons which have no strangeness content. 
The vast underprediction of these ratios suggests either an underestimate of $\Xi_c$ baryon production, or the simultaneous overestimation of both the $\Lambda_c$ and $D^0$ hadrons. Given charm production is entirely sourced from hard processes and supposing perturbative charm modelling is sound, this leaves the primary assumed source of error to be strangeness production nearby charm quarks. 
With approximately 70\% of prompt $\Xi_c$ baryons with QCD CR being sourced from junctions -- and indeed the majority of any heavy flavour baryons -- the study of strange quark production next to junctions themselves become particularly important. Thus in the following, we not only compare to the various Closepacking tunes and Strange Junctions, but compare to an additional set of parameters called ``Strange Junctions (max.)'' which is simply the equivalent of the Strange Junctions tune however with \parnew{StringFragmentation:enhanceStrangeJunction} parameter set to unity. This has been done to test the full extent of the strange junction model on strange charm baryon production specifically, and effectively means that the production of a strange quark is considered equal to that of an up/down quark directly next to a junction.
As one can see in fig.~\ref{fig:xiRatios} which is indeed reflected in the $\chi^2_\mathrm{GoF}$ values in~\tabRef{tab:GOF_xiRatios}, despite these various strangeness enhancement mechanisms in place, the $\Xi_c$ ratios remain poorly described. Even with this strange junction maximization, the $\Xi_c/D^0$ and $\Xi_c^0/\Lambda_c^+$ ratios are severely underpredicted. 


\begin{table}[tp]
    \centering
    \begin{tabular}{l| >{\centering\arraybackslash}p{1.0cm} | >{\centering\arraybackslash}p{1.0cm} | >{\centering\arraybackslash}p{1.3cm} | >{\centering\arraybackslash}p{1.0cm} | >{\centering\arraybackslash}p{1.0cm} | >{\centering\arraybackslash}p{1.0cm} | >{\centering\arraybackslash}p{1.0cm}}
        & $\frac{\Lambda_c^+}{D^0}$
        & $\frac{\Sigma_c}{D^0}$
        & $\frac{\Lambda_c^+(\leftarrow\Sigma_c)}{\Lambda_c^+}$ & $\frac{\Xi_c^0}{D^0}$ & $\frac{\Xi_c^+}{D^0}$ & $\frac{\Xi_c^0}{\Lambda_c^+}$ & $\frac{\Xi_c^{0,+}}{\Sigma_c}$ \\[3pt]
        \hline
        Monash                     & \cellcolor[HTML]{000000} \textcolor{white}{41.53}
                                   & \cellcolor[HTML]{B92D5D} \textcolor{white}{19.83}
                                   & \cellcolor[HTML]{F78220} 7.72
                                   & \cellcolor[HTML]{F78220} 8.67
                                   & \cellcolor[HTML]{FDCD06} 3.14
                                   & \cellcolor[HTML]{FDCD06} 3.19
                                   & \cellcolor[HTML]{56BB46} 0.40 \\

        QCD CR ($m_\tau = 2$)      & \cellcolor[HTML]{81C342} 1.69
                                   & \cellcolor[HTML]{6CBD45} 0.50
                                   & \cellcolor[HTML]{6CBD45} 0.64
                                   & \cellcolor[HTML]{F79717} 6.06
                                   & \cellcolor[HTML]{BDCD30} 2.19
                                   & \cellcolor[HTML]{FDB314} 5.43
                                   & \cellcolor[HTML]{FDB314} 4.00 \\

        Rope Hadronization         & \cellcolor[HTML]{81C342} 1.46
                                   & \cellcolor[HTML]{56BB46} 0.16
                                   & \cellcolor[HTML]{56BB46} 0.42
                                   & \cellcolor[HTML]{FDB314} 5.40
                                   & \cellcolor[HTML]{BDCD30} 2.20
                                   & \cellcolor[HTML]{FDB314} 5.11
                                   & \cellcolor[HTML]{FDCD06} 3.86 \\

        Closepacking (T1)          & \cellcolor[HTML]{81C342} 1.66
                                   & \cellcolor[HTML]{6CBD45} 0.97
                                   & \cellcolor[HTML]{56BB46} 0.13
                                   & \cellcolor[HTML]{FDB314} 4.54
                                   & \cellcolor[HTML]{BDCD30} 2.07
                                   & \cellcolor[HTML]{FDB314} 4.01
                                   & \cellcolor[HTML]{BDCD30} 2.67 \\

        Closepacking (T2)          & \cellcolor[HTML]{81C342} 1.75
                                   & \cellcolor[HTML]{6CBD45} 0.54
                                   & \cellcolor[HTML]{56BB46} 0.16
                                   & \cellcolor[HTML]{FDB314} 5.18
                                   & \cellcolor[HTML]{BDCD30} 2.22
                                   & \cellcolor[HTML]{FDB314} 4.97
                                   & \cellcolor[HTML]{FDCD06} 3.65 \\

        Closepacking (simple)      & \cellcolor[HTML]{BDCD30} 2.67
                                   & \cellcolor[HTML]{6CBD45} 0.70
                                   & \cellcolor[HTML]{56BB46} 0.40
                                   & \cellcolor[HTML]{FDB314} 5.31
                                   & \cellcolor[HTML]{FDCD06} 2.84
                                   & \cellcolor[HTML]{FDB314} 4.36
                                   & \cellcolor[HTML]{FDCD06} 3.72 \\

        Strange Junctions          & \cellcolor[HTML]{EF4A24} 11.26
                                   & \cellcolor[HTML]{F79717} 6.11
                                   & \cellcolor[HTML]{6CBD45} 0.79
                                   & \cellcolor[HTML]{F79717} 6.79
                                   & \cellcolor[HTML]{FDCD06} 3.05
                                   & \cellcolor[HTML]{FDCD06} 3.87
                                   & \cellcolor[HTML]{BDCD30} 2.58 \\

        Strange Junctions (max.)   & \cellcolor[HTML]{EF4A24} 12.21
                                   & \cellcolor[HTML]{FDCD06} 3.67
                                   & \cellcolor[HTML]{56BB46} 0.36
                                   & \cellcolor[HTML]{F79717} 6.15
                                   & \cellcolor[HTML]{BDCD30} 2.60
                                   & \cellcolor[HTML]{FDCD06} 3.35
                                   & \cellcolor[HTML]{BDCD30} 2.49 \\
                                   \hline
    \end{tabular}
    \caption{Goodness of fit values, $\chi^2_\textup{GoF}$, of various tunes for heavy-flavour prompt charm observables. The colour scheme gradients (see \tabRef{tab:chi2_colours}) range from green to purple/black, with green corresponding to low $\chi^2_\textup{GoF}$ and purple/black to high $\chi^2_\textup{GoF}$ values.    \label{tab:GOF_xiRatios}}
\end{table}

\FloatBarrier

To fully deduce how to well describe these ratios, a dedicated study into the production of charm hadrons would be required, including examination into the string configurations they form and the environment they sit in (\ie, the number of nearby strings and therefore closepacking strength). Such a detailed examination has been left to a future study. One should also reexamine the effect of strangeness enhancement mechanisms on the $\Lambda_b/B^0$ ratio~\cite{LHCb:2023wbo}, which appears to be significantly overpredicted by the QCD CR model~\cite{Altmann:2024odn}.

\section{Conclusions and Outlook}

In this study, we have revised and extended the so-called string closepacking model~\cite{Fischer:2016zzs} for collective hadronization effects in the PYTHIA~8 Monte Carlo event generator~\cite{Bierlich:2022pfr}. We have explored the properties of the new model in the context of fitting LHC data, and contrasted it to a set of existing physics modelling options and tunes.

The original closepacking model was introduced in the context of a model for thermal string breaks~\cite{Fischer:2016zzs}. In this work, we have: 
\begin{itemize}[itemsep=3pt]
\item extended the closepacking model to apply not only to thermal string breaks but also to conventional Schwinger-type ones; 
\item introduced sensitivity to the flux direction, so that not only the total number of strings but also their (relative) flux orientations affect the scaling of the effective string tension, adopting a baseline expectation in line with Casimir scaling;
\item proposed a mechanism --- popcorn destructive interference --- that may reduce diquark production in regular string breaks, based on an argument that nearby colour fields may affect virtual colour fluctuations on a string in a way that effectively suppresses popcorn-type diquark formation.
\end{itemize}
 A final physics modelling option we have implemented is that of Strange Junctions, which allows for the effective string tension near a string junction to be higher than far away from junctions.  

A tuning procedure outlined in \secRef{sec:tuning} was carried out, with the strength parameters in each of the closepacking, popcorn destructive interference, and strange-junction models being predominantly tuned to ALICE light hadronic ratios with respect to ALICE charged multiplicity classes (see refs.~\cite{ALICE:2017jyt, ALICE:2018pal}). 
In particular, three different tuning configurations were considered -- Closepacking Trieste 2 (T2) which was a combinatorial tune of all of the three of the above models, Closepacking (simple) which included only closepacking and popcorn destructive interference, and Strange Junctions which exclusively implemented the strange-junction model.
These tunes produced varying levels of strangeness-enhancement effects with respect to charged multiplicity, with results summarized in tabs.~\ref{tab:GOF} and~\ref{tab:GOF2}. None of the tuning configurations managed to describe all of the observables simultaneously, with particularly interesting  disparities occurring for the double-strange $\phi/\pi$ ratio and the non-strange baryon-to-meson $p/\pi$ ratio. Nonetheless, they proved to be competitive with or better than previous modelling attempts, in producing enhanced strange-hadron levels whilst avoiding egregiously overshooting proton production. We attribute this mainly to the popcorn destructive interference modelling, which --- being  developed in this work --- was not enabled in the existing Monash, QCD CR, and Rope hadronization tunes.

Additional to the light sector, \secRef{sec:heavyFlav} demonstrates how these tunes fare in describing the prompt charm hadronic ratios. Despite their enhanced strangeness production, none of the models we considered are able to reach agreement with the $\Xi_c$ ratios. We find this puzzling, and would recommend further dedicated investigations into $\Xi_c$ and/or other heavy-flavoured strange hadrons. 

Another potential future development would be to fully generalize the closepacking model to arbitrary collision processes and ensure applicability to all observables. In the current form, the closepacking and popcorn destructive interference implementations rely on the assumption that the bulk of colour flow is aligned along the $z$-axis. This beam-axis dependence of course no longer has significance in the case of $e^+e^-$ collision events, nor could it be applied to the description of hadrochemistry in high-$p_\perp$ jets. The study of cleaner string environments such as $e^+e^-\rightarrow WW$ events would provide an extremely useful baseline expectation for the closepacking model. The accurate description of jet hadrochemistry is also vital to pin down, given, \eg, its impact on jet calibrations, cf.,  \eg,~\cite{ATLAS:2022xau}.

\acknowledgments

We thank the Rudolf Peierls Centre for Theoretical Physics, University of Oxford, for hospitality and a wonderful working environment during the completion of this work. This work was supported by the Royal Society, grant no.~RSWVF-R1-231006, by the Australian Research Council Discovery Project grant no.~DP230103014, and by the Monash-Warwick Alliance for Particle Physics. This work was supported also by the Friuli Venezia Giulia region with Microgrant J93C22001380002.


\vspace*{1cm}
\appendix
\section{Diquark scaling for closepacking}
\label{app:probQQtoQ}

The probability \parold{StringFlav:probQQtoQ} scales in a more complicated manner in comparison to the other flavour probabilities given a modified string tension.
The probability of diquark compared to quark production from string breaks can be parameterized as, 

\begin{equation}
    P(qq:q) = \frac{\sum_{qq_s}\mathcal{P}_{qq_s}}{\sum_q\mathcal{P}_q} = \alpha \frac{\mathcal{P}_{ud0}}{\mathcal{P}_u},
\end{equation}
with the variable $\alpha$ here being a function of the other flavour probabilities. Given the parameters \parold{StringFlav:probStoUD}, \parold{StringFlav:probSQtoQQ}, \parold{StringFlav:probQQ1toQQ0} being represented by symbols $\rho$, $x$, and $y$ respectively, then the factor $\alpha$ can be parameterized as, 
\begin{equation}
    \alpha = \frac{1+2x\rho +9y +6x\rho y +3yx^2\rho^2}{2+\rho}.
    \label{eq:alpha}
\end{equation}
In the parameterization of $P'(qq:q)$ given in \eqRef{eq:probQQtoQmod}, $\alpha$ and $\alpha'$ take the same form as \eqRef{eq:alpha}, simply with modified flavour probabilities defined by $\kappa_0$ and $\kappa_{\text{eff}}$ respectively.
Important to note is that this scaling of the diquark probability works under the assumption that an effective string tension does not effect the probability $\mathcal{P}_{ud0}/\mathcal{P}_u$, and assumes Schwinger-type diquark-antidiquark pair production.

\clearpage
\section{Table of Casimir-scaling coefficients}
\label{app:casimirs}
Table \ref{tab:casimirs} shows the baseline string-tension enhancement factors $\kappa_p/\kappa_0$ for each string in an environment with $p$ total strings with parallel flux orientations (including the string itself) and $q$ antiparallel ones, in the absence of suppression factors. (The numbers for $q > p$ are identical, with $p \leftrightarrow q$, and
hence are not shown.) Also shown is the total (summed) string tension normalized to the tension for a single string, which agrees with the ratio of the quadratic Casimir in SU(3), $C_2(p,q)$, to the fundamental one, $C_2(1,0) = C_F = 4/3$. The last column gives the size of the corresponding multiplets in SU(3).

\begin{table}[h!]
\centering
\small\begin{tabular}{c|ccccc|cc}
\toprule
\parbox[c]{1.8cm}{\centering\# of overlapping strings} & \parbox[c]{1.1cm}{\centering \#\\ parallel} & \parbox[c]{1.1cm}{\centering \# anti\-parallel} & \parbox[c]{1cm}{\centering\Large$\frac{\kappa_p}{\kappa_0}$} & \parbox[c]{1cm}{\centering\Large$\frac{\kappa_q}{\kappa_0}$} & ~Total \Large$\frac\kappa{\kappa_0}$~ & \parbox[c]{1.3cm}{\centering Casimir ratio} & \parbox[c]{2cm}{\centering SU(3) Rep} \\[5.5mm]
$C_1 = p+q$ & $p$ & $q$ & $\frac{6+2p+q}{8}$ & $\frac{6+2q+p}{8}$ & \normalsize$\frac{p\,\kappa_p\, +\, q\,\kappa_q}{\kappa_0}$ & $\frac{C_2(p,q)}{C_2(1,0)}$ & $\frac{(p+1)(q+1)(p+q+2)}{2}$\\[2mm] 
\midrule
\vspace*{-3mm}\\
1 & 1 & 0 & 1 & -- & 1 & 1 & \bf 3 
\\[2mm]
2 & 2 & 0 & 1.25 & -- & 2.5 & 2.5 & \bf 6 \\
2 & 1 & 1 & 1.125 & 1.125 & 2.25 & 2.25 & \bf 8 
\\[2mm]
3 & 3 &	0 &	1.5	& --	& 4.5 & 4.5	& \bf 10 \\
3 & 2 &	1 & 1.375	& 1.25 & 4	& 4	& \bf 15
\\[2mm]
4&	4&	0&	1.75&	--&	7&	7&	\bf 15' \\
4&	3&	1&	1.625&	1.375&	6.25&	6.25&	\bf 24\\
4&	2&	2&	1.5&	1.5&	6&	6&	\bf 27
\\[2mm]
5&	5&	0&	2&	--&	10&	10&	\bf 21\\
5&	4&	1&	1.875&	1.5&	9&	9&	\bf 35\\
5&	3&	2&	1.75&	1.625&	8.5&	8.5&	\bf 42
\\[2mm]
6&	6&	0&	2.25&	--&	13.5&	13.5&	\bf 28\\
6&	5&	1&	2.125&	1.625&	12.25&	12.25&	\bf 48\\
6&	4&	2&	2&	1.75&	11.5&	11.5&	\bf 60\\
6&	3&	3&	1.875&	1.875&	11.25&	11.25&	\bf 64
\\[2mm]
7&	7&	0&	2.5&	--&	17.5&	17.5&\bf 	36 \\
7&	6&	1&	2.375&	1.75&	16&	16&	\bf 63 \\
7&	5&	2&	2.25&	1.875&	15&	15&	\bf 81\\
7&	4&	3&	2.125&	2&	14.5&	14.5&\bf	90
\\[2mm]
8&	8&	0&	2.75&	--&	22&	22&\bf	45 \\
8&	7&	1&	2.625&	1.875&	20.25&	20.25&\bf	80 \\
8&	6&	2&	2.5&	2&	19&	19&	\bf 105 \\
8&  5&	3&	2.375&	2.125&	18.25&	18.25&	\bf120 \\
8&	4&	4&	2.25&	2.25&	18&	18&	\bf 125
\\[2mm]
\bottomrule\end{tabular}\vspace*{-1mm}
\caption{Numerical values of the string-tension enhancement factors $\kappa_p/\kappa_0$ and $\kappa_q/\kappa_0$, for closepacking with Casimir scaling, for up to 8 total strings, with $p$ parallel flux directions and $q$ antiparallel ones, in the absence of suppression factors. The  summed string tensions, $\kappa/\kappa_0$, agree with the exact ratios of SU(3) quadratic Casimirs $C_2(p,q)/C_2(1,0)$. The corresponding SU(3) multiplets are also given. Note that there are two different 15-plets, labelled {\bf 15} and {\bf 15'} respectively.\label{tab:casimirs}}
\vspace*{-.5cm}\end{table}

\clearpage
\section{Reference Tune Parameters}
\label{app:modelParm}
Here, we list the parameter sets of the existing models and tunes that we have compared to in this work, including the Monash tune~\cite{Skands:2014pea}, QCD CR ($m_\tau = 2$)~\cite{Christiansen:2015yqa}, Rope Hadronization, and Closepacking (T1)~\cite{LorenzoThesis}. Note that older tunes normally use \texttt{StringZ:useOldAExtra = on}, to reproduce a bug in the versions they were tuned with, while the newer \texttt{off} setting was used for this work (without retuning the fragmentation parameters).

\begin{table}[h!tp]
    \centering\vspace*{-1mm}\scriptsize
    \begin{tabular}{l|p{1.7cm}  p{1.7cm} p{1.7cm} p{1.6cm}}
        & \centering \textbf{Monash} & \centering \textbf{QCD CR} ($\mathbf{m_\tau = 2}$) & \textbf{Rope} \newline \textbf{Hadron\-i\-zation} & \textbf{Close\-packing (T1)} \\
        \toprule
        \parold{StringZ:useOldAExtra}                    & = off   & = off   & = off   & = off   \\
        \parold{StringPT:sigma}                          & = 0.335 & = 0.335 & = 0.335 & = 0.335 \\
        \parold{StringZ:aLund}                           & = 0.68  & = 0.36  & = 0.68  & = 0.68  \\
        \parold{StringZ:bLund}                           & = 0.98  & = 0.56  & = 0.98  & = 0.98  \\
        \parold{StringFlav:probQQtoQ}                    & = 0.081 & = 0.078 & = 0.081 & = 0.081 \\
        \parold{StringFlav:probStoUD}                    & = 0.217 & = 0.2   & = 0.217 & = 0.217 \\
        \parold{StringFlav:probQQ1toQQ0join}             & \multicolumn{4}{c}{= \{0.5, 0.7, 0.9, 1.0\}} \\
        \midrule
        \parold{MultiPartonInteractions:pT0Ref}          & = 2.28 & = 2.15 & = 2.15 & = 2.194 \\
        \midrule
        \parold{BeamRemnants:remnantMode}                & = 0    & = 1    & = 1    & = 1 \\
        \parold{BeamRemnants:saturation}                 & --     & = 5    & = 5    & = 5 \\
        \midrule
        \parold{ColourReconnection:mode}                 & = 0    & = 1    & = 1    &  = 1    \\
        \parold{ColourReconnection:timeDilationMode}     & --     & = 2    & = 2    &  = 2    \\
        \parold{ColourReconnection:allowDoubleJunRem}    & --     & = off  & = off  &  = off  \\
        \parold{ColourReconnection:allowJunctions}       & --     & = on   & = on   &  = on   \\
        \parold{ColourReconnection:mPseudo}              & --     & = 0.3  & = 0.3  &  = 0.403 \\
        \parold{ColourReconnection:m0}                   & --     & = 0.3  & = 0.3  &  = 0.618 \\
        \parold{ColourReconnection:junctionCorrection}   & --     & = 1.20 & = 1.20 &  = 1.349 \\
        \parold{ColourReconnection:timeDilationPar}      & --     & = 0.18 & = 0.18 &  = 0.18 \\
        \midrule
        \parnew{ClosePacking:doClosePacking}             & = off  & = off  & = off  & = on \\
        \parnew{ClosePacking:enhanceStrange}             & --     & --     & --     & = 0.014 \\
        \parnew{ClosePacking:enhancePT}                  & --     & --     & --     & = 0.014 \\
        \parnew{ClosePacking:PT0}                        & --     & --     & --     & = 2.0 \\ 
        \parnew{ClosePacking:doEnhanceDiquark}           & = off  & = off  & = off  & = off \\
        \midrule
        \parnew{StringFragmentation:doStrangeJunctions}  & = off  & = off  & = off  & = on \\
        \parnew{StringFragmentation:enhanceStrangeJunction}  & --     & --     & --     & = 0.540\\
        \midrule
        \parnew{ClosePacking:baryonSup}                  & --     & --     & --     & = 0.928 \\
        \parnew{ClosePacking:parallelBaryonSup}          & --     & --     & --     & = 0 \\
        \midrule
        \parold{Ropewalk:RopeHadronization}              & = off  & = off  & = on   & = off \\ 
        \parold{Ropewalk:doShoving}                      & --     & --     & = on   & -- \\ 
        \parold{Ropewalk:tInit}                          & --     & --     & = 1.5  & -- \\ 
        \parold{Ropewalk:deltat}                         & --     & --     & = 0.05 & -- \\ 
        \parold{Ropewalk:tShove}                         & --     & --     & = 0.1  & -- \\ 
        \parold{Ropewalk:gAmplitude}                     & --     & --     & = 0.   & -- \\ 
        \parold{Ropewalk:doFlavour}                      & --     & --     & = on   & -- \\ 
        \parold{Ropewalk:r0}                             & --     & --     & = 0.5  & -- \\ 
        \parold{Ropewalk:m0}                             & --     & --     & = 0.2  & -- \\ 
        \parold{Ropewalk:beta}                           & --     & --     & = 0.1  & -- \\
        \parold{PartonVertex:setVertex}                  & --     & --     & = on   & -- \\
        \parold{PartonVertex:protonRadius}               & --     & --     & = 0.7  & -- \\
        \parold{PartonVertex:emissionWidth}              & --     & --     & = 0.1  & -- \\
        \midrule
        \parold{ParticleDecays:limitTau0}                & = on   & = on   & = on & = on \\
        \parold{ParticleDecays:tau0Max}                  & = 10   & = 10   & = 10 & = 10 \\
        \bottomrule
    \end{tabular}\vspace*{-1mm}
    \caption{Parameter values for the following tunes: Monash, QCD CR ($m_\tau = 2$), Rope Hadronization and Closepacking (T1). Parameters for the new tunes developed in this work are listed in \tabRef{tab:tunes}.
    \label{tab:command_cards}}
\vspace*{-10mm}\end{table}
\FloatBarrier

\section{Further Plots}
\label{app:morePlots}

\begin{figure}[h!]
    \centering
    \subfloat{\includegraphics[width=0.49\linewidth]{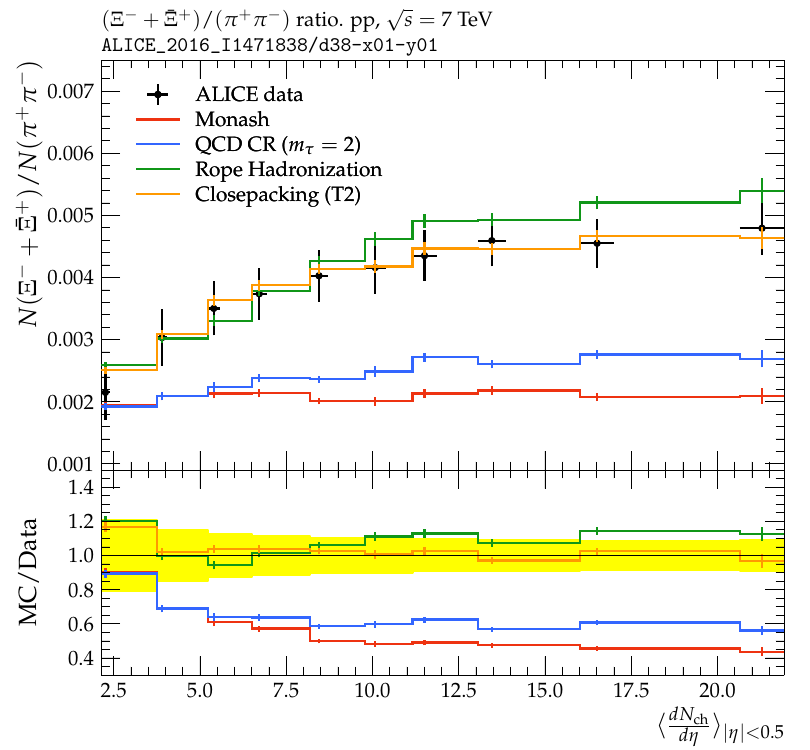}\label{fig:XiPi_old}} \hfill
    \subfloat{\includegraphics[width=0.49\linewidth]{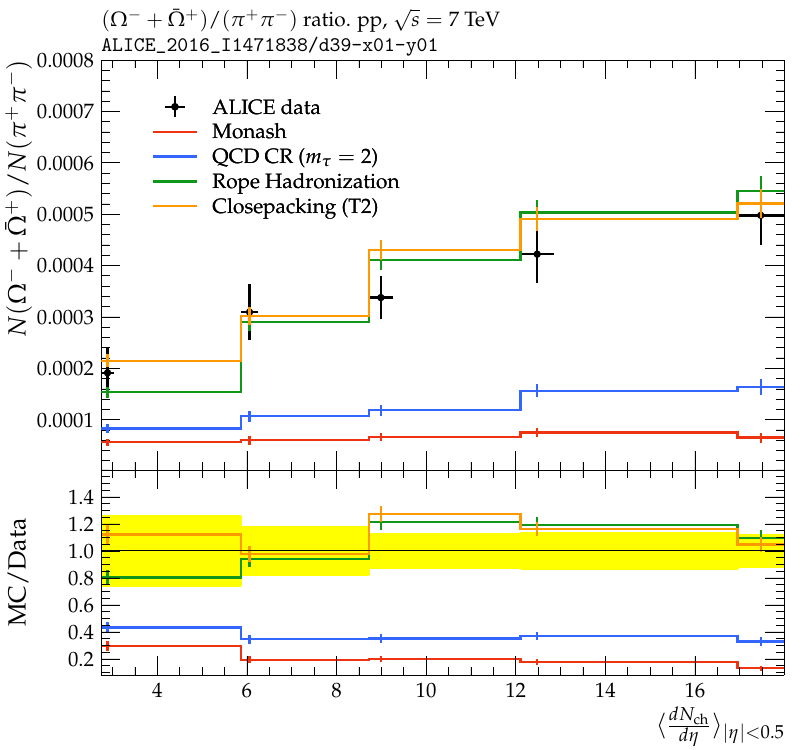}\label{fig:OmegaPi_old}} \\
    \subfloat{\includegraphics[width=0.49\linewidth]{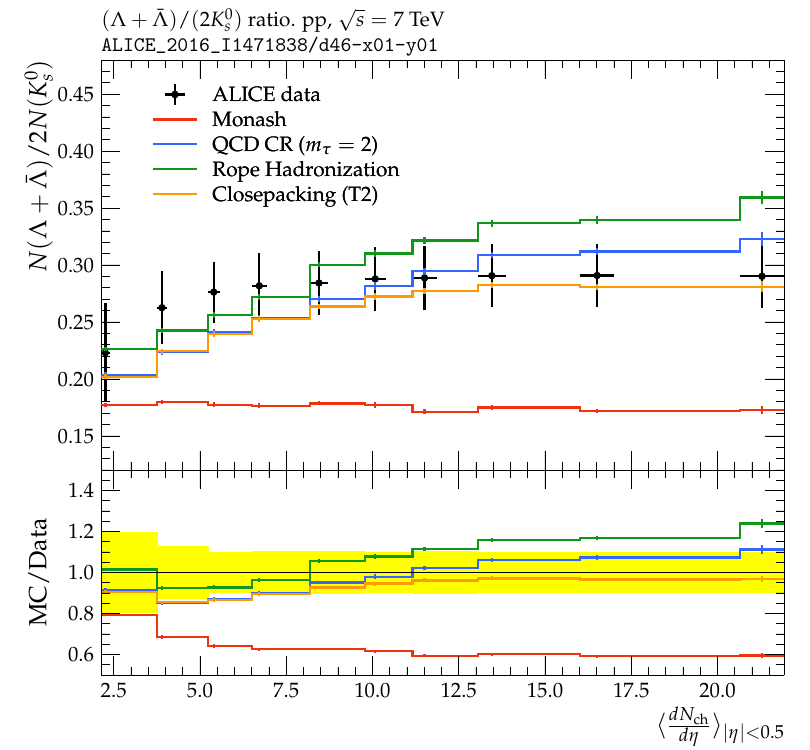}\label{fig:LambdaKs_old}} \hfill
    \subfloat{\includegraphics[width=0.49\linewidth]{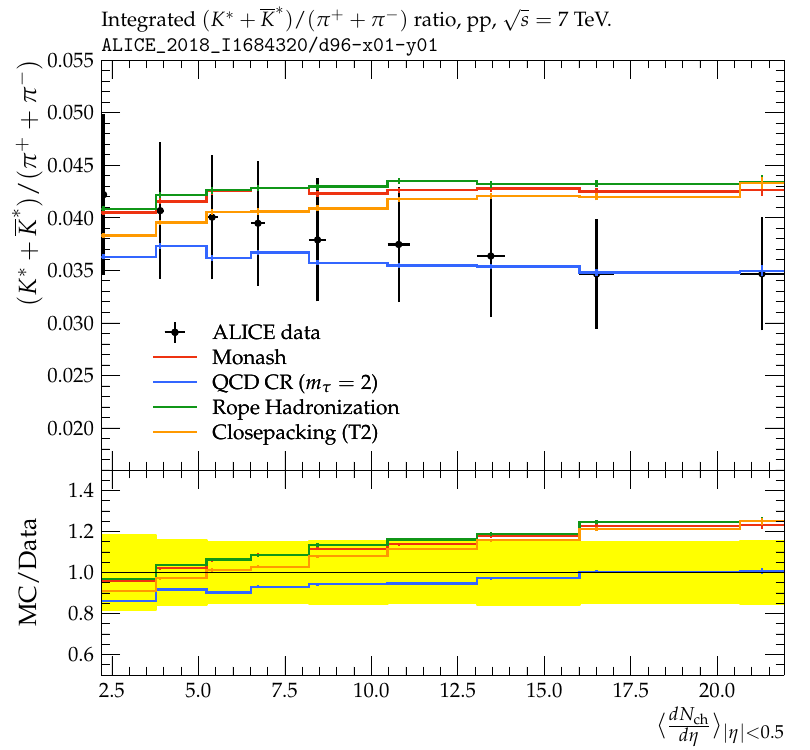}\label{fig:KstarPi_old}} \\

    \caption{Hadron-to-pion ratios with respect to ALICE midrapidity charged multiplicity classes. \Py simulations for parameters in \tabRef{tab:command_cards} of 7~TeV INEL>0 events are run with a lifetime cut of $\tau_{max} = 10$mm/c, and no $p_\perp$ cuts on final-state particles. Top left: $\Xi/\pi$~\cite{ALICE:2017jyt}. Top right: $\Omega/\pi$~\cite{ALICE:2017jyt}. Bottom left: $\Lambda/2K_S^0$~\cite{ALICE:2017jyt}. Bottom right: $K^*/\pi$~\cite{ALICE:2018pal}.}
    \label{fig:ratioApp_old}
\end{figure}

\begin{figure}[tp]
    \centering
    \subfloat{\includegraphics[width=0.49\linewidth]{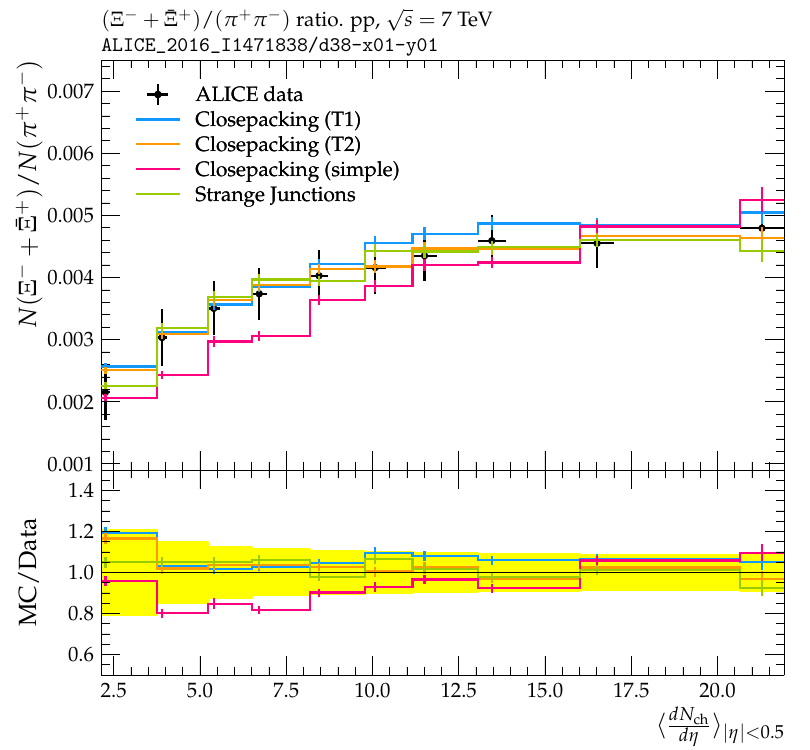}\label{fig:XiPi_new}} \hfill
    \subfloat{\includegraphics[width=0.49\linewidth]{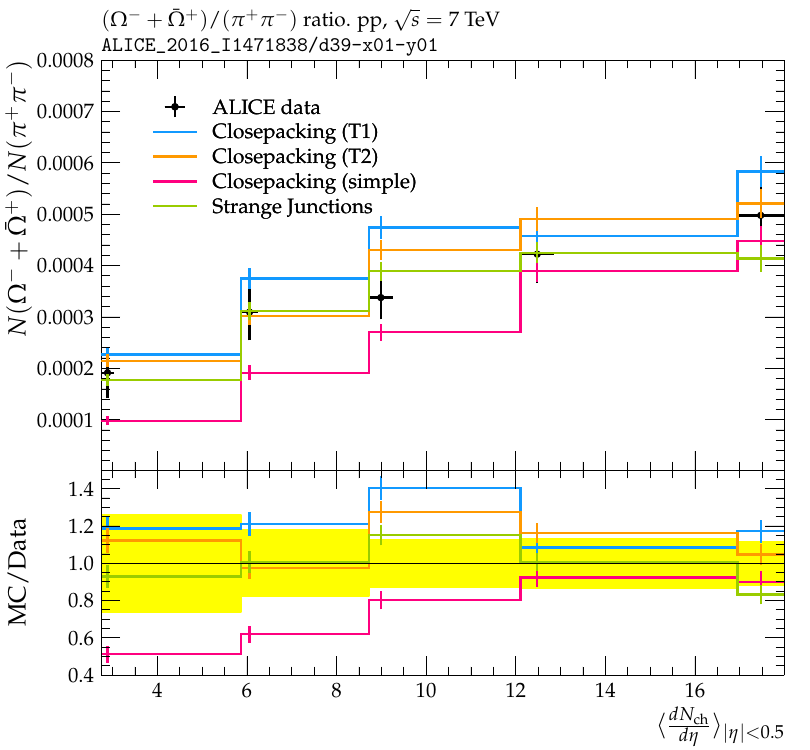}\label{fig:OmegaPi_new}} \\
    \subfloat{\includegraphics[width=0.49\linewidth]{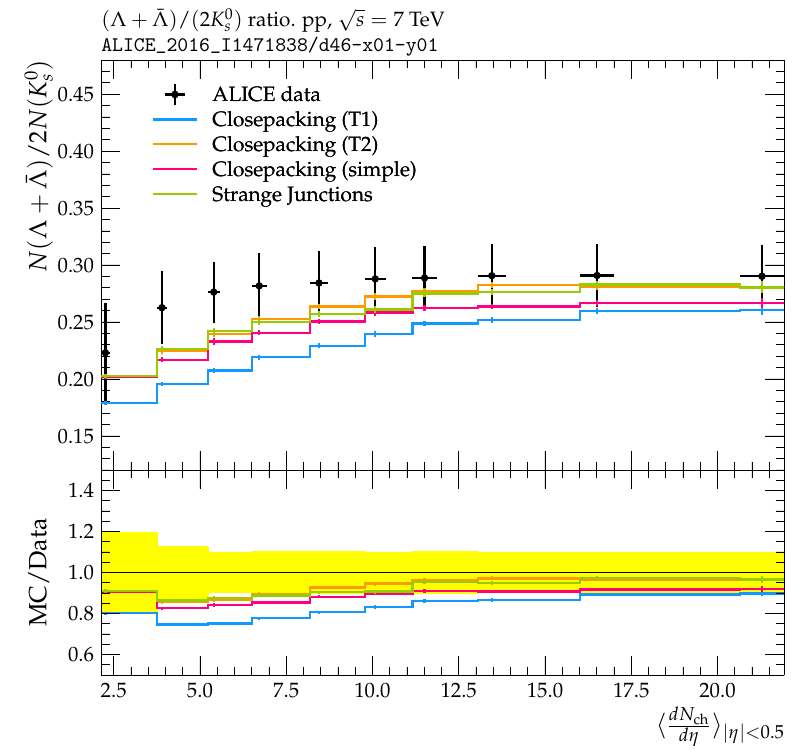}\label{fig:LambdaKs_new}} \hfill
    \subfloat{\includegraphics[width=0.49\linewidth]{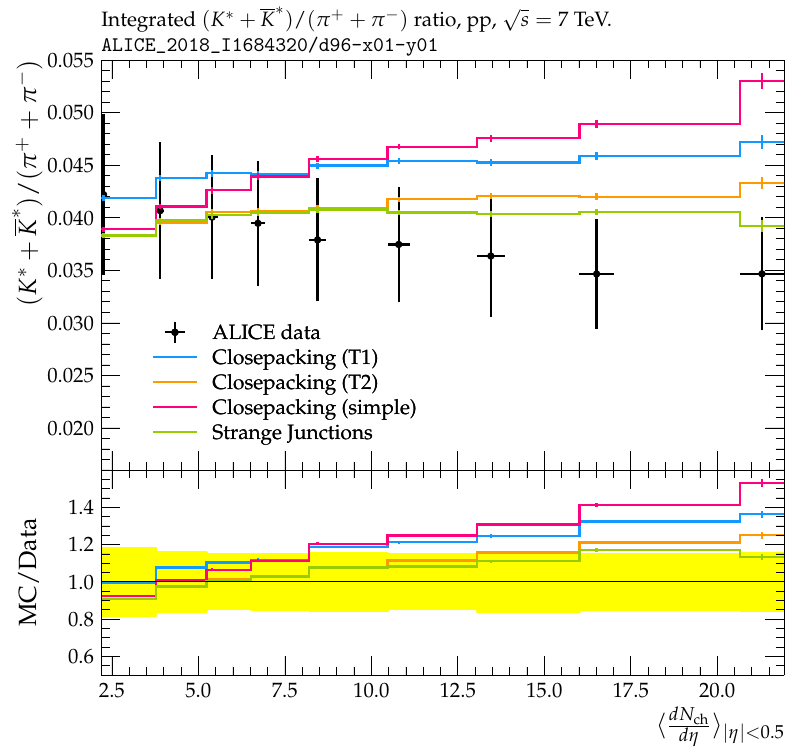}\label{fig:KstarPi_new}} \\

    \caption{Hadron-to-pion ratios with respect to ALICE midrapidity charged multiplicity classes. \Py simulations for parameters in \tabRef{tab:tunes} of 7~TeV INEL>0 events are run with a lifetime cut of $\tau_{max} = 10$mm/c, and no $p_\perp$ cuts on final-state particles. Top left: $\Xi/\pi$~\cite{ALICE:2017jyt}. Top right: $\Omega/\pi$~\cite{ALICE:2017jyt}. Bottom left: $\Lambda/2K_S^0$~\cite{ALICE:2017jyt}. Bottom right: $K^*/\pi$~\cite{ALICE:2018pal}.}
    \label{fig:ratioApp_new}
\end{figure}

\clearpage
\section{Colour map for Goodness-of-Fit assessment}
Table \ref{tab:chi2_colours} shows the colour scheme, based on $\chi^2_\textup{GoF}$ values, adopted in this work.

\begin{table}[h!]
    \centering
    \caption{Reduced $\chi^2_\textup{GoF}$ colour map for Goodness of Fit (GoF) assessment, with equivalent standard deviations $\sigma \equiv \sqrt{\chi^2_\textup{GoF}}$, and $p$ values (for a normal distribution) indicating the probability that a purely stochastic fluctuation could yield a higher $\chi^2_\textup{GoF}$.}
    \label{tab:chi2_colours}
    \begin{tabular}{|c|c|c|c|c|}
        \hline
        \textbf{Colour} &
        \textbf{Hex Code} \rule[-1.5ex]{0pt}{0pt} &
        \textbf{$\mathbf{\sigma}_\textup{GoF}$}
        \rule{0pt}{3ex} &
        \textbf{$\mathbf{\chi^2_\textup{GoF}}$ }
        \rule{0pt}{3ex} &
        \textbf{$\mathbf{p}_\textup{GoF}$} 
        \rule{0pt}{3ex}
        \\
\hline
\cellcolor[HTML]{40B748} & \texttt{\#40B748} 
& $ ~~~0.00$ -- $0.33~~~ $  &  $ ~~~0.00$ -- $0.11~~~ $  &  $ ~~~0.74 $ -- $ 1.00~~~ $ 
\\
\cellcolor[HTML]{56BB46} & \texttt{\#56BB46} 
& $ 0.33$ -- $0.67 $  &  $ 0.11$ -- $0.44 $  &  $ 0.50$ -- $0.74 $ 
\\
\cellcolor[HTML]{6CBD45} & \texttt{\#6CBD45}
& $ 0.67$ -- $1.00 $  &  $ 0.44$ -- $1.00 $  &  $ 0.32$ -- $0.50 $ 
\\
\hline
\cellcolor[HTML]{81C342} & \texttt{\#81C342} 
& $ 1.00$ -- $1.33 $  &  $ 1.00$ -- $1.78 $  &  $ 0.18$ -- $0.32 $
\\
\cellcolor[HTML]{BDCD30} & \texttt{\#BDCD30}
& $ 1.33$ -- $1.67 $  &  $ 1.78$ -- $2.78 $  &  $ 0.096$ -- $0.18 $
\\
\cellcolor[HTML]{FDCD06} & \texttt{\#FDCD06}
& $ 1.67$ -- $2.00 $  &  $ 2.78$ -- $4.00 $  &  $ 0.046$ -- $0.096 $
\\
\hline
\cellcolor[HTML]{FDB314} & \texttt{\#FDB314} 
& $ 2.00$ -- $2.33 $  &  $ 4.00$ -- $5.44 $  &  $ 0.020$ -- $0.046 $
\\
\cellcolor[HTML]{F79717} & \texttt{\#F79717}
& $ 2.33$ -- $2.67 $  &  $ 5.44$ -- $7.11 $  &  $ 0.0077$ -- $0.020 $
  \\
\cellcolor[HTML]{F78220} & \texttt{\#F78220} 
& $ 2.67$ -- $3.00 $  &  $ 7.11$ -- $9.00 $  &  $ 0.0027$ -- $0.0077 $    \\
\hline
\cellcolor[HTML]{F16820} & \texttt{\#F16820} 
& $ 3.00$ -- $3.33 $  &  $ 9.00$ -- $11.11 $  &  
$ [0.86$ -- $2.7]\times 10^{-3}$ 
\\
\cellcolor[HTML]{EF4A24} & \texttt{\#EF4A24}
& $ 3.33$ -- $3.67 $  &  $ 11.11$ -- $13.44 $  &  
$ [0.25$ -- $0.86]\times10^{-3} $
\\
\cellcolor[HTML]{FF0000} & \texttt{\#FF0000}
& $ 3.67$ -- $4.00 $  &  $ 13.44$ -- $16.00 $  &  
$ [0.63$ -- $2.5]\times10^{-4} $
\\
\hline
\cellcolor[HTML]{E63B7A} & \texttt{\#E63B7A}
& $ 4.00$ -- $4.33 $  &  $ 16.00$ -- $18.78 $  &  $ [1.5$ -- $6.3]\times10^{-5} $
\\
\cellcolor[HTML]{B92D5D} & \texttt{\#B92D5D} 
& $ 4.33$ -- $4.67 $  &  $ 18.78$ -- $21.78 $  &  $ [0.3$ -- $1.5]\times10^{-5} $
\\
\cellcolor[HTML]{9A244F} & \texttt{\#9A244F}
& $ 4.67$ -- $5.00 $  &  $ 21.78$ -- $25.00 $  &  $ [1$ -- $3]\times10^{-6} $
\\
\hline
\cellcolor[HTML]{000000} & \texttt{\#000000} 
& $\ge 5.00$ & $\ge 25.00$ & $\le 10^{-6}$ \\
        \hline
    \end{tabular}
\end{table}

\section{Tuning Observables}
\label{app:observables}

The observables used for tuning, along with their assigned weights, order of polynomial, and versions of \Py used are listed in this section, for each tuning step. The Rivet analyses utilized for the tuning include\footnote{Rivet 3 naming conventions.}:
\begin{itemize}[noitemsep]
    \item \lstinline|ALICE_2016_I1471838| \cite{ALICE:2017jyt}
    \item \lstinline|ALICE_2018_I1684320|\footnote{Note that this Rivet analysis is not yet published and publicly available.} \cite{ALICE:2018pal}
    \item \lstinline|ATLAS_2010_S8918562| \cite{ATLAS:2010jvh} 
    \item \lstinline|CMS_2010_S8656010| \cite{CMS:2010tjh}
    \item \lstinline|CMS_2011_S8884919| \cite{CMS:2010qvf}
    \item \lstinline|CMS_2011_S8978280| \cite{CMS:2011jlm}
    \item \lstinline|LHCB_2011_I917009| \cite{LHCb:2011ioc}
\end{itemize}

\subsection{Tuning Step 1}
Weights were set to 1 for all observables used in the first tuning step, as reported in Table \ref{tab:tune_obs_01}. Simulations were run using \Py 8.315 with an unpublished closepacking code update, and 4th order polynomials were used for interpolation.

\begin{table}[htbp]
    \centering
    \scriptsize
    \caption{Observables and corresponding weights used in the first tuning step.}
    \label{tab:tune_obs_01}
    \begin{tabularx}{\textwidth}{l|p{5cm}|X|c}
    \toprule
    \textbf{Rivet path} & \textbf{Description} & \textbf{Axis} & \textbf{Weight} \\
    \midrule
        \lstinline[basicstyle=\tiny\ttfamily]|/ATLAS_2010_S8918562/d03-x01-y01|
        & Charged particle $\eta$ at 7$\,$TeV, track $p_\perp > 500\,\mathrm{MeV}$, for $N_\mathrm{ch} \geq 1$
        & y-axis: $1/N_\mathrm{ev} \, \mathrm{d}N_\mathrm{ch}/\mathrm{d}\eta$ \newline
          x-axis: $\eta$
        & 1.0 \\
        \midrule 
        
        \lstinline[basicstyle=\tiny\ttfamily]|/ATLAS_2010_S8918562/d05-x01-y01|
        & Charged particle $\eta$ at 7$\,$TeV, track $p_\perp > 100\,\mathrm{MeV}$, for $N_\mathrm{ch} \geq 2$
        & y-axis: $1/N_\mathrm{ev} \, \mathrm{d}N_\mathrm{ch}/\mathrm{d}\eta$ \newline
          x-axis: $\eta$
        & 1.0 \\
        \midrule 
        
        \lstinline[basicstyle=\tiny\ttfamily]|/ATLAS_2010_S8918562/d07-x01-y01|
        & Charged particle $\eta$ at 7$\,$TeV, track $p_\perp > 500\,\mathrm{MeV}$, for $N_\mathrm{ch} \geq 6$
        & y-axis: $1/N_\mathrm{ev} \, \mathrm{d}N_\mathrm{ch}/\mathrm{d}\eta$ \newline
          x-axis: $\eta$
        & 1.0 \\
        \midrule 
        
        \lstinline[basicstyle=\tiny\ttfamily]|/ATLAS_2010_S8918562/d23-x01-y01| & Charged $\langle p_\perp \rangle$ vs. $N_\mathrm{ch}$ at 7$\,$TeV, track $p_\perp > 500\,\mathrm{MeV}$, for $N_\mathrm{ch} \geq 1$ & 
        y-axis: $\langle p_\perp \rangle$ $\;$ [GeV] \newline
        x-axis: $N_\mathrm{ch}$
        & 1.0 \\
        \midrule 
        
        \lstinline[basicstyle=\tiny\ttfamily]|/ATLAS_2010_S8918562/d25-x01-y01|
        & Charged $\langle p_\perp \rangle$ vs. $N_\mathrm{ch}$ at 7$\,$TeV, track $p_\perp > 100\,\mathrm{MeV}$, for $N_\mathrm{ch} \geq 2$
        & y-axis: $\langle p_\perp \rangle$ $\;$ [GeV] \newline
          x-axis: $N_\mathrm{ch}$
        & 1.0 \\
        \midrule 
        
        \lstinline[basicstyle=\tiny\ttfamily]|/ATLAS_2010_S8918562/d27-x01-y01|
        & Charged particle $\eta$ at 7$\,$TeV, track $p_\perp > 100\,\mathrm{MeV}$, for $N_\mathrm{ch} \geq 20$
        & y-axis: $1/N_\mathrm{ev} \, \mathrm{d}N_\mathrm{ch}/\mathrm{d}\eta$ \newline
          x-axis: $\eta$
        & 1.0 \\
    \bottomrule
    \end{tabularx}
\end{table} 
\normalsize
\FloatBarrier

\subsection{Tuning Step 2}
Weights were set to 1 for all observables used in the second tuning step, as reported in Table \ref{tab:tune_obs_02}. Other configurations were also tested. For instance, in one test, a weight of 2 was assigned only to the $\phi/\pi$ ratio. In another configuration, the same weight was applied to $\phi/\pi$, $K^{*0}/\pi$ and $p/\pi$. However, the configuration with all weights set to one ultimately performed best in terms of resulting $\chi^2_\textup{GoF}$ values.
Simulations were run using \Py 8.316, using 5th order polynomials for interpolation.

{
\setlength{\belowrulesep}{0pt} 
\begin{xltabular}{\textwidth}{>{\scriptsize}l|>{\scriptsize}p{5cm}|>{\scriptsize}X|>{\scriptsize}c}
    \caption{Observables and corresponding weights used in the second tuning step.}
    \label{tab:tune_obs_02}
    \\
    \toprule
    \textbf{Rivet path} & \textbf{Description} & \textbf{Axis} & \textbf{Weight} \\
    \midrule
    \endhead

    \multicolumn{4}{r}{\scriptsize\textit{Continued on next page}} \\ 
    \endfoot

    \multicolumn{4}{r}{\scriptsize\textit{End of table.}} \\
    \endlastfoot
    \lstinline[basicstyle=\tiny\ttfamily]|/ALICE_2016_I1471838/d36-x01-y01|
    & $2K^0_s/(\pi^+ \pi^-)$ ratio
    & y-axis: $2N(K^0_s)/N(\pi^+\pi^-)$ \newline
      x-axis: $\langle \frac{dN_\mathrm{ch}}{d\eta} \rangle_{|\eta| < 0.5}$
    & 1.0 \\
    \midrule

    \lstinline[basicstyle=\tiny\ttfamily]|/ALICE_2016_I1471838/d37-x01-y01|
    & $(\Lambda + \bar{\Lambda})/(\pi^+ \pi^-)$ ratio
    & y-axis: $N(\Lambda + \bar{\Lambda})/N(\pi^+\pi^-)$ \newline
      x-axis: $\langle \frac{dN_\mathrm{ch}}{d\eta} \rangle_{|\eta| < 0.5}$
    & 1.0 \\
    \midrule 
    
    \lstinline[basicstyle=\tiny\ttfamily]|/ALICE_2016_I1471838/d38-x01-y01|
    & $(\Xi^- + \bar{\Xi}^+)/(\pi^+ \pi^-)$ ratio
    & y-axis: $N(\Xi^- + \bar{\Xi}^+)/N(\pi^+\pi^-)$ \newline
      x-axis: $\langle \frac{dN_\mathrm{ch}}{d\eta} \rangle_{|\eta| < 0.5}$
    & 1.0 \\
    \midrule 
    
    \lstinline[basicstyle=\tiny\ttfamily]|/ALICE_2016_I1471838/d39-x01-y01|
    & $(\Omega^- + \bar{\Omega}^+)/(\pi^+ \pi^-)$ ratio
    & y-axis: $N(\Omega^- + \bar{\Omega}^+)/N(\pi^+\pi^-)$ \newline
      x-axis: $\langle \frac{dN_\mathrm{ch}}{d\eta} \rangle_{|\eta| < 0.5}$
    & 1.0 \\
    \midrule 
    
    \lstinline[basicstyle=\tiny\ttfamily]|/ALICE_2016_I1471838/d46-x01-y01|
    & $(\Lambda + \bar{\Lambda})/(2K^0_s)$ ratio
    & y-axis: $N(\Lambda + \bar{\Lambda})/2N(K^0_s)$ \newline
      x-axis: $\langle \frac{dN_\mathrm{ch}}{d\eta} \rangle_{|\eta| < 0.5}$
    & 1.0 \\
    \midrule 
    
    \lstinline[basicstyle=\tiny\ttfamily]|/ALICE_2016_I1471838/d47-x01-y01|
    & $(p + \bar{p})/(\pi^+ + \pi^-)$ ratio
    & y-axis: $N(p + \bar{p})/N(\pi^+ + \pi^-)$ \newline
      x-axis: $\langle \frac{dN_\mathrm{ch}}{d\eta} \rangle_{|\eta| < 0.5}$
    & 1.0 \\
    \midrule 
    
    \lstinline[basicstyle=\tiny\ttfamily]|/ALICE_2018_I1684320/d96-x01-y01|
    & Integrated $(K^* + \overline{K}^*)/(\pi^+ + \pi^-)$ ratio, pp, $\sqrt{s}=7$ TeV.
    & y-axis: $(K^* + \overline{K}^*)/(\pi^+ + \pi^-)$ \newline
      x-axis: $\langle \frac{dN_\mathrm{ch}}{d\eta} \rangle_{|\eta| < 0.5}$
    & 1.0 \\
    \midrule 
    
    \lstinline[basicstyle=\tiny\ttfamily]|/ALICE_2018_I1684320/d97-x01-y01|
    & Integrated $\phi/(\pi^+ + \pi^-)$ ratio, pp, $\sqrt{s}=7$ TeV.
    & y-axis: $\phi/(\pi^+ + \pi^-)$ \newline
      x-axis: $\langle \frac{dN_\mathrm{ch}}{d\eta} \rangle_{|\eta| < 0.5}$
    & 1.0 \\
    \midrule 
    
    \lstinline[basicstyle=\tiny\ttfamily]|/CMS_2011_S8978280/d05-x01-y02|
    & $\Xi^-$ rapidity distribution at $\sqrt{s}$ = 7 TeV
    & y-axis: $(1/N_\mathrm{NSD})\; dN/dy$ \newline
      x-axis: $\Xi^-$ $|y|$
    & 1.0 \\
    \midrule 
    
    \lstinline[basicstyle=\tiny\ttfamily]|/CMS_2011_S8978280/d09-x01-y02|
    & $\Lambda / \mathrm{K}^0_\mathrm{S}$ versus rapidity at $\sqrt{s}$ = 7 TeV
    & y-axis: $N(\Lambda)\, /\, N(\mathrm{K}^0_\mathrm{S})$ \newline
      x-axis: $|y|$
    & 1.0 \\
    \midrule 
    
    \lstinline[basicstyle=\tiny\ttfamily]|/CMS_2011_S8978280/d10-x01-y02|
    & $\Xi^- / \Lambda$ versus rapidity at $\sqrt{s}$ = 7 TeV
    & y-axis: $N(\Xi^-)\, /\, N(\Lambda)$ \newline
      x-axis: $|y|$
    & 1.0 \\
    \midrule 
    
    \lstinline[basicstyle=\tiny\ttfamily]|/LHCB_2011_I917009/d12-x01-y01|
    & $\bar{\Lambda}/K_{s}^{0}$ ratio at $\sqrt{s}=7$ TeV ($0.15 < p_\perp < 2.5$ GeV/$c$)
    & y-axis: $\bar{\Lambda}/K_{s}^{0}$ \newline
      x-axis: Rapidity (y)
    & 1.0 \\
    \midrule 
    
    \lstinline[basicstyle=\tiny\ttfamily]|/LHCB_2011_I917009/d16-x01-y01|
    & $\bar{\Lambda}/K_{s}^{0}$ ratio at $\sqrt{s}=7$ TeV ($0.15 < p_\perp < 2.50$ GeV/$c$)
    & y-axis: $\bar{\Lambda}/K_{s}^{0}$ \newline
      x-axis: Rapidity loss ($\Delta y$)
    & 1.0 \\
    
    \bottomrule    
\end{xltabular} 
}
\normalsize
\FloatBarrier

\subsection{Tuning step 3}
In this third step, weights differing from unity are used to account for the different number of observables in each category. In particular, weight 6 is assigned to pseudorapidity distributions, and weight 8 is assigned to the $\langle p_\perp\rangle$ vs $n_{ch}$ distributions, as reported in Table \ref{tab:tune_obs_03}. Simulations were run using \Py 8.316, using 5th order polynomials for interpolation.

{
\setlength{\belowrulesep}{0pt} 
\begin{xltabular}{\textwidth}{>{\scriptsize}l|>{\scriptsize}p{5cm}|>{\scriptsize}X|>{\scriptsize}c}
    \caption{Observables and corresponding weights used in the third tuning step.
    \label{tab:tune_obs_03}}
    \\
    \toprule
    \textbf{Rivet path} & \textbf{Description} & \textbf{Axis} & \textbf{Weight} \\
    \midrule
    \endhead

    \multicolumn{4}{r}{\scriptsize\textit{Continued on next page}} \\ 
    \endfoot

    \multicolumn{4}{r}{\scriptsize\textit{End of table.}} \\
    \endlastfoot
    
    \lstinline[basicstyle=\tiny\ttfamily]|/ALICE_2016_I1471838/d01-x01-y01|
    & $K^0_s, p_\perp$ spectrum V0M Class I
    & y-axis: $1/N \frac{d^2N}{dydp_\perp}$ \newline
      x-axis: $p_\perp$ [GeV]
    & 1.0 \\
    \midrule 

    \lstinline[basicstyle=\tiny\ttfamily]|/ALICE_2016_I1471838/d02-x01-y01| & $K^0_s, p_\perp$ spectrum V0M Class II & 
    y-axis: $1/N \frac{d^2N}{dydp_\perp}$ \newline
    x-axis: $p_\perp$ [GeV]
     & 1.0 \\
    \midrule
    
    \lstinline[basicstyle=\tiny\ttfamily]|/ALICE_2016_I1471838/d03-x01-y01| & $K^0_s, p_\perp$ spectrum V0M Class III & 
    y-axis: $1/N \frac{d^2N}{dydp_\perp}$ \newline
    x-axis: $p_\perp$ [GeV]
     & 1.0 \\
    \midrule
    
    \lstinline[basicstyle=\tiny\ttfamily]|/ALICE_2016_I1471838/d04-x01-y01| & $K^0_s, p_\perp$ spectrum V0M Class IV & 
    y-axis: $1/N \frac{d^2N}{dydp_\perp}$ \newline
    x-axis: $p_\perp$ [GeV]
     & 1.0 \\
    \midrule
    
    \lstinline[basicstyle=\tiny\ttfamily]|/ALICE_2016_I1471838/d05-x01-y01| & $K^0_s, p_\perp$ spectrum V0M Class V & 
    y-axis: $1/N \frac{d^2N}{dydp_\perp}$ \newline
    x-axis: $p_\perp$ [GeV]
     & 1.0 \\
    \midrule
    
    \lstinline[basicstyle=\tiny\ttfamily]|/ALICE_2016_I1471838/d06-x01-y01| & $K^0_s, p_\perp$ spectrum V0M Class VI & 
    y-axis: $1/N \frac{d^2N}{dydp_\perp}$ \newline
    x-axis: $p_\perp$ [GeV]
     & 1.0 \\
    \midrule
    
    \lstinline[basicstyle=\tiny\ttfamily]|/ALICE_2016_I1471838/d07-x01-y01| & $K^0_s, p_\perp$ spectrum V0M Class VII & 
    y-axis: $1/N \frac{d^2N}{dydp_\perp}$ \newline
    x-axis: $p_\perp$ [GeV]
     & 1.0 \\
    \midrule
    
    \lstinline[basicstyle=\tiny\ttfamily]|/ALICE_2016_I1471838/d08-x01-y01| & $K^0_s, p_\perp$ spectrum V0M Class VIII & 
    y-axis: $1/N \frac{d^2N}{dydp_\perp}$ \newline
    x-axis: $p_\perp$ [GeV]
     & 1.0 \\
    \midrule
    
    \lstinline[basicstyle=\tiny\ttfamily]|/ALICE_2016_I1471838/d09-x01-y01| & $K^0_s, p_\perp$ spectrum V0M Class IX & 
    y-axis: $1/N \frac{d^2N}{dydp_\perp}$ \newline
    x-axis: $p_\perp$ [GeV]
     & 1.0 \\
    \midrule
    
    \lstinline[basicstyle=\tiny\ttfamily]|/ALICE_2016_I1471838/d10-x01-y01| & $K^0_s, p_\perp$ spectrum V0M Class X & 
    y-axis: $1/N \frac{d^2N}{dydp_\perp}$ \newline
    x-axis: $p_\perp$ [GeV]
     & 1.0 \\
    \midrule
    
    \lstinline[basicstyle=\tiny\ttfamily]|/ALICE_2016_I1471838/d11-x01-y01| & $\Lambda + \bar{\Lambda}, p_\perp$ spectrum V0M Class I & 
    y-axis: $1/N \frac{d^2N}{dydp_\perp}$ \newline
    x-axis: $p_\perp$ [GeV]
     & 1.0 \\
    \midrule
    
    \lstinline[basicstyle=\tiny\ttfamily]|/ALICE_2016_I1471838/d12-x01-y01| & $\Lambda + \bar{\Lambda}, p_\perp$ spectrum V0M Class II & 
    y-axis: $1/N \frac{d^2N}{dydp_\perp}$ \newline
    x-axis: $p_\perp$ [GeV]
     & 1.0 \\
    \midrule
    
    \lstinline[basicstyle=\tiny\ttfamily]|/ALICE_2016_I1471838/d13-x01-y01| & $\Lambda + \bar{\Lambda}, p_\perp$ spectrum V0M Class III & 
    y-axis: $1/N \frac{d^2N}{dydp_\perp}$ \newline
    x-axis: $p_\perp$ [GeV]
     & 1.0 \\
    \midrule
    
    \lstinline[basicstyle=\tiny\ttfamily]|/ALICE_2016_I1471838/d14-x01-y01| & $\Lambda + \bar{\Lambda}, p_\perp$ spectrum V0M Class IV & 
    y-axis: $1/N \frac{d^2N}{dydp_\perp}$ \newline
    x-axis: $p_\perp$ [GeV]
     & 1.0 \\
    \midrule
    
    \lstinline[basicstyle=\tiny\ttfamily]|/ALICE_2016_I1471838/d15-x01-y01| & $\Lambda + \bar{\Lambda}, p_\perp$ spectrum V0M Class V & 
    y-axis: $1/N \frac{d^2N}{dydp_\perp}$ \newline
    x-axis: $p_\perp$ [GeV]
     & 1.0 \\
    \midrule
    
    \lstinline[basicstyle=\tiny\ttfamily]|/ALICE_2016_I1471838/d16-x01-y01| & $\Lambda + \bar{\Lambda}, p_\perp$ spectrum V0M Class VI & 
    y-axis: $1/N \frac{d^2N}{dydp_\perp}$ \newline
    x-axis: $p_\perp$ [GeV]
     & 1.0 \\
    \midrule
    
     \lstinline[basicstyle=\tiny\ttfamily]|/ALICE_2016_I1471838/d17-x01-y01| & $\Lambda + \bar{\Lambda}, p_\perp$ spectrum V0M Class VII & 
    y-axis: $1/N \frac{d^2N}{dydp_\perp}$ \newline
    x-axis: $p_\perp$ [GeV]
     & 1.0 \\
    \midrule
    
    \lstinline[basicstyle=\tiny\ttfamily]|/ALICE_2016_I1471838/d18-x01-y01| & $\Lambda + \bar{\Lambda}, p_\perp$ spectrum V0M Class VIII & 
    y-axis: $1/N \frac{d^2N}{dydp_\perp}$ \newline
    x-axis: $p_\perp$ [GeV]
     & 1.0 \\
    \midrule
    
    \lstinline[basicstyle=\tiny\ttfamily]|/ALICE_2016_I1471838/d19-x01-y01| & $\Lambda + \bar{\Lambda}, p_\perp$ spectrum V0M Class IX & 
    y-axis: $1/N \frac{d^2N}{dydp_\perp}$ \newline
    x-axis: $p_\perp$ [GeV]
     & 1.0 \\
    \midrule
    
    \lstinline[basicstyle=\tiny\ttfamily]|/ALICE_2016_I1471838/d20-x01-y01| & $\Lambda + \bar{\Lambda}, p_\perp$ spectrum V0M Class X & 
    y-axis: $1/N \frac{d^2N}{dydp_\perp}$ \newline
    x-axis: $p_\perp$ [GeV]
     & 1.0 \\
    \midrule
    
    \lstinline[basicstyle=\tiny\ttfamily]|/ALICE_2016_I1471838/d21-x01-y01| & $\Xi^- + \bar{\Xi}^+, p_\perp$ spectrum V0M Class I & 
    y-axis: $1/N \frac{d^2N}{dydp_\perp}$ \newline
    x-axis: $p_\perp$ [GeV]
     & 1.0 \\
    \midrule
    
    \lstinline[basicstyle=\tiny\ttfamily]|/ALICE_2016_I1471838/d22-x01-y01| & $\Xi^- + \bar{\Xi}^+, p_\perp$ spectrum V0M Class II & 
    y-axis: $1/N \frac{d^2N}{dydp_\perp}$ \newline
    x-axis: $p_\perp$ [GeV]
     & 1.0 \\
    \midrule
    
    \lstinline[basicstyle=\tiny\ttfamily]|/ALICE_2016_I1471838/d23-x01-y01| & $\Xi^- + \bar{\Xi}^+, p_\perp$ spectrum V0M Class III & 
    y-axis: $1/N \frac{d^2N}{dydp_\perp}$ \newline
    x-axis: $p_\perp$ [GeV]
     & 1.0 \\
    \midrule
    
    \lstinline[basicstyle=\tiny\ttfamily]|/ALICE_2016_I1471838/d24-x01-y01| & $\Xi^- + \bar{\Xi}^+, p_\perp$ spectrum V0M Class IV & 
    y-axis: $1/N \frac{d^2N}{dydp_\perp}$ \newline
    x-axis: $p_\perp$ [GeV]
     & 1.0 \\
    \midrule
    
    \lstinline[basicstyle=\tiny\ttfamily]|/ALICE_2016_I1471838/d25-x01-y01| & $\Xi^- + \bar{\Xi}^+, p_\perp$ spectrum V0M Class V & 
    y-axis: $1/N \frac{d^2N}{dydp_\perp}$ \newline
    x-axis: $p_\perp$ [GeV]
     & 1.0 \\
    \midrule
    
    \lstinline[basicstyle=\tiny\ttfamily]|/ALICE_2016_I1471838/d26-x01-y01| & $\Xi^- + \bar{\Xi}^+, p_\perp$ spectrum V0M Class VI & 
    y-axis: $1/N \frac{d^2N}{dydp_\perp}$ \newline
    x-axis: $p_\perp$ [GeV]
     & 1.0 \\
    \midrule
    
    \lstinline[basicstyle=\tiny\ttfamily]|/ALICE_2016_I1471838/d27-x01-y01| & $\Xi^- + \bar{\Xi}^+, p_\perp$ spectrum V0M Class VII & 
    y-axis: $1/N \frac{d^2N}{dydp_\perp}$ \newline
    x-axis: $p_\perp$ [GeV]
     & 1.0 \\
    \midrule
    
    \lstinline[basicstyle=\tiny\ttfamily]|/ALICE_2016_I1471838/d28-x01-y01| & $\Xi^- + \bar{\Xi}^+, p_\perp$ spectrum V0M Class VIII & 
    y-axis: $1/N \frac{d^2N}{dydp_\perp}$ \newline
    x-axis: $p_\perp$ [GeV]
     & 1.0 \\
    \midrule
    
    \lstinline[basicstyle=\tiny\ttfamily]|/ALICE_2016_I1471838/d29-x01-y01| & $\Xi^- + \bar{\Xi}^+, p_\perp$ spectrum V0M Class IX & 
    y-axis: $1/N \frac{d^2N}{dydp_\perp}$ \newline
    x-axis: $p_\perp$ [GeV]
     & 1.0 \\
    \midrule
    
    \lstinline[basicstyle=\tiny\ttfamily]|/ALICE_2016_I1471838/d30-x01-y01| & $\Xi^- + \bar{\Xi}^+, p_\perp$ spectrum V0M Class X & 
    y-axis: $1/N \frac{d^2N}{dydp_\perp}$ \newline
    x-axis: $p_\perp$ [GeV]
     & 1.0 \\
    \midrule
    
    \lstinline[basicstyle=\tiny\ttfamily]|/ALICE_2016_I1471838/d31-x01-y01| & $\Omega^- + \bar{\Omega}^+, p_\perp$ spectrum V0M Class I+II & 
    y-axis: $1/N \frac{d^2N}{dydp_\perp}$ \newline
    x-axis: $p_\perp$ [GeV]
     & 1.0 \\
    \midrule
    
    \lstinline[basicstyle=\tiny\ttfamily]|/ALICE_2016_I1471838/d32-x01-y01| & $\Omega^- + \bar{\Omega}^+, p_\perp$ spectrum V0M Class III+IV & 
    y-axis: $1/N \frac{d^2N}{dydp_\perp}$ \newline
    x-axis: $p_\perp$ [GeV]
     & 1.0 \\
    \midrule
    
    \lstinline[basicstyle=\tiny\ttfamily]|/ALICE_2016_I1471838/d33-x01-y01| & $\Omega^- + \bar{\Omega}^+, p_\perp$ spectrum V0M Class V+VI & 
    y-axis: $1/N \frac{d^2N}{dydp_\perp}$ \newline
    x-axis: $p_\perp$ [GeV]
     & 1.0 \\
    \midrule
    
    \lstinline[basicstyle=\tiny\ttfamily]|/ALICE_2016_I1471838/d34-x01-y01| & $\Omega^- + \bar{\Omega}^+, p_\perp$ spectrum V0M Class VI+VIII & 
    y-axis: $1/N \frac{d^2N}{dydp_\perp}$ \newline
    x-axis: $p_\perp$ [GeV]
     & 1.0 \\
    \midrule
    
    \lstinline[basicstyle=\tiny\ttfamily]|/ALICE_2016_I1471838/d35-x01-y01| & $\Omega^- + \bar{\Omega}^+, p_\perp$ spectrum V0M Class IX+X & 
    y-axis: $1/N \frac{d^2N}{dydp_\perp}$ \newline
    x-axis: $p_\perp$ [GeV]
     & 1.0 \\
    \midrule
    
    \lstinline[basicstyle=\tiny\ttfamily]|/ALICE_2018_I1684320/d01-x01-y01| & Charged particles, $p_\perp$ spectra, $|\eta| < 0.5$, V0M Class I, pp, $\sqrt{s}=7$ TeV & 
    y-axis: $1/N_\mathrm{evt}\,\frac{d^2 N}{d\eta \, dp_\perp}$ \,[$(\mathrm{GeV}/c)^{-1}$] \newline
    x-axis: $p_\perp\,[\mathrm{GeV}/c]$
     & 1.0 \\
    \midrule
    
    \lstinline[basicstyle=\tiny\ttfamily]|/ALICE_2018_I1684320/d02-x01-y01| & Charged particles, $p_\perp$ spectra, $|\eta| < 0.5$, V0M Class II, pp, $\sqrt{s}=7$ TeV & 
    y-axis: $1/N_\mathrm{evt}\,\frac{d^2 N}{d\eta \, dp_\perp}$ \,[$(\mathrm{GeV}/c)^{-1}$] \newline
    x-axis: $p_\perp\,[\mathrm{GeV}/c]$
     & 1.0 \\
    \midrule
    
    \lstinline[basicstyle=\tiny\ttfamily]|/ALICE_2018_I1684320/d03-x01-y01| & Charged particles, $p_\perp$ spectra, $|\eta| < 0.5$, V0M Class III, pp, $\sqrt{s}=7$ TeV & 
    y-axis: $1/N_\mathrm{evt}\,\frac{d^2 N}{d\eta \, dp_\perp}$ \,[$(\mathrm{GeV}/c)^{-1}$] \newline
    x-axis: $p_\perp\,[\mathrm{GeV}/c]$
     & 1.0 \\
    \midrule
    
    \lstinline[basicstyle=\tiny\ttfamily]|/ALICE_2018_I1684320/d04-x01-y01| & Charged particles, $p_\perp$ spectra, $|\eta| < 0.5$, V0M Class IV, pp, $\sqrt{s}=7$ TeV & 
    y-axis: $1/N_\mathrm{evt}\,\frac{d^2 N}{d\eta \, dp_\perp}$ \,[$(\mathrm{GeV}/c)^{-1}$] \newline
    x-axis: $p_\perp\,[\mathrm{GeV}/c]$
     & 1.0 \\
    \midrule
    
    \lstinline[basicstyle=\tiny\ttfamily]|/ALICE_2018_I1684320/d05-x01-y01| & Charged particles, $p_\perp$ spectra, $|\eta| < 0.5$, V0M Class V, pp, $\sqrt{s}=7$ TeV & 
    y-axis: $1/N_\mathrm{evt}\,\frac{d^2 N}{d\eta \, dp_\perp}$ \,[$(\mathrm{GeV}/c)^{-1}$] \newline
    x-axis: $p_\perp\,[\mathrm{GeV}/c]$
     & 1.0 \\
    \midrule
    
    \lstinline[basicstyle=\tiny\ttfamily]|/ALICE_2018_I1684320/d06-x01-y01| & Charged particles, $p_\perp$ spectra, $|\eta| < 0.5$, V0M Class VI, pp, $\sqrt{s}=7$ TeV & 
    y-axis: $1/N_\mathrm{evt}\,\frac{d^2 N}{d\eta \, dp_\perp}$ \,[$(\mathrm{GeV}/c)^{-1}$] \newline
    x-axis: $p_\perp\,[\mathrm{GeV}/c]$
     & 1.0 \\
    \midrule
    
    \lstinline[basicstyle=\tiny\ttfamily]|/ALICE_2018_I1684320/d07-x01-y01| & Charged particles, $p_\perp$ spectra, $|\eta| < 0.5$, V0M Class VII, pp, $\sqrt{s}=7$ TeV & 
    y-axis: $1/N_\mathrm{evt}\,\frac{d^2 N}{d\eta \, dp_\perp}$ \,[$(\mathrm{GeV}/c)^{-1}$] \newline
    x-axis: $p_\perp\,[\mathrm{GeV}/c]$
     & 1.0 \\
    \midrule
    
    \lstinline[basicstyle=\tiny\ttfamily]|/ALICE_2018_I1684320/d08-x01-y01| & Charged particles, $p_\perp$ spectra, $|\eta| < 0.5$, V0M Class VIII, pp, $\sqrt{s}=7$ TeV & 
    y-axis: $1/N_\mathrm{evt}\,\frac{d^2 N}{d\eta \, dp_\perp}$ \,[$(\mathrm{GeV}/c)^{-1}$] \newline
    x-axis: $p_\perp\,[\mathrm{GeV}/c]$
     & 1.0 \\
    \midrule
    
    \lstinline[basicstyle=\tiny\ttfamily]|/ALICE_2018_I1684320/d09-x01-y01| & Charged particles, $p_\perp$ spectra, $|\eta| < 0.5$, V0M Class IX, pp, $\sqrt{s}=7$ TeV & 
    y-axis: $1/N_\mathrm{evt}\,\frac{d^2 N}{d\eta \, dp_\perp}$ \,[$(\mathrm{GeV}/c)^{-1}$] \newline
    x-axis: $p_\perp\,[\mathrm{GeV}/c]$
     & 1.0 \\
    \midrule
    
    \lstinline[basicstyle=\tiny\ttfamily]|/ALICE_2018_I1684320/d10-x01-y01| & Charged particles, $p_\perp$ spectra, $|\eta| < 0.5$, V0M Class X, pp, $\sqrt{s}=7$ TeV & 
    y-axis: $1/N_\mathrm{evt}\,\frac{d^2 N}{d\eta \, dp_\perp}$ \,[$(\mathrm{GeV}/c)^{-1}$] \newline
    x-axis: $p_\perp\,[\mathrm{GeV}/c]$
     & 1.0 \\
    \midrule
    
    \lstinline[basicstyle=\tiny\ttfamily]|/ALICE_2018_I1684320/d11-x01-y01| & $\pi^+ + \pi^-$, $p_\perp$ spectra, V0M Class I, pp, $\sqrt{s}=7$ TeV & 
    y-axis: $1/N_\mathrm{evt}\,\frac{d^2 N}{dy \, dp_\perp}$ \,[$(\mathrm{GeV}/c)^{-1}$] \newline
    x-axis: $p_\perp\,[\mathrm{GeV}/c]$
     & 1.0 \\
    \midrule
    
    \lstinline[basicstyle=\tiny\ttfamily]|/ALICE_2018_I1684320/d12-x01-y01| & $\pi^+ + \pi^-$, $p_\perp$ spectra, V0M Class II, pp, $\sqrt{s}=7$ TeV & 
    y-axis: $1/N_\mathrm{evt}\,\frac{d^2 N}{dy \, dp_\perp}$ \,[$(\mathrm{GeV}/c)^{-1}$] \newline
    x-axis: $p_\perp\,[\mathrm{GeV}/c]$
     & 1.0 \\
    \midrule
    
    \lstinline[basicstyle=\tiny\ttfamily]|/ALICE_2018_I1684320/d13-x01-y01| & $\pi^+ + \pi^-$, $p_\perp$ spectra, V0M Class III, pp, $\sqrt{s}=7$ TeV & 
    y-axis: $1/N_\mathrm{evt}\,\frac{d^2 N}{dy \, dp_\perp}$ \,[$(\mathrm{GeV}/c)^{-1}$] \newline
    x-axis: $p_\perp\,[\mathrm{GeV}/c]$
     & 1.0 \\
    \midrule
    
     \lstinline[basicstyle=\tiny\ttfamily]|/ALICE_2018_I1684320/d14-x01-y01| & $\pi^+ + \pi^-$, $p_\perp$ spectra, V0M Class IV, pp, $\sqrt{s}=7$ TeV & 
    y-axis: $1/N_\mathrm{evt}\,\frac{d^2 N}{dy \, dp_\perp}$ \,[$(\mathrm{GeV}/c)^{-1}$] \newline
    x-axis: $p_\perp\,[\mathrm{GeV}/c]$
     & 1.0 \\
    \midrule
    
    \lstinline[basicstyle=\tiny\ttfamily]|/ALICE_2018_I1684320/d15-x01-y01| & $\pi^+ + \pi^-$, $p_\perp$ spectra, V0M Class V, pp, $\sqrt{s}=7$ TeV & 
    y-axis: $1/N_\mathrm{evt}\,\frac{d^2 N}{dy \, dp_\perp}$ \,[$(\mathrm{GeV}/c)^{-1}$] \newline
    x-axis: $p_\perp\,[\mathrm{GeV}/c]$
     & 1.0 \\
    \midrule
    
    \lstinline[basicstyle=\tiny\ttfamily]|/ALICE_2018_I1684320/d16-x01-y01| & $\pi^+ + \pi^-$, $p_\perp$ spectra, V0M Class VI, pp, $\sqrt{s}=7$ TeV & 
    y-axis: $1/N_\mathrm{evt}\,\frac{d^2 N}{dy \, dp_\perp}$ \,[$(\mathrm{GeV}/c)^{-1}$] \newline
    x-axis: $p_\perp\,[\mathrm{GeV}/c]$
     & 1.0 \\
    \midrule
    
    \lstinline[basicstyle=\tiny\ttfamily]|/ALICE_2018_I1684320/d17-x01-y01| & $\pi^+ + \pi^-$, $p_\perp$ spectra, V0M Class VII, pp, $\sqrt{s}=7$ TeV & 
    y-axis: $1/N_\mathrm{evt}\,\frac{d^2 N}{dy \, dp_\perp}$ \,[$(\mathrm{GeV}/c)^{-1}$] \newline
    x-axis: $p_\perp\,[\mathrm{GeV}/c]$
     & 1.0 \\
    \midrule
    
    \lstinline[basicstyle=\tiny\ttfamily]|/ALICE_2018_I1684320/d18-x01-y01| & $\pi^+ + \pi^-$, $p_\perp$ spectra, V0M Class VIII, pp, $\sqrt{s}=7$ TeV & 
    y-axis: $1/N_\mathrm{evt}\,\frac{d^2 N}{dy \, dp_\perp}$ \,[$(\mathrm{GeV}/c)^{-1}$] \newline
    x-axis: $p_\perp\,[\mathrm{GeV}/c]$
     & 1.0 \\
    \midrule
    
    \lstinline[basicstyle=\tiny\ttfamily]|/ALICE_2018_I1684320/d19-x01-y01| & $\pi^+ + \pi^-$, $p_\perp$ spectra, V0M Class IX, pp, $\sqrt{s}=7$ TeV & 
    y-axis: $1/N_\mathrm{evt}\,\frac{d^2 N}{dy \, dp_\perp}$ \,[$(\mathrm{GeV}/c)^{-1}$] \newline
    x-axis: $p_\perp\,[\mathrm{GeV}/c]$
     & 1.0 \\
    \midrule
    
    \lstinline[basicstyle=\tiny\ttfamily]|/ALICE_2018_I1684320/d20-x01-y01| & $\pi^+ + \pi^-$, $p_\perp$ spectra, V0M Class X, pp, $\sqrt{s}=7$ TeV & 
    y-axis: $1/N_\mathrm{evt}\,\frac{d^2 N}{dy \, dp_\perp}$ \,[$(\mathrm{GeV}/c)^{-1}$] \newline
    x-axis: $p_\perp\,[\mathrm{GeV}/c]$
     & 1.0 \\
    \midrule
    
    \lstinline[basicstyle=\tiny\ttfamily]|/ALICE_2018_I1684320/d21-x01-y01| & $K^+ + K^-$, $p_\perp$ spectra, V0M Class I, pp, $\sqrt{s}=7$ TeV & 
    y-axis: $1/N_\mathrm{evt}\,\frac{d^2 N}{dy \, dp_\perp}$ \,[$(\mathrm{GeV}/c)^{-1}$] \newline
    x-axis: $p_\perp\,[\mathrm{GeV}/c]$
     & 1.0 \\
    \midrule
    
    \lstinline[basicstyle=\tiny\ttfamily]|/ALICE_2018_I1684320/d22-x01-y01| & $K^+ + K^-$, $p_\perp$ spectra, V0M Class II, pp, $\sqrt{s}=7$ TeV & 
    y-axis: $1/N_\mathrm{evt}\,\frac{d^2 N}{dy \, dp_\perp}$ \,[$(\mathrm{GeV}/c)^{-1}$] \newline
    x-axis: $p_\perp\,[\mathrm{GeV}/c]$
     & 1.0 \\
    \midrule
    
    \lstinline[basicstyle=\tiny\ttfamily]|/ALICE_2018_I1684320/d23-x01-y01| & $K^+ + K^-$, $p_\perp$ spectra, V0M Class III, pp, $\sqrt{s}=7$ TeV & 
    y-axis: $1/N_\mathrm{evt}\,\frac{d^2 N}{dy \, dp_\perp}$ \,[$(\mathrm{GeV}/c)^{-1}$] \newline
    x-axis: $p_\perp\,[\mathrm{GeV}/c]$
     & 1.0 \\
    \midrule
    
    \lstinline[basicstyle=\tiny\ttfamily]|/ALICE_2018_I1684320/d24-x01-y01| & $K^+ + K^-$, $p_\perp$ spectra, V0M Class IV, pp, $\sqrt{s}=7$ TeV & 
    y-axis: $1/N_\mathrm{evt}\,\frac{d^2 N}{dy \, dp_\perp}$ \,[$(\mathrm{GeV}/c)^{-1}$] \newline
    x-axis: $p_\perp\,[\mathrm{GeV}/c]$
     & 1.0 \\
    \midrule
    
    \lstinline[basicstyle=\tiny\ttfamily]|/ALICE_2018_I1684320/d25-x01-y01| & $K^+ + K^-$, $p_\perp$ spectra, V0M Class V, pp, $\sqrt{s}=7$ TeV & 
    y-axis: $1/N_\mathrm{evt}\,\frac{d^2 N}{dy \, dp_\perp}$ \,[$(\mathrm{GeV}/c)^{-1}$] \newline
    x-axis: $p_\perp\,[\mathrm{GeV}/c]$
     & 1.0 \\
    \midrule
    
    \lstinline[basicstyle=\tiny\ttfamily]|/ALICE_2018_I1684320/d26-x01-y01| & $K^+ + K^-$, $p_\perp$ spectra, V0M Class VI, pp, $\sqrt{s}=7$ TeV & 
    y-axis: $1/N_\mathrm{evt}\,\frac{d^2 N}{dy \, dp_\perp}$ \,[$(\mathrm{GeV}/c)^{-1}$] \newline
    x-axis: $p_\perp\,[\mathrm{GeV}/c]$
     & 1.0 \\
    \midrule
    
    \lstinline[basicstyle=\tiny\ttfamily]|/ALICE_2018_I1684320/d27-x01-y01| & $K^+ + K^-$, $p_\perp$ spectra, V0M Class VII, pp, $\sqrt{s}=7$ TeV & 
    y-axis: $1/N_\mathrm{evt}\,\frac{d^2 N}{dy \, dp_\perp}$ \,[$(\mathrm{GeV}/c)^{-1}$] \newline
    x-axis: $p_\perp\,[\mathrm{GeV}/c]$
     & 1.0 \\
    \midrule
    
    \lstinline[basicstyle=\tiny\ttfamily]|/ALICE_2018_I1684320/d28-x01-y01| & $K^+ + K^-$, $p_\perp$ spectra, V0M Class VIII, pp, $\sqrt{s}=7$ TeV & 
    y-axis: $1/N_\mathrm{evt}\,\frac{d^2 N}{dy \, dp_\perp}$ \,[$(\mathrm{GeV}/c)^{-1}$] \newline
    x-axis: $p_\perp\,[\mathrm{GeV}/c]$
     & 1.0 \\
    \midrule
    
    \lstinline[basicstyle=\tiny\ttfamily]|/ALICE_2018_I1684320/d29-x01-y01| & $K^+ + K^-$, $p_\perp$ spectra, V0M Class IX, pp, $\sqrt{s}=7$ TeV & 
    y-axis: $1/N_\mathrm{evt}\,\frac{d^2 N}{dy \, dp_\perp}$ \,[$(\mathrm{GeV}/c)^{-1}$] \newline
    x-axis: $p_\perp\,[\mathrm{GeV}/c]$
     & 1.0 \\
    \midrule
    
    \lstinline[basicstyle=\tiny\ttfamily]|/ALICE_2018_I1684320/d30-x01-y01| & $K^+ + K^-$, $p_\perp$ spectra, V0M Class X, pp, $\sqrt{s}=7$ TeV & 
    y-axis: $1/N_\mathrm{evt}\,\frac{d^2 N}{dy \, dp_\perp}$ \,[$(\mathrm{GeV}/c)^{-1}$] \newline
    x-axis: $p_\perp\,[\mathrm{GeV}/c]$
     & 1.0 \\
    \midrule
    
    \lstinline[basicstyle=\tiny\ttfamily]|/ALICE_2018_I1684320/d31-x01-y01| & $p + \overline{p}$, $p_\perp$ spectra, V0M Class I, pp, $\sqrt{s}=7$ TeV & 
    y-axis: $1/N_\mathrm{evt}\,\frac{d^2 N}{dy \, dp_\perp}$ \,[$(\mathrm{GeV}/c)^{-1}$] \newline
    x-axis: $p_\perp\,[\mathrm{GeV}/c]$
     & 1.0 \\
    \midrule
    
    \lstinline[basicstyle=\tiny\ttfamily]|/ALICE_2018_I1684320/d32-x01-y01| & $p + \overline{p}$, $p_\perp$ spectra, V0M Class II, pp, $\sqrt{s}=7$ TeV & 
    y-axis: $1/N_\mathrm{evt}\,\frac{d^2 N}{dy \, dp_\perp}$ \,[$(\mathrm{GeV}/c)^{-1}$] \newline
    x-axis: $p_\perp\,[\mathrm{GeV}/c]$
     & 1.0 \\
    \midrule
    
    \lstinline[basicstyle=\tiny\ttfamily]|/ALICE_2018_I1684320/d33-x01-y01| & $p + \overline{p}$, $p_\perp$ spectra, V0M Class III, pp, $\sqrt{s}=7$ TeV & 
    y-axis: $1/N_\mathrm{evt}\,\frac{d^2 N}{dy \, dp_\perp}$ \,[$(\mathrm{GeV}/c)^{-1}$] \newline
    x-axis: $p_\perp\,[\mathrm{GeV}/c]$
     & 1.0 \\
    \midrule
    
    \lstinline[basicstyle=\tiny\ttfamily]|/ALICE_2018_I1684320/d34-x01-y01| & $p + \overline{p}$, $p_\perp$ spectra, V0M Class VI, pp, $\sqrt{s}=7$ TeV & 
    y-axis: $1/N_\mathrm{evt}\,\frac{d^2 N}{dy \, dp_\perp}$ \,[$(\mathrm{GeV}/c)^{-1}$] \newline
    x-axis: $p_\perp\,[\mathrm{GeV}/c]$
     & 1.0 \\
    \midrule
    
    \lstinline[basicstyle=\tiny\ttfamily]|/ALICE_2018_I1684320/d35-x01-y01| & $p + \overline{p}$, $p_\perp$ spectra, V0M Class V, pp, $\sqrt{s}=7$ TeV & 
    y-axis: $1/N_\mathrm{evt}\,\frac{d^2 N}{dy \, dp_\perp}$ \,[$(\mathrm{GeV}/c)^{-1}$] \newline
    x-axis: $p_\perp\,[\mathrm{GeV}/c]$
     & 1.0 \\
    \midrule
    
    \lstinline[basicstyle=\tiny\ttfamily]|/ALICE_2018_I1684320/d36-x01-y01| & $p + \overline{p}$, $p_\perp$ spectra, V0M Class VI, pp, $\sqrt{s}=7$ TeV & 
    y-axis: $1/N_\mathrm{evt}\,\frac{d^2 N}{dy \, dp_\perp}$ \,[$(\mathrm{GeV}/c)^{-1}$] \newline
    x-axis: $p_\perp\,[\mathrm{GeV}/c]$
     & 1.0 \\
    \midrule
    
    \lstinline[basicstyle=\tiny\ttfamily]|/ALICE_2018_I1684320/d37-x01-y01| & $p + \overline{p}$, $p_\perp$ spectra, V0M Class VII, pp, $\sqrt{s}=7$ TeV & 
    y-axis: $1/N_\mathrm{evt}\,\frac{d^2 N}{dy \, dp_\perp}$ \,[$(\mathrm{GeV}/c)^{-1}$] \newline
    x-axis: $p_\perp\,[\mathrm{GeV}/c]$
     & 1.0 \\
    \midrule
    
    \lstinline[basicstyle=\tiny\ttfamily]|/ALICE_2018_I1684320/d38-x01-y01| & $p + \overline{p}$, $p_\perp$ spectra, V0M Class VIII, pp, $\sqrt{s}=7$ TeV & 
    y-axis: $1/N_\mathrm{evt}\,\frac{d^2 N}{dy \, dp_\perp}$ \,[$(\mathrm{GeV}/c)^{-1}$] \newline
    x-axis: $p_\perp\,[\mathrm{GeV}/c]$
     & 1.0 \\
    \midrule
    
    \lstinline[basicstyle=\tiny\ttfamily]|/ALICE_2018_I1684320/d39-x01-y01| & $p + \overline{p}$, $p_\perp$ spectra, V0M Class IX, pp, $\sqrt{s}=7$ TeV & 
    y-axis: $1/N_\mathrm{evt}\,\frac{d^2 N}{dy \, dp_\perp}$ \,[$(\mathrm{GeV}/c)^{-1}$] \newline
    x-axis: $p_\perp\,[\mathrm{GeV}/c]$
     & 1.0 \\
    \midrule
    
    \lstinline[basicstyle=\tiny\ttfamily]|/ALICE_2018_I1684320/d40-x01-y01| & $p + \overline{p}$, $p_\perp$ spectra, V0M Class X, pp, $\sqrt{s}=7$ TeV & 
    y-axis: $1/N_\mathrm{evt}\,\frac{d^2 N}{dy \, dp_\perp}$ \,[$(\mathrm{GeV}/c)^{-1}$] \newline
    x-axis: $p_\perp\,[\mathrm{GeV}/c]$
     & 1.0 \\
    \midrule
    
    \lstinline[basicstyle=\tiny\ttfamily]|/ALICE_2018_I1684320/d41-x01-y01| & $K^* + \overline{K}^*$, $p_\perp$ spectra, V0M Class I, pp, $\sqrt{s}=7$ TeV & 
    y-axis: $1/N_\mathrm{evt}\,\frac{d^2 N}{dy \, dp_\perp}$ \,[$(\mathrm{GeV}/c)^{-1}$] \newline
    x-axis: $p_\perp\,[\mathrm{GeV}/c]$
     & 1.0 \\
    \midrule
    
    \lstinline[basicstyle=\tiny\ttfamily]|/ALICE_2018_I1684320/d42-x01-y01| & $K^* + \overline{K}^*$, $p_\perp$ spectra, V0M Class II, pp, $\sqrt{s}=7$ TeV & 
    y-axis: $1/N_\mathrm{evt}\,\frac{d^2 N}{dy \, dp_\perp}$ \,[$(\mathrm{GeV}/c)^{-1}$] \newline
    x-axis: $p_\perp\,[\mathrm{GeV}/c]$
     & 1.0 \\
    \midrule
    
    \lstinline[basicstyle=\tiny\ttfamily]|/ALICE_2018_I1684320/d43-x01-y01| & $K^* + \overline{K}^*$, $p_\perp$ spectra, V0M Class II, pp, $\sqrt{s}=7$ TeV & 
    y-axis: $1/N_\mathrm{evt}\,\frac{d^2 N}{dy \, dp_\perp}$ \,[$(\mathrm{GeV}/c)^{-1}$] \newline
    x-axis: $p_\perp\,[\mathrm{GeV}/c]$
     & 1.0 \\
    \midrule
    
    \lstinline[basicstyle=\tiny\ttfamily]|/ALICE_2018_I1684320/d44-x01-y01| & $K^* + \overline{K}^*$, $p_\perp$ spectra, V0M Class IV+V, pp, $\sqrt{s}=7$ TeV & 
    y-axis: $1/N_\mathrm{evt}\,\frac{d^2 N}{dy \, dp_\perp}$ \,[$(\mathrm{GeV}/c)^{-1}$] \newline
    x-axis: $p_\perp\,[\mathrm{GeV}/c]$
     & 1.0 \\
    \midrule
    
    \lstinline[basicstyle=\tiny\ttfamily]|/ALICE_2018_I1684320/d45-x01-y01| & $K^* + \overline{K}^*$, $p_\perp$ spectra, V0M Class VI, pp, $\sqrt{s}=7$ TeV & 
    y-axis: $1/N_\mathrm{evt}\,\frac{d^2 N}{dy \, dp_\perp}$ \,[$(\mathrm{GeV}/c)^{-1}$] \newline
    x-axis: $p_\perp\,[\mathrm{GeV}/c]$
     & 1.0 \\
    \midrule
    
     \lstinline[basicstyle=\tiny\ttfamily]|/ALICE_2018_I1684320/d46-x01-y01| & $K^* + \overline{K}^*$, $p_\perp$ spectra, V0M Class VII, pp, $\sqrt{s}=7$ TeV & 
    y-axis: $1/N_\mathrm{evt}\,\frac{d^2 N}{dy \, dp_\perp}$ \,[$(\mathrm{GeV}/c)^{-1}$] \newline
    x-axis: $p_\perp\,[\mathrm{GeV}/c]$
     & 1.0 \\
    \midrule
    
    \lstinline[basicstyle=\tiny\ttfamily]|/ALICE_2018_I1684320/d47-x01-y01| & $K^* + \overline{K}^*$, $p_\perp$ spectra, V0M Class VIII, pp, $\sqrt{s}=7$ TeV & 
    y-axis: $1/N_\mathrm{evt}\,\frac{d^2 N}{dy \, dp_\perp}$ \,[$(\mathrm{GeV}/c)^{-1}$] \newline
    x-axis: $p_\perp\,[\mathrm{GeV}/c]$
     & 1.0 \\
    \midrule
    
    \lstinline[basicstyle=\tiny\ttfamily]|/ALICE_2018_I1684320/d48-x01-y01| & $K^* + \overline{K}^*$, $p_\perp$ spectra, V0M Class IX, pp, $\sqrt{s}=7$ TeV & 
    y-axis: $1/N_\mathrm{evt}\,\frac{d^2 N}{dy \, dp_\perp}$ \,[$(\mathrm{GeV}/c)^{-1}$] \newline
    x-axis: $p_\perp\,[\mathrm{GeV}/c]$
     & 1.0 \\
    \midrule
    
    \lstinline[basicstyle=\tiny\ttfamily]|/ALICE_2018_I1684320/d49-x01-y01| & $K^* + \overline{K}^*$, $p_\perp$ spectra, V0M Class X, pp, $\sqrt{s}=7$ TeV & 
    y-axis: $1/N_\mathrm{evt}\,\frac{d^2 N}{dy \, dp_\perp}$ \,[$(\mathrm{GeV}/c)^{-1}$] \newline
    x-axis: $p_\perp\,[\mathrm{GeV}/c]$
     & 1.0 \\
    \midrule
    
    \lstinline[basicstyle=\tiny\ttfamily]|/ALICE_2018_I1684320/d50-x01-y01| & $\phi$, $p_\perp$ spectra, V0M Class I, pp, $\sqrt{s}=7$ TeV & 
    y-axis: $1/N_\mathrm{evt}\,\frac{d^2 N}{dy \, dp_\perp}$ \,[$(\mathrm{GeV}/c)^{-1}$] \newline
    x-axis: $p_\perp\,[\mathrm{GeV}/c]$
     & 1.0 \\
    \midrule
    
    \lstinline[basicstyle=\tiny\ttfamily]|/ALICE_2018_I1684320/d51-x01-y01| & $\phi$, $p_\perp$ spectra, V0M Class II, pp, $\sqrt{s}=7$ TeV & 
    y-axis: $1/N_\mathrm{evt}\,\frac{d^2 N}{dy \, dp_\perp}$ \,[$(\mathrm{GeV}/c)^{-1}$] \newline
    x-axis: $p_\perp\,[\mathrm{GeV}/c]$
     & 1.0 \\
    \midrule
    
    \lstinline[basicstyle=\tiny\ttfamily]|/ALICE_2018_I1684320/d52-x01-y01| & $\phi$, $p_\perp$ spectra, V0M Class III, pp, $\sqrt{s}=7$ TeV & 
    y-axis: $1/N_\mathrm{evt}\,\frac{d^2 N}{dy \, dp_\perp}$ \,[$(\mathrm{GeV}/c)^{-1}$] \newline
    x-axis: $p_\perp\,[\mathrm{GeV}/c]$
     & 1.0 \\
    \midrule
    
    \lstinline[basicstyle=\tiny\ttfamily]|/ALICE_2018_I1684320/d53-x01-y01| & $\phi$, $p_\perp$ spectra, V0M Class IV+V, pp, $\sqrt{s}=7$ TeV & 
    y-axis: $1/N_\mathrm{evt}\,\frac{d^2 N}{dy \, dp_\perp}$ \,[$(\mathrm{GeV}/c)^{-1}$] \newline
    x-axis: $p_\perp\,[\mathrm{GeV}/c]$
     & 1.0 \\
    \midrule
    
    \lstinline[basicstyle=\tiny\ttfamily]|/ALICE_2018_I1684320/d54-x01-y01| & $\phi$, $p_\perp$ spectra, V0M Class VI, pp, $\sqrt{s}=7$ TeV & 
    y-axis: $1/N_\mathrm{evt}\,\frac{d^2 N}{dy \, dp_\perp}$ \,[$(\mathrm{GeV}/c)^{-1}$] \newline
    x-axis: $p_\perp\,[\mathrm{GeV}/c]$
     & 1.0 \\
    \midrule
    
    \lstinline[basicstyle=\tiny\ttfamily]|/ALICE_2018_I1684320/d55-x01-y01| & $\phi$, $p_\perp$ spectra, V0M Class VII, pp, $\sqrt{s}=7$ TeV & 
    y-axis: $1/N_\mathrm{evt}\,\frac{d^2 N}{dy \, dp_\perp}$ \,[$(\mathrm{GeV}/c)^{-1}$] \newline
    x-axis: $p_\perp\,[\mathrm{GeV}/c]$
     & 1.0 \\
    \midrule
    
    \lstinline[basicstyle=\tiny\ttfamily]|/ALICE_2018_I1684320/d56-x01-y01| & $\phi$, $p_\perp$ spectra, V0M Class VIII, pp, $\sqrt{s}=7$ TeV & 
    y-axis: $1/N_\mathrm{evt}\,\frac{d^2 N}{dy \, dp_\perp}$ \,[$(\mathrm{GeV}/c)^{-1}$] \newline
    x-axis: $p_\perp\,[\mathrm{GeV}/c]$
     & 1.0 \\
    \midrule
    
    \lstinline[basicstyle=\tiny\ttfamily]|/ALICE_2018_I1684320/d57-x01-y01| & $\phi$, $p_\perp$ spectra, V0M Class IX, pp, $\sqrt{s}=7$ TeV & 
    y-axis: $1/N_\mathrm{evt}\,\frac{d^2 N}{dy \, dp_\perp}$ \,[$(\mathrm{GeV}/c)^{-1}$] \newline
    x-axis: $p_\perp\,[\mathrm{GeV}/c]$
     & 1.0 \\
    \midrule
    
    \lstinline[basicstyle=\tiny\ttfamily]|/ALICE_2018_I1684320/d58-x01-y01| & $\phi$, $p_\perp$ spectra, V0M Class X, pp, $\sqrt{s}=7$ TeV & 
    y-axis: $1/N_\mathrm{evt}\,\frac{d^2 N}{dy \, dp_\perp}$ \,[$(\mathrm{GeV}/c)^{-1}$] \newline
    x-axis: $p_\perp\,[\mathrm{GeV}/c]$
     & 1.0 \\
    \midrule
    
    \lstinline[basicstyle=\tiny\ttfamily]|/ATLAS_2010_S8918562/d03-x01-y01| & Charged particle $\eta$ at 7$\,$TeV, track $p_\perp > 500\,\mathrm{MeV}$, for $N_\mathrm{ch} \geq 1$ & 
    y-axis: $1/N_\mathrm{ev} \, \mathrm{d}N_\mathrm{ch}/\mathrm{d}\eta$ \newline
    x-axis: $\eta$
     & 6.0 \\
    \midrule
    
    \lstinline[basicstyle=\tiny\ttfamily]|/ATLAS_2010_S8918562/d05-x01-y01| & Charged particle $\eta$ at 7$\,$TeV, track $p_\perp > 100\,\mathrm{MeV}$, for $N_\mathrm{ch} \geq 2$ & 
    y-axis: $1/N_\mathrm{ev} \, \mathrm{d}N_\mathrm{ch}/\mathrm{d}\eta$ \newline
    x-axis: $\eta$
     & 6.0 \\
    \midrule 
    
    \lstinline[basicstyle=\tiny\ttfamily]|/ATLAS_2010_S8918562/d07-x01-y01| & Charged particle $\eta$ at 7$\,$TeV, track $p_\perp > 500\,\mathrm{MeV}$, for $N_\mathrm{ch} \geq 6$ & 
    y-axis: $1/N_\mathrm{ev} \, \mathrm{d}N_\mathrm{ch}/\mathrm{d}\eta$ \newline
    x-axis: $\eta$
     & 6.0 \\
    \midrule
    
    \lstinline[basicstyle=\tiny\ttfamily]|/ATLAS_2010_S8918562/d10-x01-y01| & Charged particle $p_\perp$ at 7$\,$TeV, track $p_\perp > 500\,\mathrm{MeV}$, for $N_\mathrm{ch} \geq 1$ & 
    y-axis: $1/N_\mathrm{ev} \, 1/2\pi{}p_\perp \, \mathrm{d}\sigma/\mathrm{d}\eta\mathrm{d}p_\perp$ \newline
    x-axis: $p_\perp$ $\;$ [GeV]
     & 1.0 \\
    \midrule 
    
    \lstinline[basicstyle=\tiny\ttfamily]|/ATLAS_2010_S8918562/d12-x01-y01| & Charged particle $p_\perp$ at 7$\,$TeV, track $p_\perp > 100\,\mathrm{MeV}$, for $N_\mathrm{ch} \geq 2$ & 
    y-axis: $1/N_\mathrm{ev} \, 1/2\pi{}p_\perp \, \mathrm{d}\sigma/\mathrm{d}\eta\mathrm{d}p_\perp$ \newline
    x-axis: $p_\perp$ $\;$ [GeV]
     & 1.0 \\
    \midrule 
    
    \lstinline[basicstyle=\tiny\ttfamily]|/ATLAS_2010_S8918562/d14-x01-y01| & Charged particle $p_\perp$ at 7$\,$TeV, track $p_\perp > 500\,\mathrm{MeV}$, for $N_\mathrm{ch} \geq 6$ & 
    y-axis: $1/N_\mathrm{ev} \, 1/2\pi{}p_\perp \, \mathrm{d}\sigma/\mathrm{d}\eta\mathrm{d}p_\perp$ \newline
    x-axis: $p_\perp$ $\;$ [GeV]
     & 1.0 \\
    \midrule 
    
    \lstinline[basicstyle=\tiny\ttfamily]|/ATLAS_2010_S8918562/d23-x01-y01| & Charged $\langle p_\perp \rangle$ vs. $N_\mathrm{ch}$ at 7$\,$TeV, track $p_\perp > 500\,\mathrm{MeV}$, for $N_\mathrm{ch} \geq 1$ & 
    y-axis: $\langle p_\perp \rangle$ $\;$ [GeV] \newline
    x-axis: $N_\mathrm{ch}$
     & 8.0 \\
    \midrule 
    
    \lstinline[basicstyle=\tiny\ttfamily]|/ATLAS_2010_S8918562/d25-x01-y01| & Charged $\langle p_\perp \rangle$ vs. $N_\mathrm{ch}$ at 7$\,$TeV, track $p_\perp > 100\,\mathrm{MeV}$, for $N_\mathrm{ch} \geq 2$ & 
    y-axis: $\langle p_\perp \rangle$ $\;$ [GeV] \newline
    x-axis: $N_\mathrm{ch}$
     & 8.0 \\
    \midrule 
    
    \lstinline[basicstyle=\tiny\ttfamily]|/CMS_2010_S8656010/d01-x01-y03| & Charged hadron $p_\perp$ for $|\eta|=0.5$ at $\sqrt{s} = 7\,\mathrm{TeV}$ & 
    y-axis: $d^2N_\mathrm{ch}\,/d\eta\, dp_\perp\; [(\mathrm{GeV}/c)^{-1}]$ \newline
    x-axis: $p_\perp$ [GeV/$c$]
     & 1.0 \\
    \midrule 
    
    \lstinline[basicstyle=\tiny\ttfamily]|/CMS_2010_S8656010/d01-x01-y04| & Charged hadron $p_\perp$ for $|\eta|=0.7$ at $\sqrt{s} = 7\,\mathrm{TeV}$ & 
    y-axis: $d^2N_\mathrm{ch}\,/d\eta\, dp_\perp\; [(\mathrm{GeV}/c)^{-1}]$ \newline
    x-axis: $p_\perp$ [GeV/$c$]
     & 1.0 \\
    \midrule 
    
    \lstinline[basicstyle=\tiny\ttfamily]|/CMS_2010_S8656010/d02-x01-y01| & Charged hadron $p_\perp$ for $|\eta|=0.9$ at $\sqrt{s} = 7\,\mathrm{TeV}$ & 
    y-axis: $d^2N_\mathrm{ch}\,/d\eta\, dp_\perp\; [(\mathrm{GeV}/c)^{-1}]$ \newline
    x-axis: $p_\perp$ [GeV/$c$]
     & 1.0 \\
    \midrule 
    
    \lstinline[basicstyle=\tiny\ttfamily]|/CMS_2010_S8656010/d02-x01-y02| & Charged hadron $p_\perp$ for $|\eta|=1.1$ at $\sqrt{s} = 7\,\mathrm{TeV}$ & 
    y-axis: $d^2N_\mathrm{ch}\,/d\eta\, dp_\perp\; [(\mathrm{GeV}/c)^{-1}]$ \newline
    x-axis: $p_\perp$ [GeV/$c$]
     & 1.0 \\
    \midrule 
    
    \lstinline[basicstyle=\tiny\ttfamily]|/CMS_2010_S8656010/d02-x01-y03| & Charged hadron $p_\perp$ for $|\eta|=1.3$ at $\sqrt{s} = 7\,\mathrm{TeV}$ & 
    y-axis: $d^2N_\mathrm{ch}\,/d\eta\, dp_\perp\; [(\mathrm{GeV}/c)^{-1}]$ \newline
    x-axis: $p_\perp$ [GeV/$c$]
     & 1.0 \\
    \midrule 
    
    \lstinline[basicstyle=\tiny\ttfamily]|/CMS_2010_S8656010/d02-x01-y04| & Charged hadron $p_\perp$ for $|\eta|=1.5$ at $\sqrt{s} = 7\,\mathrm{TeV}$ & 
    y-axis: $d^2N_\mathrm{ch}\,/d\eta\, dp_\perp\; [(\mathrm{GeV}/c)^{-1}]$ \newline
    x-axis: $p_\perp$ [GeV/$c$]
     & 1.0 \\
    \midrule 
    
    \lstinline[basicstyle=\tiny\ttfamily]|/CMS_2010_S8656010/d03-x01-y01| & Charged hadron $p_\perp$ for $|\eta|=1.7$ at $\sqrt{s} = 7\,\mathrm{TeV}$ & 
    y-axis: $d^2N_\mathrm{ch}\,/d\eta\, dp_\perp\; [(\mathrm{GeV}/c)^{-1}]$ \newline
    x-axis: $p_\perp$ [GeV/$c$]
     & 1.0 \\
    \midrule 
    
    \lstinline[basicstyle=\tiny\ttfamily]|/CMS_2010_S8656010/d03-x01-y02| & Charged hadron $p_\perp$ for $|\eta|=1.9$ at $\sqrt{s} = 7\,\mathrm{TeV}$ & 
    y-axis: $d^2N_\mathrm{ch}\,/d\eta\, dp_\perp\; [(\mathrm{GeV}/c)^{-1}]$ \newline
    x-axis: $p_\perp$ [GeV/$c$]
     & 1.0 \\
    \midrule 
    
    \lstinline[basicstyle=\tiny\ttfamily]|/CMS_2010_S8656010/d03-x01-y03| & Charged hadron $p_\perp$ for $|\eta|=2.1$ at $\sqrt{s} = 7\,\mathrm{TeV}$ & 
    y-axis: $d^2N_\mathrm{ch}\,/d\eta\, dp_\perp\; [(\mathrm{GeV}/c)^{-1}]$ \newline
    x-axis: $p_\perp$ [GeV/$c$]
     & 1.0 \\
    \midrule 
    
    \lstinline[basicstyle=\tiny\ttfamily]|/CMS_2010_S8656010/d03-x01-y04| & Charged hadron $p_\perp$ for $|\eta|=2.3$ at $\sqrt{s} = 7\,\mathrm{TeV}$ & 
    y-axis: $d^2N_\mathrm{ch}\,/d\eta\, dp_\perp\; [(\mathrm{GeV}/c)^{-1}]$ \newline
    x-axis: $p_\perp$ [GeV/$c$]
     & 1.0 \\
    \midrule 
    
    \lstinline[basicstyle=\tiny\ttfamily]|/CMS_2010_S8656010/d04-x01-y01|
    & Charged hadron $p_\perp$ for $|\eta|<2.4$ at $\sqrt{s} = 7\,\mathrm{TeV}$
    & y-axis: $(1/2\pi p_\perp)\, d^2N_\mathrm{ch}/d\eta\, dp_\perp$ $[(\mathrm{GeV}/c)^{-2}]$ \newline
      x-axis: $p_\perp$ [GeV/$c$]
    & 1.0 \\
    \midrule
    
    \lstinline[basicstyle=\tiny\ttfamily]|/CMS_2010_S8656010/d05-x01-y01|
    & Charged hadron $\eta$ integrated over $p_\perp$ at $\sqrt{s} = 7\,\mathrm{TeV}$
    & y-axis: $dN_\mathrm{ch}\,/d\eta$ \newline
      x-axis: $\eta$
    & 6.0 \\
    \midrule
    
    \lstinline[basicstyle=\tiny\ttfamily]|/CMS_2011_S8884919/d25-x01-y01|
    & Mean $p_\perp$ vs charged hadron multiplicity, $|\eta| < 2.4$, $\sqrt{s} = 7\,\mathrm{TeV}$
    & y-axis: $\langle p_\perp \rangle$ [GeV] \newline
      x-axis: n
    & 8.0 \\
    \bottomrule    
\end{xltabular} 
}
\normalsize
\FloatBarrier

\clearpage
\bibliographystyle{JHEP}
\bibliography{bibliography}

@article{Bali:2000un,
    author = "Bali, Gunnar Singh",
    title = "{Casimir scaling of SU(3) static potentials}",
    eprint = "hep-lat/0006022",
    archivePrefix = "arXiv",
    reportNumber = "GUTPA-00-06-02",
    doi = "10.1103/PhysRevD.62.114503",
    journal = "Phys. Rev. D",
    volume = "62",
    pages = "114503",
    year = "2000"
}

@article{Sjostrand:2002ip,
    author = "Sj{\"o}strand, T. and Skands, Peter",
    title = "{Baryon number violation and string topologies}",
    eprint = "hep-ph/0212264",
    archivePrefix = "arXiv",
    reportNumber = "LU-TP-02-46",
    doi = "10.1016/S0550-3213(03)00193-7",
    journal = "Nucl. Phys. B",
    volume = "659",
    pages = "243",
    year = "2003"
}

@article{Altmann:2024odn,
    author = "Altmann, Javira and Skands, Peter",
    title = "{String junctions revisited}",
    eprint = "2404.12040",
    archivePrefix = "arXiv",
    primaryClass = "hep-ph",
    doi = "10.1007/JHEP07(2024)238",
    journal = "JHEP",
    volume = "07",
    pages = "238",
    year = "2024"
}

@article{Christiansen:2015yqa,
    author = "Christiansen, Jesper R. and Skands, Peter",
    title = "{String Formation Beyond Leading Colour}",
    eprint = "1505.01681",
    archivePrefix = "arXiv",
    primaryClass = "hep-ph",
    reportNumber = "COEPP-MN-15-1, LU-TP-15-16, MCNET-15-09",
    doi = "10.1007/JHEP08(2015)003",
    journal = "JHEP",
    volume = "08",
    pages = "003",
    year = "2015"
}

@article{Altmann:2024kwx,
    author = "Altmann, J. and Dubla, A. and Greco, V. and Rossi, A. and Skands, P.",
    title = "{Towards the understanding of heavy quarks hadronization: from leptonic to heavy-ion collisions}",
    eprint = "2405.19137",
    archivePrefix = "arXiv",
    primaryClass = "hep-ph",
    doi = "10.1140/epjc/s10052-024-13641-5",
    journal = "Eur. Phys. J. C",
    volume = "85",
    number = "1",
    pages = "16",
    year = "2025"
}

@article{Nayak:2005pf,
    author = "Nayak, Gouranga C.",
    title = "{Non-perturbative quark-antiquark production from a constant chromo-electric field via the Schwinger mechanism}",
    eprint = "hep-ph/0510052",
    archivePrefix = "arXiv",
    reportNumber = "YITP-SB-05-33",
    doi = "10.1103/PhysRevD.72.125010",
    journal = "Phys. Rev. D",
    volume = "72",
    pages = "125010",
    year = "2005"
}

@article{Schwinger:1951nm,
    author = "Schwinger, Julian S.",
    editor = "Milton, K. A.",
    title = "{On gauge invariance and vacuum polarization}",
    doi = "10.1103/PhysRev.82.664",
    journal = "Phys. Rev.",
    volume = "82",
    pages = "664--679",
    year = "1951"
}

@article{CMS:2011jlm,
    author = "Khachatryan, Vardan and others",
    collaboration = "CMS",
    title = "{Strange Particle Production in $pp$ Collisions at $\sqrt{s}=0.9$ and 7 TeV}",
    eprint = "1102.4282",
    archivePrefix = "arXiv",
    primaryClass = "hep-ex",
    reportNumber = "CERN-PH-EP-2010-094, CMS-QCD-10-007",
    doi = "10.1007/JHEP05(2011)064",
    journal = "JHEP",
    volume = "05",
    pages = "064",
    year = "2011"
}

@article{ALICE:2021rzj,
    author = "Acharya, Shreyasi and others",
    collaboration = "ALICE",
    title = "{Measurement of prompt D$^{0}$, $\Lambda_{c}^{+}$, and $\Sigma_{c}^{0,++}$(2455) production in pp collisions at $\sqrt{s} = 13$ TeV}",
    eprint = "2106.08278",
    archivePrefix = "arXiv",
    primaryClass = "hep-ex",
    reportNumber = "CERN-EP-2021-103",
    month = "6",
    year = "2021"
}

@article{LHCb:2021xyh,
    author = "Aaij, Roel and others",
    collaboration = "LHCb",
    title = "{Observation of a $\Lambda_b^0-\overline{\Lambda}_b^0$ production asymmetry in proton-proton collisions at $\sqrt{s} = 7 \textrm{ and } 8\,\textrm{TeV}$}",
    eprint = "2107.09593",
    archivePrefix = "arXiv",
    primaryClass = "hep-ex",
    reportNumber = "LHCb-PAPER-2021-016, CERN-EP-2021-121",
    doi = "10.1007/JHEP10(2021)060",
    journal = "JHEP",
    volume = "10",
    pages = "060",
    year = "2021"
}

@article{ALICE:2021bli,
    author = "Acharya, Shreyasi and others",
    collaboration = "ALICE",
    title = "{Measurement of the cross sections of $\Xi^0_{\rm c}$ and $\Xi^+_{\rm c}$ baryons and branching-fraction ratio BR($\Xi^0_{\rm c} \rightarrow \Xi^-{\rm e}^+\nu_{\rm e}$)/BR($\Xi^0_{\rm c} \rightarrow \Xi^-\pi^+$) in pp collisions at 13 TeV}",
    eprint = "2105.05187",
    archivePrefix = "arXiv",
    primaryClass = "nucl-ex",
    reportNumber = "CERN-EP-2021-084",
    month = "5",
    year = "2021"
}

@article{CMS:2010tjh,
    author = "Khachatryan, Vardan and others",
    collaboration = "CMS",
    title = "{Transverse-momentum and pseudorapidity distributions of charged hadrons in $pp$ collisions at $\sqrt{s}=7$ TeV}",
    eprint = "1005.3299",
    archivePrefix = "arXiv",
    primaryClass = "hep-ex",
    reportNumber = "CERN-PH-EP-2010-009, CMS-QCD-10-006",
    doi = "10.1103/PhysRevLett.105.022002",
    journal = "Phys. Rev. Lett.",
    volume = "105",
    pages = "022002",
    year = "2010"
}

@article{CMS:2010qvf,
    author = "Khachatryan, Vardan and others",
    collaboration = "CMS",
    title = "{Charged Particle Multiplicities in $pp$ Interactions at $\sqrt{s}=0.9$, 2.36, and 7 TeV}",
    eprint = "1011.5531",
    archivePrefix = "arXiv",
    primaryClass = "hep-ex",
    reportNumber = "CERN-PH-EP-2010-048, CMS-QCD-10-004",
    doi = "10.1007/JHEP01(2011)079",
    journal = "JHEP",
    volume = "01",
    pages = "079",
    year = "2011"
}

@article{Komargodski:2024swh,
    author = "Komargodski, Zohar and Zhong, Siwei",
    title = "{Baryon junction and string interactions}",
    eprint = "2405.12005",
    archivePrefix = "arXiv",
    primaryClass = "hep-th",
    doi = "10.1103/PhysRevD.110.056018",
    journal = "Phys. Rev. D",
    volume = "110",
    number = "5",
    pages = "056018",
    year = "2024"
}

@article{Buckley:2009bj,
    author = "Buckley, Andy and Hoeth, Hendrik and Lacker, Heiko and Schulz, Holger and von Seggern, Jan Eike",
    title = "{Systematic event generator tuning for the LHC}",
    eprint = "0907.2973",
    archivePrefix = "arXiv",
    primaryClass = "hep-ph",
    reportNumber = "IPPP-09-52, DCPT-104-22, LU-TP-09-18, HU-EP-09-33, MCNET-09-14",
    doi = "10.1140/epjc/s10052-009-1196-7",
    journal = "Eur. Phys. J. C",
    volume = "65",
    pages = "331--357",
    year = "2010"
}

@article{Rivet3,
    author = "Bierlich, Christian and others",
    title = "{Robust Independent Validation of Experiment and Theory: Rivet version 3}",
    eprint = "1912.05451",
    archivePrefix = "arXiv",
    primaryClass = "hep-ph",
    reportNumber = "MCnet-19-26",
    doi = "10.21468/SciPostPhys.8.2.026",
    journal = "SciPost Phys.",
    volume = "8",
    pages = "026",
    year = "2020"
}

@article{Karneyeu:2013aha,
    author = "Karneyeu, A. and Mijovic, L. and Prestel, S. and Skands, P.",
    title = "{MCPLOTS: a particle physics resource based on volunteer computing}",
    eprint = "1306.3436",
    archivePrefix = "arXiv",
    primaryClass = "hep-ph",
    reportNumber = "CERN-PH-TH-2013-105, DESY-13-104, LU-TP-13-23, NSF-KITP-13-.116",
    doi = "10.1140/epjc/s10052-014-2714-9",
    journal = "Eur. Phys. J. C",
    volume = "74",
    pages = "2714",
    year = "2014"
}

@article{Korneeva:2024oho,
    author = "Korneeva, Natalia and Karneyeu, Anton and Skands, Peter",
    title = "{Event-generator validation with MCPLOTS and LHC@home}",
    eprint = "2401.10621",
    archivePrefix = "arXiv",
    primaryClass = "hep-ph",
    doi = "10.1140/epjp/s13360-024-05353-2",
    journal = "Eur. Phys. J. Plus",
    volume = "139",
    number = "7",
    pages = "653",
    year = "2024"
}

@article{Werner:2007bf,
    author = "Werner, Klaus",
    title = "{Core-corona separation in ultra-relativistic heavy ion collisions}",
    eprint = "0704.1270",
    archivePrefix = "arXiv",
    primaryClass = "nucl-th",
    doi = "10.1103/PhysRevLett.98.152301",
    journal = "Phys. Rev. Lett.",
    volume = "98",
    pages = "152301",
    year = "2007"
}

@article{Pierog:2013ria,
    author = "Pierog, T. and Karpenko, Iu. and Katzy, J. M. and Yatsenko, E. and Werner, K.",
    title = "{EPOS LHC: Test of collective hadronization with data measured at the CERN Large Hadron Collider}",
    eprint = "1306.0121",
    archivePrefix = "arXiv",
    primaryClass = "hep-ph",
    reportNumber = "DESY-13-125",
    doi = "10.1103/PhysRevC.92.034906",
    journal = "Phys. Rev. C",
    volume = "92",
    number = "3",
    pages = "034906",
    year = "2015"
}

@article{Pierog:2019opp,
    author = "Pierog, Tanguy and Guiot, Benjamin and Karpenko, Iurii and Sophys, Gabriel and Stefaniak, Maria and Werner, Klaus",
    editor = "Lhenry-Yvon, I. and Biteau, J. and Biteau, O. and Ghia, P.",
    title = "{EPOS 3 and Air Showers}",
    doi = "10.1051/epjconf/201921002008",
    journal = "EPJ Web Conf.",
    volume = "210",
    pages = "02008",
    year = "2019"
}

@article{ALICE:2023sgl,
    author = "Acharya, Shreyasi and others",
    collaboration = "ALICE",
    title = "{Charm production and fragmentation fractions at midrapidity in pp collisions at $ \sqrt{\textrm{s}} $ = 13 TeV}",
    eprint = "2308.04877",
    archivePrefix = "arXiv",
    primaryClass = "hep-ex",
    reportNumber = "CERN-EP-2023-162",
    doi = "10.1007/JHEP12(2023)086",
    journal = "JHEP",
    volume = "12",
    pages = "086",
    year = "2023"
}

@article{Sjostrand:2004pf,
    author = "Sj{\"o}strand, T. and Skands, Peter",
    title = "{Multiple interactions and the structure of beam remnants}",
    eprint = "hep-ph/0402078",
    archivePrefix = "arXiv",
    reportNumber = "LU-TP-04-08",
    doi = "10.1088/1126-6708/2004/03/053",
    journal = "JHEP",
    volume = "03",
    pages = "053",
    year = "2004"
}

@article{Fieg:2023kld,
    author = {Fieg, Max and Kling, Felix and Schulz, Holger and Sj{\"o}strand, Torbj{\"o}rn},
    title = "{Tuning PYTHIA for forward physics experiments}",
    eprint = "2309.08604",
    archivePrefix = "arXiv",
    primaryClass = "hep-ph",
    reportNumber = "DESY-23-133",
    doi = "10.1103/PhysRevD.109.016010",
    journal = "Phys. Rev. D",
    volume = "109",
    number = "1",
    pages = "016010",
    year = "2024"
}

@article{Skands:2010ak,
      author         = "Skands, Peter",
      title          = "{Tuning Monte Carlo Generators: The Perugia Tunes}",
      journal        = "Phys. Rev.",
      volume         = "D82",
      year           = "2010",
      pages          = "074018",
      doi            = "10.1103/PhysRevD.82.074018",
      eprint         = "1005.3457",
      archivePrefix  = "arXiv",
      primaryClass   = "hep-ph",
      reportNumber   = "MCNET-10-08, CERN-PH-TH-2010-113",
      SLACcitation   = "%%CITATION = ARXIV:1005.3457;%%"
}

@article{LHCb:2023wbo,
    author = "Aaij, Roel and others",
    collaboration = "LHCb",
    title = "{Enhanced Production of $\Lambda^0_{\rm b}$ Baryons in High-Multiplicity pp Collisions at $\sqrt{s}$ = 13\,\,TeV}",
    eprint = "2310.12278",
    archivePrefix = "arXiv",
    primaryClass = "hep-ex",
    reportNumber = "LHCb-PAPER-2023-027, CERN-EP-2023-208",
    doi = "10.1103/PhysRevLett.132.081901",
    journal = "Phys. Rev. Lett.",
    volume = "132",
    number = "8",
    pages = "081901",
    year = "2024"
}

@article{Andersson:2001yu,
    author = "Andersson, Bo and Mohanty, Sandipan and S{\"o}derberg, Fredrik",
    title = "{The Lund fragmentation process for a multigluon string according to the area law}",
    eprint = "hep-ph/0106185",
    archivePrefix = "arXiv",
    reportNumber = "LU-TP-01-22",
    doi = "10.1007/s100520100757",
    journal = "Eur. Phys. J. C",
    volume = "21",
    pages = "631--647",
    year = "2001"
}

@article{Andersson:1983ia,
    author = "Andersson, Bo and Gustafson, G. and Ingelman, G. and Sj{\"o}strand, T.",
    title = "{Parton Fragmentation and String Dynamics}",
    reportNumber = "LU-TP-83-10",
    doi = "10.1016/0370-1573(83)90080-7",
    journal = "Phys. Rept.",
    volume = "97",
    pages = "31--145",
    year = "1983"
}

@article{TheATLAScollaboration:2014rfk,
    collaboration = "ATLAS",
    title = "{ATLAS Pythia 8 tunes to 7 TeV data}",
    journal = "ATL-PHYS-PUB-2014-021",
    month = "11",
    year = "2014"
}

@article{CMS:2022awf,
    author = "Tumasyan, Armen and others",
    collaboration = "CMS",
    title = "{CMS Pythia~8 colour reconnection tunes based on underlying-event data}",
    eprint = "2205.02905",
    archivePrefix = "arXiv",
    primaryClass = "hep-ex",
    reportNumber = "CMS-GEN-17-002, CERN-EP-2022-005",
    doi = "10.1140/epjc/s10052-023-11630-8",
    journal = "Eur. Phys. J. C",
    volume = "83",
    number = "7",
    pages = "587",
    year = "2023"
}

@article{Bierlich:2022pfr,
    author = "Bierlich, Christian and others",
    title = "{A comprehensive guide to the physics and usage of PYTHIA 8.3}",
    eprint = "2203.11601",
    archivePrefix = "arXiv",
    primaryClass = "hep-ph",
    reportNumber = "LU-TP 22-16, MCNET-22-04, FERMILAB-PUB-22-227-SCD",
    doi = "10.21468/SciPostPhysCodeb.8",
    journal = "SciPost Phys. Codeb.",
    volume = "2022",
    pages = "8",
    year = "2022"
}

@article{Sjostrand:2006za,
      author         = "Sj{\"o}strand, Torbjorn and Mrenna, Stephen and Skands, Peter
                        Z.",
      title          = "{PYTHIA 6.4 Physics and Manual}",
      journal        = "JHEP",
      volume         = "05",
      year           = "2006",
      pages          = "026",
      doi            = "10.1088/1126-6708/2006/05/026",
      eprint         = "hep-ph/0603175",
      archivePrefix  = "arXiv",
      primaryClass   = "hep-ph",
      reportNumber   = "FERMILAB-PUB-06-052-CD-T, LU-TP-06-13",
      SLACcitation   = "%%CITATION = HEP-PH/0603175;%%"
}

@article{Sjostrand:1987su,
    author = "Sj{\"o}strand, Torbjorn and van Zijl, Maria",
    title = "{A Multiple Interaction Model for the Event Structure in Hadron Collisions}",
    reportNumber = "LU-TP-87-5",
    doi = "10.1103/PhysRevD.36.2019",
    journal = "Phys. Rev. D",
    volume = "36",
    pages = "2019",
    year = "1987"
}

@article{Sjostrand:2004ef,
    author = "Sj{\"o}strand, T. and Skands, Peter",
    title = "{Transverse-momentum-ordered showers and interleaved multiple interactions}",
    eprint = "hep-ph/0408302",
    archivePrefix = "arXiv",
    reportNumber = "LU-TP-04-29",
    doi = "10.1140/epjc/s2004-02084-y",
    journal = "Eur. Phys. J. C",
    volume = "39",
    pages = "129--154",
    year = "2005"
}

@article{Skands:2014pea,
    author = "Skands, Peter and Carrazza, Stefano and Rojo, Juan",
    title = "{Tuning PYTHIA 8.1: the Monash 2013 Tune}",
    eprint = "1404.5630",
    archivePrefix = "arXiv",
    primaryClass = "hep-ph",
    reportNumber = "CERN-PH-TH-2014-069, MCNET-14-08, OUTP-14-05P",
    doi = "10.1140/epjc/s10052-014-3024-y",
    journal = "Eur. Phys. J. C",
    volume = "74",
    number = "8",
    pages = "3024",
    year = "2014"
}

@article{ALICE:2023wbx,
    author = "Acharya, Shreyasi and others",
    collaboration = "ALICE",
    title = "{Study of flavor dependence of the baryon-to-meson ratio in proton-proton collisions at $\sqrt{s}$ = 13 TeV}",
    eprint = "2308.04873",
    archivePrefix = "arXiv",
    primaryClass = "hep-ex",
    reportNumber = "CERN-EP-2023-159",
    doi = "10.1103/PhysRevD.108.112003",
    journal = "Phys. Rev. D",
    volume = "108",
    number = "11",
    pages = "112003",
    year = "2023"
}

@article{Hamacher:1995df,
    author = "Hamacher, Klaus and Weierstall, Michael",
    collaboration = "DELPHI",
    title = "{The Next round of hadronic generator tuning heavily based on identified particle data}",
    eprint = "hep-ex/9511011",
    archivePrefix = "arXiv",
    reportNumber = "WU-B-95-07, DELPHI-95-80-PHYS-515",
    month = "6",
    journal = "",
    year = "1995"
}

@article{Jueid:2023vrb,
    author = "Jueid, Adil and Kip, Jochem and de Austri, Roberto Ruiz and Skands, Peter",
    title = "{The Strong Force meets the Dark Sector: a robust estimate of QCD uncertainties for anti-matter dark matter searches}",
    eprint = "2303.11363",
    archivePrefix = "arXiv",
    primaryClass = "hep-ph",
    reportNumber = "CTPU-PTC-23-08",
    journal = "",
    month = "3",
    year = "2023"
}

@article{Amoroso:2018qga,
    author = "Amoroso, Simone and Caron, Sascha and Jueid, Adil and Ruiz de Austri, Roberto and Skands, Peter",
    title = "{Estimating QCD uncertainties in Monte Carlo event generators for gamma-ray dark matter searches}",
    eprint = "1812.07424",
    archivePrefix = "arXiv",
    primaryClass = "hep-ph",
    doi = "10.1088/1475-7516/2019/05/007",
    journal = "JCAP",
    volume = "05",
    pages = "007",
    year = "2019"
}

@article{Aamodt:2008zz,
      author         = "Aamodt, K. and others",
      title          = "{The ALICE experiment at the CERN LHC}",
      collaboration  = "ALICE",
      journal        = "JINST",
      volume         = "3",
      year           = "2008",
      pages          = "S08002",
      doi            = "10.1088/1748-0221/3/08/S08002",
      SLACcitation   = "%%CITATION = JINST,3,S08002;%%"
}

@article{Eden:1996xi,
    author = "Eden, Patrik and Gustafson, Gosta",
    title = "{Baryon production in the string fragmentation picture}",
    eprint = "hep-ph/9606454",
    archivePrefix = "arXiv",
    reportNumber = "LU-TP-96-20",
    doi = "10.1007/s002880050445",
    journal = "Z. Phys. C",
    volume = "75",
    pages = "41--49",
    year = "1997"
}

@article{Andersson:1984af,
    author = "Andersson, Bo and Gustafson, G. and Sj{\"o}strand, T.",
    title = "{Baryon Production in Jet Fragmentation and $\Upsilon$ Decay}",
    reportNumber = "LU-TP-84-9",
    doi = "10.1088/0031-8949/32/6/003",
    journal = "Phys. Scripta",
    volume = "32",
    pages = "574",
    year = "1985"
}

@article{ALICE:2017jyt,
    author = "Adam, Jaroslav and others",
    collaboration = "ALICE",
    title = "{Enhanced production of multi-strange hadrons in high-multiplicity proton-proton collisions}",
    eprint = "1606.07424",
    archivePrefix = "arXiv",
    primaryClass = "nucl-ex",
    reportNumber = "CERN-EP-2016-153",
    doi = "10.1038/nphys4111",
    journal = "Nature Phys.",
    volume = "13",
    pages = "535--539",
    year = "2017"
}

@article{ALICE:2018pal,
    author = "Acharya, Shreyasi and others",
    collaboration = "ALICE",
    title = "{Multiplicity dependence of light-flavor hadron production in pp collisions at $\sqrt{s}$ = 7 TeV}",
    eprint = "1807.11321",
    archivePrefix = "arXiv",
    primaryClass = "nucl-ex",
    reportNumber = "CERN-EP-2018-209",
    doi = "10.1103/PhysRevC.99.024906",
    journal = "Phys. Rev. C",
    volume = "99",
    number = "2",
    pages = "024906",
    year = "2019"
}

@article{Bierlich:2017sxk,
    author = "Bierlich, Christian",
    editor = "Mischke, A. and Kuijer, P.",
    title = "{Rope Hadronization and Strange Particle Production}",
    eprint = "1710.04464",
    archivePrefix = "arXiv",
    primaryClass = "nucl-th",
    reportNumber = "LU-TP-17-31, MCNET-17-15",
    doi = "10.1051/epjconf/201817114003",
    journal = "EPJ Web Conf.",
    volume = "171",
    pages = "14003",
    year = "2018"
}

@article{Fischer:2016zzs,
    author = {Fischer, Nadine and Sj\"ostrand, Torbj\"orn},
    title = "{Thermodynamical String Fragmentation}",
    eprint = "1610.09818",
    archivePrefix = "arXiv",
    primaryClass = "hep-ph",
    reportNumber = "LU-TP-16-57, COEPP-MN-16-25, MCNET-16-40",
    doi = "10.1007/JHEP01(2017)140",
    journal = "JHEP",
    volume = "01",
    pages = "140",
    year = "2017"
}

@article{Bierlich:2016vgw,
    author = {Bierlich, Christian and Gustafson, G\"osta and L\"onnblad, Leif},
    title = "{A shoving model for collectivity in hadronic collisions}",
    eprint = "1612.05132",
    archivePrefix = "arXiv",
    primaryClass = "hep-ph",
    journal = "",
    reportNumber = "MCNET-16-48, LU-TP-16-64",
    month = "12",
    year = "2016"
}

@article{Bierlich:2020naj,
    author = {Bierlich, Christian and Chakraborty, Smita and Gustafson, G\"osta and L\"onnblad, Leif},
    title = "{Setting the string shoving picture in a new frame}",
    eprint = "2010.07595",
    archivePrefix = "arXiv",
    primaryClass = "hep-ph",
    reportNumber = "LU-TP 20-48, MCnet-20-22",
    doi = "10.1007/JHEP03(2021)270",
    journal = "JHEP",
    volume = "03",
    pages = "270",
    year = "2021"
}

@article{Bierlich:2024odg,
    author = "Bierlich, Christian",
    title = "{String Interactions as a Source of Collective Behaviour}",
    eprint = "2401.07585",
    archivePrefix = "arXiv",
    primaryClass = "hep-ph",
    reportNumber = "MCnet-24-01",
    doi = "10.3390/universe10010046",
    journal = "Universe",
    volume = "10",
    number = "1",
    pages = "46",
    year = "2024"
}

@article{Eichten:1978tg,
    author = "Eichten, E. and Gottfried, K. and Kinoshita, T. and Lane, K. D. and Yan, Tung-Mow",
    title = "{Charmonium: The Model}",
    reportNumber = "CLNS-375",
    doi = "10.1103/PhysRevD.17.3090",
    journal = "Phys. Rev. D",
    volume = "17",
    pages = "3090",
    year = "1978",
    note = "[Erratum: Phys.Rev.D 21, 313 (1980)]"
}

@article{Bali:1992ab,
    author = "Bali, G. S. and Schilling, K.",
    title = "{Static quark - anti-quark potential: Scaling behavior and finite size effects in SU(3) lattice gauge theory}",
    reportNumber = "WUB-92-02",
    doi = "10.1103/PhysRevD.46.2636",
    journal = "Phys. Rev. D",
    volume = "46",
    pages = "2636--2646",
    year = "1992"
}

@article{Sjostrand:1986ep,
    author = "Sj{\"o}strand, Torbjorn and van Zijl, Maria",
    title = "{Multiple Parton-parton Interactions in an Impact Parameter Picture}",
    reportNumber = "LU-TP-86-25",
    doi = "10.1016/0370-2693(87)90722-2",
    journal = "Phys. Lett. B",
    volume = "188",
    pages = "149--154",
    year = "1987"
}

@article{ALICE:2011ac,
    author = "Abelev, Betty and others",
    collaboration = "ALICE",
    title = "{Underlying Event measurements in $pp$ collisions at $\sqrt{s}=0.9$ and 7 TeV with the ALICE experiment at the LHC}",
    eprint = "1112.2082",
    archivePrefix = "arXiv",
    primaryClass = "hep-ex",
    reportNumber = "CERN-PH-EP-2011-204",
    doi = "10.1007/JHEP07(2012)116",
    journal = "JHEP",
    volume = "07",
    pages = "116",
    year = "2012"
}

@article{CDF:2004jod,
    author = "Acosta, D. and others",
    collaboration = "CDF",
    title = "{The underlying event in hard interactions at the Tevatron $\bar{p}p$ collider}",
    eprint = "hep-ex/0404004",
    archivePrefix = "arXiv",
    reportNumber = "FERMILAB-PUB-04-041-E",
    doi = "10.1103/PhysRevD.70.072002",
    journal = "Phys. Rev. D",
    volume = "70",
    pages = "072002",
    year = "2004"
}

@article{CMS:2010rux,
    author = "Khachatryan, Vardan and others",
    collaboration = "CMS",
    title = "{First Measurement of the Underlying Event Activity at the LHC with $\sqrt{s} = 0.9$ TeV}",
    eprint = "1006.2083",
    archivePrefix = "arXiv",
    primaryClass = "hep-ex",
    reportNumber = "CERN-PH-EP-2010-014, CMS-QCD-10-001",
    doi = "10.1140/epjc/s10052-010-1453-9",
    journal = "Eur. Phys. J. C",
    volume = "70",
    pages = "555--572",
    year = "2010"
}

@article{ATLAS:2010kmf,
    author = "Aad, Georges and others",
    collaboration = "ATLAS",
    title = "{Measurement of underlying event characteristics using charged particles in pp collisions at $\sqrt{s} = 900 GeV$ and 7 TeV with the ATLAS detector}",
    eprint = "1012.0791",
    archivePrefix = "arXiv",
    primaryClass = "hep-ex",
    reportNumber = "CERN-PH-EP-2010-063",
    doi = "10.1103/PhysRevD.83.112001",
    journal = "Phys. Rev. D",
    volume = "83",
    pages = "112001",
    year = "2011"
}

@article{Field:2000dy,
    author = "Field, R.",
    editor = "Gan, K. K. and Kass, R.",
    collaboration = "CDF",
    title = "{The Underlying Event in Large Transverse Momentum Charged Jet and Z-Boson Production}",
    reportNumber = "FERMILAB-CONF-00-289-E, CDF-5463",
    journal = "Int. J. Mod. Phys. A",
    volume = "16S1A",
    pages = "250--254",
    year = "2001"
}

@article{AxialFieldSpectrometer:1984fhw,
    author = "{\r{A}}kesson, T. and others",
    collaboration = "Axial Field Spectrometer",
    title = "{Properties of Jets in High $e(T$) Events Produced in $p p$ Collisions at $\sqrt{s}=63$-{GeV}}",
    reportNumber = "CERN-EP-84-56, BNL-34317",
    doi = "10.1007/BF01571952",
    journal = "Z. Phys. C",
    volume = "25",
    pages = "13",
    year = "1984"
}

@article{Buckley:2023xqh,
    author = "Buckley, Andy and Corpe, Louie and Filipovich, Matthew and Gutschow, Christian and Rozinsky, Nick and Thor, Simon and Yeh, Yoran and Yellen, Jamie",
    title = "{Consistent, multidimensional differential histogramming and summary statistics with YODA 2}",
    eprint = "2312.15070",
    journal = "SciPost Phys. Codeb.",
    archivePrefix = "arXiv",
    primaryClass = "hep-ph",
    reportNumber = "MCNET-23-21",
    doi = "10.21468/SciPostPhysCodeb.45",
    month = "12",
    year = "2023"
}

@article{Jueid:2022qjg,
    author = "Jueid, Adil and Kip, Jochem and de Austri, Roberto Ruiz and Skands, Peter",
    title = "{Impact of QCD uncertainties on antiproton spectra from dark-matter annihilation}",
    eprint = "2202.11546",
    archivePrefix = "arXiv",
    primaryClass = "hep-ph",
    reportNumber = "CTPU-PTC-22-22, KIAS-Q22004",
    doi = "10.1088/1475-7516/2023/04/068",
    journal = "JCAP",
    volume = "04",
    pages = "068",
    year = "2023"
}

@article{UA1:1983hhd,
    author = "Arnison, G. and others",
    collaboration = "UA1",
    title = "{Hadronic Jet Production at the CERN Proton - anti-Proton Collider}",
    reportNumber = "CERN-EP/83-118",
    doi = "10.1016/0370-2693(83)90254-X",
    journal = "Phys. Lett. B",
    volume = "132",
    pages = "214",
    year = "1983"
}

@article{UA5:1988osh,
    author = "Ansorge, R. E. and others",
    collaboration = "UA5",
    title = "{Charged Particle Correlations in $\bar{P} P$ Collisions at c.m. Energies of 200-{GeV}, 546-{GeV} and 900-{GeV}}",
    doi = "10.1007/BF01579906",
    journal = "Z. Phys. C",
    volume = "37",
    pages = "191--213",
    year = "1988"
}

@article{UA5:1985fid,
    author = "Alner, G. J. and others",
    collaboration = "UA5",
    title = "{Scaling Violations in Multiplicity Distributions at 200-GeV and 900-GeV}",
    reportNumber = "CERN-EP/85-197",
    doi = "10.1016/0370-2693(86)91304-3",
    journal = "Phys. Lett. B",
    volume = "167",
    pages = "476--480",
    year = "1986"
}

@article{Uhlig:1977dc,
    author = "Uhlig, S. and Derado, I. and Meinke, R. and Preissner, H.",
    title = "{Observation of Charged Particle Correlations Between the Forward and Backward Hemispheres in $p p$ Collisions at ISR Energies}",
    reportNumber = "MPI-PAE/Exp-El-66",
    doi = "10.1016/0550-3213(78)90254-7",
    journal = "Nucl. Phys. B",
    volume = "132",
    pages = "15--28",
    year = "1978"
}

@article{L3:2004cdh,
    author = "Achard, P. and others",
    collaboration = "L3",
    title = "{Studies of hadronic event structure in $e^{+} e^{-}$ annihilation from 30-GeV to 209-GeV with the L3 detector}",
    eprint = "hep-ex/0406049",
    archivePrefix = "arXiv",
    reportNumber = "CERN-PH-EP-2004-024, CERN-EP-PH-2004-024",
    doi = "10.1016/j.physrep.2004.07.002",
    journal = "Phys. Rept.",
    volume = "399",
    pages = "71--174",
    year = "2004"
}

@article{SLD:2003ogn,
    author = "Abe, Koya and others",
    collaboration = "SLD",
    title = "{Production of $\pi^+$, $\pi^-$, $K^+$, $K^-$, p and $\bar{\rm p}$                                                                                        in Light ($uds$), $c$ and $b$ Jets from $Z^0$ Decays}",
    eprint = "hep-ex/0310017",
    archivePrefix = "arXiv",
    reportNumber = "SLAC-PUB-9949",
    doi = "10.1103/PhysRevD.69.072003",
    journal = "Phys. Rev. D",
    volume = "69",
    pages = "072003",
    year = "2004"
}

@article{OPAL:1994zan,
    author = "Akers, R. and others",
    collaboration = "OPAL",
    title = "{Measurement of the production rates of charged hadrons in e+ e- annihilation at the Z0}",
    reportNumber = "CERN-PPE-94-49, CERN-PPE-94-049",
    doi = "10.1007/BF01411010",
    journal = "Z. Phys. C",
    volume = "63",
    pages = "181--196",
    year = "1994"
}

@article{DELPHI:2000ahn,
    author = "Abreu, P. and others",
    collaboration = "DELPHI",
    title = "{Charged and identified particles in the hadronic decay of W bosons and in e+ e- ---\ensuremath{>} q anti-q from 130-GeV to 200-GeV}",
    eprint = "hep-ex/0103031",
    archivePrefix = "arXiv",
    reportNumber = "CERN-EP-2000-023",
    doi = "10.1007/s100520000528",
    journal = "Eur. Phys. J. C",
    volume = "18",
    pages = "203--228",
    year = "2000",
    note = "[Erratum: Eur.Phys.J.C 25, 493 (2002)]"
}

@article{ALEPH:2003obs,
    author = "Heister, A. and others",
    collaboration = "ALEPH",
    title = "{Studies of QCD at e+ e- centre-of-mass energies between 91-GeV and 209-GeV}",
    reportNumber = "CERN-EP-2003-084",
    doi = "10.1140/epjc/s2004-01891-4",
    journal = "Eur. Phys. J. C",
    volume = "35",
    pages = "457--486",
    year = "2004"
}

@article{ParticleDataGroup:2024cfk,
    author = "Navas, S. and others",
    collaboration = "Particle Data Group",
    title = "{Review of particle physics}",
    doi = "10.1103/PhysRevD.110.030001",
    journal = "Phys. Rev. D",
    volume = "110",
    number = "3",
    pages = "030001",
    year = "2024"
}

@article{ATLAS:2010jvh,
    author = "Aad, G. and others",
    collaboration = "ATLAS",
    title = "{Charged-particle multiplicities in pp interactions measured with the ATLAS detector at the LHC}",
    eprint = "1012.5104",
    archivePrefix = "arXiv",
    primaryClass = "hep-ex",
    reportNumber = "CERN-PH-EP-2010-079",
    doi = "10.1088/1367-2630/13/5/053033",
    journal = "New J. Phys.",
    volume = "13",
    pages = "053033",
    year = "2011"
}

@article{LHCb:2011ioc,
    author = "Aaij, R. and others",
    collaboration = "LHCb",
    title = "{Measurement of $V^0$ production ratios in $pp$ collisions at $\sqrt{s} = 0.9$ and $7~\rm{TeV}$}",
    eprint = "1107.0882",
    archivePrefix = "arXiv",
    primaryClass = "hep-ex",
    reportNumber = "CERN-PH-EP-2011-082, LHCB-PAPER-2011-005",
    doi = "10.1007/JHEP08(2011)034",
    journal = "JHEP",
    volume = "08",
    pages = "034",
    year = "2011"
}

@article{ATLAS:2022xau,
    collaboration = "ATLAS",
    title = "{Dependence of the Jet Energy Scale on the Particle Content of Hadronic Jets in the ATLAS Detector Simulation}",
    reportNumber = "ATL-PHYS-PUB-2022-021",
    year = "2022"
}

@mastersthesis{LorenzoThesis,
  title        = {Tuning of {PYTHIA} event generator with
strange hadron measurements from {ALICE}
and test on strange jet production},
  author       = {Lorenzo Bernardinis},
  year         = 2024,
  school       = {Universit\`a degli studi di Trieste},
  type         = "{Master's thesis}"
}

\end{document}